\RequirePackage[l2tabu,orthodox]{nag}
\documentclass[a4paper,10pt,DIV=10]{scrartcl}
\pdfoutput=1
\usepackage[T1]{fontenc}
\usepackage[utf8]{inputenc}
\DeclareUnicodeCharacter{2212}{-}
\usepackage{mathtools}
\usepackage{microtype}
\usepackage{pdflscape}

\usepackage{lmodern}
\usepackage{fixmath}
\usepackage{textgreek}
\usepackage{amssymb}

\usepackage[main=english]{babel}
\usepackage{csquotes}
\usepackage[style=phys,articletitle=true,biblabel=brackets,eprint=true,sortcites=true,maxnames=5,backend=biber]{biblatex}
\usepackage{physics}
\usepackage{eqnarray}
\usepackage{booktabs}
\usepackage{multirow}
\usepackage{longtable}
\usepackage[nolist]{acronym}
\usepackage{siunitx}[2021-05-17]
\usepackage[usenames,svgnames,table]{xcolor}
\usepackage{tikz}
\usepackage[outline]{contour}
\usepackage{authblk}
\usepackage{adjustbox}  
\usepackage{csvsimple} 
\usepackage{hyperref}
\hypersetup{
  unicode=True,
  linkcolor=FireBrick,
  citecolor=ForestGreen,
  urlcolor=Navy,
}
\usepackage[ocgcolorlinks]{ocgx2}

\setkomafont{captionlabel}{\sffamily\bfseries}
\KOMAoption{toc}{index,bibliography,flat}

\NewDocumentCommand\cpi{}{\text{\textpi}}                        
\NewDocumentCommand\I{}{\mathrm{i}}
\NewDocumentCommand\e{ m }{\mathrm{e}^{#1}}
\DeclareMathOperator{\diag}{diag}
\NewDocumentCommand\Nf{}{N_\mathrm{f}}                        
\NewDocumentCommand\Nc{}{N_\mathrm{c}}                        
\NewDocumentCommand\SU{}{\mathrm{SU}}
\NewDocumentCommand\aHVP{}{a_\mu^{\mathrm{HVP,LO}}}
\NewDocumentCommand\Dhad{}{\Delta_{\mathrm{had}}}
\NewDocumentCommand\Dalphahad{}{\Delta\alpha_{\mathrm{had}}}
\NewDocumentCommand\DalphaIIhad{}{\Delta\alpha_{2,\mathrm{had}}}
\NewDocumentCommand\dalpfive{ m }{\Delta\alpha^{(5)}_{\mathrm{had}}(#1)}
\NewDocumentCommand\thetaW{}{\theta_{\mathrm{W}}}
\NewDocumentCommand\sIIW{}{\sin^2\thetaW}
\NewDocumentCommand\sIIWh{}{\sin^2\hat{\theta}_{\mathrm{W}}}
\NewDocumentCommand\SPi{}{\bar{\Pi}}
\NewDocumentCommand\tauint{}{\tau_{\mathrm{int}}}
\NewDocumentCommand\tcut{}{t_{\mathrm{cut}}}
\NewDocumentCommand\ti{}{t_i}
\NewDocumentCommand\tnotsym{}{t_0^{\mathrm{sym}}}

\NewDocumentCommand\con{}{\mathrm{con}}
\NewDocumentCommand\dis{}{\mathrm{dis}}
\NewDocumentCommand\tran{}{{\mkern-1.5mu\mathsf{T}}}
\NewDocumentCommand\impr{}{\mathrm{I}}
\NewDocumentCommand\ren{}{R}
\NewDocumentCommand\local{}{\mathrm{L}}
\NewDocumentCommand\cons{}{\mathrm{C}}
\NewDocumentCommand\locloc{}{\mathrm{LL}}
\NewDocumentCommand\consloc{}{\mathrm{CL}}
\NewDocumentCommand\ie{}{i.e.}
\NewDocumentCommand\eg{}{e.g.}

\makeatletter
\newcommand*{\textoverline}[1]{$\overline{\hbox{#1}}\m@th$}
\makeatother
\NewDocumentCommand\MSbar{}{\textoverline{\scshape ms}}
\DeclareSIUnit\fm{\femto\metre}

\usetikzlibrary{decorations.markings}
\usetikzlibrary{decorations.pathreplacing}
\usetikzlibrary{patterns}
\usetikzlibrary{fadings}

\tikzset{%
  ->-/.style={decoration={markings,mark=at position #1 with {\arrow{>}}},postaction={decorate}},
  ->-/.default=0.5
}

\def\pqcdptxt{{pQCD\textsuperscript{$\prime$}}}
\def\pqcdpeq{\mathrm{\pqcdptxt}}

\DeclareSourcemap{
  \maps[datatype=bibtex]{
    \map[overwrite=true]{
      \step[fieldset=abstract, null]
    }
  }
}
\begin{acronym}
  \acro{SM}{Standard Model}
  \acro{BSM}{beyond the Standard Model}
  \acro{QFT}{quantum field theory}
  \acro{QED}{quantum electrodynamics}
  \acro{QCD}{quantum chromodynamics}
  \acro{RG}{renormalization group}
  \acro{MC}{Monte Carlo}
  \acro{MDU}{molecular dynamics unit}
  \acro{HVP}{hadronic vacuum polarization}
  \acro{HLbL}{hadronic light-by-light scattering}
  \acro{TMR}{time-momentum representation}
  \acro{CCS}{Lorentz-covariant coordinate space}
  \acro{FFT}{fast Fourier transform}
  \acro{CLS}{Coordinated Lattice Simulations}
  \acro{BC}{boundary condition}
  \acro{OBC}{open boundary condition}
  \acro{PBC}{periodic boundary condition}
  \acro{GS}{Gounaris-Sakurai}
  \acro{SN}[S/N]{signal-to-noise ratio}
  \acro{chiPT}[\textchi PT]{chiral perturbation theory}
  \acro{NLO}{next-to-leading order}
  \acro{NNLO}[N\textsuperscript{2}LO]{next-to-next-to-leading order}
  \acro{HP}{Hansen-Patella}
	\acro{FSE}{finite-size effects}
	\acro{MLLGS}[MLL-GS]{Meyer-Lellouch-Lüscher Gounaris-Sakurai}
	\acro{SM}{Standard Model}
	\acro{BMWc}{Budapest-Marseille-Wuppertal collaboration}
\end{acronym}

\title{The hadronic running of the electromagnetic coupling and the
electroweak mixing angle from lattice QCD}

\author[a,b]{Marco~Cè}
\author[c]{Antoine~Gérardin}
\author[d]{Georg~von~Hippel}
\author[d,e,f]{Harvey~B.~Meyer}
\author[d,e,f,g]{Kohtaroh~Miura}
\author[d]{Konstantin~Ottnad}
\author[h]{Andreas~Risch}
\author[d,e,f]{Teseo~San~José}
\author[d]{Jonas~Wilhelm}
\author[d,e,f,b]{Hartmut~Wittig}
\affil[a]{Albert Einstein Center for Fundamental Physics (AEC) and Institut für Theoretische Physik, Universität Bern, Sidlerstrasse 5, 3012 Bern, Switzerland}
\affil[b]{Department of Theoretical Physics, CERN, 1211 Geneva 23, Switzerland}
\affil[c]{Aix-Marseille Université, Université de Toulon, CNRS, CPT, Marseille, France}
\affil[d]{PRISMA\textsuperscript{+} Cluster of Excellence and Institut für Kernphysik, Johannes Gutenberg-Universität Mainz, 55099 Mainz, Germany}
\affil[e]{Helmholtz-Institut Mainz, Johannes Gutenberg-Universität Mainz, 55099 Mainz, Germany}
\affil[f]{GSI Helmholtzzentrum für Schwerionenforschung, Planckstraße 1, 64291 Darmstadt, Germany}
\affil[g]{Kobayashi-Maskawa Institute for the Origin of Particles and the Universe, Nagoya University, Japan}
\affil[h]{John von Neumann-Institut für Computing NIC, Deutsches Elektronen-Synchrotron DESY, Platanenallee 6, 15738 Zeuthen, Germany}
\date{}

\begin{document}

\maketitle

\begin{abstract}
  We compute the hadronic running of the electromagnetic and weak
couplings in lattice QCD with $\Nf=2+1$ flavors of $\order{a}$
improved Wilson fermions. Using two different discretizations of the
vector current, we compute the quark-connected and -disconnected
contributions to the \ac{HVP} functions $\SPi^{\gamma\gamma}$ and
$\SPi^{\gamma Z}$ for Euclidean squared momenta
$Q^2\leq\SI{7}{\GeV\squared}$. Gauge field ensembles at four values of
the lattice spacing and several values of the pion mass, including its
physical value, are used to extrapolate the results to the physical
point. The ability to perform an exact flavor decomposition allows us
to present the most precise determination to date of the
$\SU(3)$-flavor-suppressed \ac{HVP} function $\SPi^{08}$ that enters the running
of $\sIIW$. Our results for $\SPi^{\gamma\gamma}$, $\SPi^{\gamma Z}$
and $\SPi^{08}$ are presented in terms of rational functions for
continuous values of $Q^2$ below \SI{7}{\GeV\squared}.
We observe a tension of up to $\num{3.5}$ standard deviation between
our lattice results for $\dalpfive{-Q^2}$ and estimates based on the
$R$-ratio for space-like momenta in the range
\num{3}--\SI{7}{\GeV\squared}. The tension is, however, strongly
diminished when translating our result to the $Z$ pole, by employing
the Euclidean split technique and perturbative QCD, which yields
$\dalpfive{M_Z^2}=\num{0.02773(15)}$ and agrees with results based on
the $R$-ratio within the quoted uncertainties.

\vspace*{0.5cm}
\begin{flushright}
  MITP-22-019\\
  CERN-TH-2022-035\\
  DESY-22-050
\end{flushright}
\end{abstract}

\cleardoublepage
\DeclareTOCStyleEntry[beforeskip=1.5ex]{section}{section}
\tableofcontents

\cleardoublepage
\acresetall\acused{QCD}
\section{Introduction}

Precision observables play a crucial role in the search for physics
\ac{BSM}. They allow for exploring the limits of the \ac{SM} and
constrain possible extensions in a way that is complementary to direct
searches in experiments at high-energy colliders. This requires
both the theoretical prediction and the corresponding
experimental result to be determined to a high level of precision. The
evaluation of the \ac{SM} prediction is particularly challenging when
the quantity of interest receives significant contributions from
hadronic effects. Indeed, because of the growth of the strong coupling
in the low-energy domain, perturbative methods fail to describe the
strong interactions at typical hadronic scales, in contrast to the
high-energy regime and the electroweak sector. Lattice \ac{QCD} has
emerged as one of the leading methods to compute these non-perturbative \ac{QCD}
contributions from first principles. Lattice calculations have reached
sub-percent precision for many observables that are now routinely used
in precision tests of the \ac{SM}~\cite{Aoki:2019cca,Aoki:2021kgd}.

The prominent example of the muon anomalous magnetic moment, $a_\mu$,
is an apt illustration of the importance of precision observables. The
measurement of $a_\mu$ by the E989 experiment at
Fermilab~\cite{Abi:2021gix}, when combined with the earlier
experimental determination at BNL~\cite{Bennett:2006fi}, produces a
tension of 4.2$\sigma$ with the theoretical prediction summarized in
the 2020 White Paper~\cite{Aoyama:2020ynm} by the \emph{Muon $g-2$
Theory Initiative}. Given that the uncertainty of the \ac{SM}
prediction is dominated by the \ac{HVP} and, to a lesser extent, the
\ac{HLbL} contribution, it is clear that efforts to reduce the
theoretical error must focus on hadronic effects.
In fact, the recent lattice calculation of the HVP contribution by the
\ac{BMWc}~\cite{Borsanyi:2020mff} suggests a strongly reduced tension
of the SM prediction for $a_\mu$ with the experiment, and further
lattice calculations are underway to confirm or refute these
findings.

In this paper, we present results for two closely related observables
that play a central role in \ac{SM} tests, namely the energy
dependence (running) of the electromagnetic coupling, $\alpha$, as
well as that of the electroweak mixing angle, $\sIIW$. The former is
an important input quantity for electroweak precision tests, while the
running of the mixing angle is susceptible to the effects of \ac{BSM}
physics, particularly at low energies~\cite{Zyla:2020zbs}. As in the case of $a_\mu$, the
overall precision of both quantities is limited by hadronic effects.
We employ the same methodology as in our earlier lattice \ac{QCD}
calculation of $\aHVP$ \cite{Gerardin:2019rua}, to compute the
hadronic vacuum polarization functions $\SPi^{\gamma\gamma}$ and
$\SPi^{\gamma Z}$ that are relevant for the running of $\alpha$ and
$\sIIW$, respectively. A key advantage of the lattice approach is the
ability to perform an exact valence flavor decomposition of the various
contributions. This has allowed us to determine the isoscalar ($I=0$)
contribution $\SPi^{08}$ to the vacuum polarization function
$\SPi^{\gamma Z}$ with much higher accuracy compared to the standard
approach based on dispersion theory and experimentally determined
hadronic cross sections.

We present our main results for the \ac{HVP} functions
$\SPi^{\gamma\gamma}$, $\SPi^{\gamma Z}$ and $\SPi^{08}$ as continuous
rational functions in the Euclidean squared momentum $Q^2$ up to
$Q^2\leq\SI{7}{\GeV\squared}$ (see
eqs.~\eqref{eq:approx_gg}, \eqref{eq:approx_Zg} and
\eqref{eq:approx_08}), together with the corresponding correlation
matrices. By employing the Euclidean split technique (or Adler function approach)
\cite{Eidelman:1998vc,Jegerlehner:2008rs} we can combine our lattice result for the
hadronic running of the QED coupling with perturbative QCD to
translate it to the time-like momentum region. At the scale of the
$Z$~boson mass we obtain
\begin{equation}
  \dalpfive{M_Z^2} = \num{0.02773(15)} \,,
\end{equation}
which agrees with corresponding results derived from dispersion theory
and the experimentally measured $R$-ratio
\cite{Keshavarzi:2019abf, Davier:2019can, Jegerlehner:2019lxt} within
errors.

This paper is organized as follows: In section~\ref{sec:running} we
review the main definitions relating to the running of the
electromagnetic and weak couplings. Our methodology to compute the
\ac{HVP} contribution to the running of $\alpha$ and $\sIIW$ in
lattice QCD, including the treatment of the different sources of
systematic errors, is discussed in section~\ref{sec:method}, with
section~\ref{sec:lat_res} describing the details of the lattice
computation and the results on individual gauge ensembles. In
section~\ref{sec:extrapolation} we discuss the extrapolation of our
lattice results to the continuum limit and physical pion and kaon masses for a
range of values of $Q^2$, quoting the complete statistical and
systematic error estimate. A detailed discussion of our results,
including their comparison with phenomenological estimates is
presented in section~\ref{sec:comparison}. We end with a short summary
and conclusions. Further details on the auxiliary calculation of
pseudoscalar meson observables, error estimation, phenomenological
models, as well as extended tables of results at the physical point
are relegated to several appendices. Readers who are not interested in
the technical aspects of the lattice calculation can skip
section~\ref{sec:method} and go directly to
sections~\ref{sec:extrapolated_running} and~\ref{sec:comparison}.

\section{The running of electroweak couplings}
\label{sec:running}
\subsection{The electromagnetic coupling}
\label{sec:alpha_intro}

The first quantity that we consider is the electromagnetic coupling
$\alpha\equiv e^2/(4\cpi)$. The value that is relevant for
interactions at energies much smaller than the electron mass, such as
in Thomson scattering, is the fine-structure constant, which is one of
the most precisely known quantities in experimental physics, with a
precision of up to \num{81} parts per trillion in the most recent
measurement~\cite{Morel:2020dww}. In the rest of this paper we use as reference value in the Thomson limit (that is, for $q^2\to 0$) the current world average, slightly
less precise but still better than a part per billion, of
$\alpha=1/\num{137.035 999 084(21)}$~\cite{Zyla:2020zbs}.

This contrasts with the \SI{7}{\percent} larger value that is relevant
for physics at or around the $Z$ pole. This value can both be measured
in collider experiments and predicted from the fine-structure constant
using the theoretical knowledge of the \ac{RG} running with energy.
Choosing to work in the \MSbar\ scheme, \ac{RG} running predicts
$\hat{\alpha}^{(5)}(M_Z)=1/\num{127.952(9)}$~\cite{Zyla:2020zbs}.
Alternatively, an effective coupling can be defined at any time-like
momentum transfer $q^2$ in the \emph{on-shell} scheme,
\begin{equation}
\label{eq:Dalpha}
  \alpha(q^2) = \frac{\alpha}{1-\Delta\alpha(q^2)} \,,
\end{equation}
in terms of the function $\Delta\alpha(q^2)$. While the leptonic
contribution to $\Delta\alpha(q^2)$ can be computed in perturbation
theory, the contribution from the quarks at low energies is
non-perturbative and encoded in the subtracted \ac{HVP} function,\footnote{
  The conventional choice of taking the real part of $\SPi(q^2)$ and discarding the imaginary part simplifies the conversion between the on-shell scheme and the \MSbar\ one, given by eq.~(10.10) of ref.~\cite{Zyla:2020zbs}.
  However, to define $\alpha(q^2)$ as a physical observable also the subleading imaginary part should be included~\cite{Aoyama:2020ynm}.
  See also the discussion around eq.~(2.11) in ref.~\cite{Colangelo:2018mtw}.
  For space-like $q^2<0$ accessible on the lattice, $\SPi(q^2)$ is real and there is no issue around the imaginary part.
}
\begin{equation}
\label{eq:Dhad_alpha}
  \Dalphahad(q^2) = 4\cpi\alpha \Re\SPi(q^2) \,, \qquad \SPi(q^2) = \Pi(q^2)-\Pi(0) \,.
\end{equation}
The standard approach to determine $\Dalphahad$ proceeds by
invoking the optical theorem, which links the \ac{HVP} function to the
$R$-ratio, \ie\ the total hadronic cross section
$\sigma(e^+e^-\to\text{hadrons})$ normalized by
$\sigma(e^+e^-\to\mu^+\mu^-)$, and evaluating a dispersion integral. 
A compilation of precise experimental data for the $R$-ratio $R(s)$ as
a function of the squared center-of-mass energy $s=q^2$ has been used
in the most recent
efforts~\cite{Davier:2019can,Jegerlehner:2019lxt,Keshavarzi:2019abf},
resulting in
$\Dalphahad^{(5)}(M_Z^2)=\num{0.02766(7)}$~\cite{Zyla:2020zbs}, which
constitutes the main uncertainty in the value of $\alpha(M_Z^2)$.

Lattice QCD allows for an \emph{ab initio}, non-perturbative
calculation of $\Dalphahad$ that avoids the dependence on experimental $R$-ratio data.
Since the lattice formulation realizes only space-like momenta in a
straightforward manner, the link to $\Dalphahad$ is provided by the
Adler function $D(Q^2)$~\cite{Adler:1974gd}, as advocated in
refs.~\cite{Eidelman:1998vc,Jegerlehner:1999hg,Jegerlehner:2003ip,Jegerlehner:2008rs}.
It is defined in terms of the derivative of $\SPi(-Q^2)$ with respect
to the space-like squared four-momentum $Q^2=-q^2$ and can also be written
as a dispersion integral over the $R$-ratio, \ie
\begin{equation}
\label{eq:adler_def}
  D(Q^2) = 12\cpi^2 Q^2 \dv{\Pi(-Q^2)}{Q^2} = Q^2 \int_0^\infty \dd{s} \frac{R(s)}{(s+Q^2)^2} \,,
\end{equation}
On the other hand, the \ac{HVP} function $\SPi(-Q^2)$ can be
represented in terms of a current
correlator~\cite{Burger:2015lqa,Francis:2015grz,Borsanyi:2017zdw},
\begin{equation}
  (Q_\mu Q_\nu-\delta_{\mu\nu}Q^2)\Pi(-Q^2) = \Pi_{\mu\nu}(Q) = \int\dd[4]{x} \e{\I Q\cdot x} \ev{j_\mu^\gamma(x) j_\nu^\gamma(0)} ,
\end{equation}
with the electromagnetic current $j_\mu^\gamma$ of the quarks given by
\begin{equation}
\label{eq:emcurrent}
  j_\mu^\gamma = \frac{2}{3}\bar{u}\gamma_\mu u -
\frac{1}{3}\bar{d}\gamma_\mu d - \frac{1}{3}\bar{s}\gamma_\mu s +
\frac{2}{3}\bar{c}\gamma_\mu c+\ldots \,.
\end{equation}
The determination of $\Dalphahad$ is closely related to that of the
leading \ac{HVP} contribution to the anomalous magnetic moment of the
muon, $\aHVP$. Both quantities can be evaluated either via a
dispersion integral using experimental data for the $R$-ratio or via a
first-principles approach based on a lattice calculation of the
\ac{HVP} function $\SPi(-Q^2)$.

The correlation between $\Dalphahad$ and $\aHVP$ implies that any
evaluation of $\aHVP$ also provides a constraint on $\Dalphahad$. 
While enormous progress has been achieved in recent years concerning
\emph{ab initio} calculations
of $\aHVP$ in lattice
QCD~\cite{Meyer:2018til,Aoyama:2020ynm,FermilabLattice:2017wgj,Borsanyi:2017zdw,Blum:2018mom,Giusti:2019xct,Shintani:2019wai,FermilabLattice:2019ugu,Gerardin:2019rua,Aubin:2019usy,Giusti:2019hkz}, the current \ac{SM} estimate is based on
dispersion theory using the experimentally measured
$R$-ratio~\cite{Aoyama:2020ynm,Davier:2017zfy,Keshavarzi:2018mgv,Colangelo:2018mtw,Hoferichter:2019mqg,Davier:2019can,Keshavarzi:2019abf}, which achieves an overall uncertainty
at the level of \SI{0.6}{\percent}. However, the recent lattice determination by
\ac{BMWc}~\cite{Borsanyi:2020mff}, which is the first to claim a level
of precision similar to that obtained from the $R$-ratio, favors a
larger value for $\aHVP$ compared to the phenomenological estimate.
While such a higher value for $\aHVP$ would reduce the tension between
the \ac{SM} and the experimental measurement, it would, via the
correlation with $\Dalphahad$, further increase the already observed
slight tension with global electroweak fits~\cite{Passera:2008jk,Crivellin:2020zul,Keshavarzi:2020bfy,Malaescu:2020zuc,Colangelo:2020lcg}.
Recent investigations, considering also the global fit predictions of
$M_W$ and the electroweak mixing angle, have concluded that an increase
in the values of $\aHVP$ and $\Dalphahad(M_Z^2)$ is still compatible
with global electroweak fits, provided that the $R$-ratio is enhanced
by \SI{+9}{\percent} in the region below
$\approx\SI{0.7}{\GeV}$~\cite{Keshavarzi:2020bfy}.
This seems an unlikely possibility, given the high precision
that hadronic cross sections have been measured with.

The precise size of the increase on $\Dalphahad(M_Z^2)$ which would
correspond to the lattice result in ref.~\cite{Borsanyi:2020mff} has
not been precisely estimated. An independent lattice determination of
$\Dalphahad(-Q^2)$ over an interval of $Q^2$ in the low-energy regime,
as described in this paper, can help to resolve this puzzle. We will
return to an in-depth discussion of this issue in
section~\ref{sec:comparison}.

\subsection{The electroweak mixing angle}

The electroweak sector of the \ac{SM} is characterized by two gauge
couplings, $g$ and $g'$, for the $\mathrm{SU}(2)_L$ weak isospin and
$\mathrm{U}(1)_Y$ weak hypercharge gauge interactions, respectively.
The electromagnetic coupling $\alpha=e^2/(4\cpi)$ is a linear
combination of $g$ and $g'$ parametrized by the electroweak mixing
angle (or Weinberg angle) $\thetaW$ defined through~\cite{Glashow:1961tr,Zyla:2020zbs}
\begin{equation}
\label{eq:Dhad_sIIW}
  e = g \sin\thetaW = g' \cos\thetaW \,, \qquad \sIIW = \frac{g'^2}{g^2+g'^2} \,.
\end{equation}
Just as the couplings in the interacting quantum field theory are
renormalization scheme and energy dependent, so is the precise
definition of $\sIIW$ beyond tree level. For instance, since the angle
enters the $W$ and $Z$ boson mass ratio, which is known precisely from
collider experiments, one choice of scheme employs the tree-level
formula $\sIIW=1-M_W^2/M_Z^2$ to all orders of perturbation theory,
which results in the \emph{on-shell} value of
$\sIIW=\num{0.22337(10)}$~\cite{Zyla:2020zbs}. Another widely used
convention is the \emph{effective coupling}
$\sin^2\theta_{\mathrm{eff}}^f$ for the $Z$-boson coupling to the
fermion $f$, which is an input to the global electroweak fit mentioned
in section~\ref{sec:alpha_intro}. Finally, in the \MSbar\ definition
of $\sIIWh(\mu)$, one substitutes the \MSbar\ couplings
$\hat{g}(\mu)$, $\hat{g}'(\mu)$ at $\mu=M_Z$ into
eq.~\eqref{eq:Dhad_sIIW}, which gives the sub-permil precision value
$\sIIWh(M_Z)=\num{0.23121(4)}$~\cite{Zyla:2020zbs}.

There is a growing interest in experiments that probe precision
electroweak observables at momentum transfers $q^2\ll M_Z^2$, such as
measurements of cross sections of neutrino scattering and
parity-violating lepton scattering, as well as nuclear weak charges in
atomic parity violation experiments. These experiments are sensitive
to modifications of the \ac{RG} running of the mixing angle by
\ac{BSM} physics. A $q^2$-dependent definition of the mixing angle
that is appropriate for low-energy experiments is obtained by applying
a form factor $\hat{\kappa}$ to the \MSbar\ $Z$-pole
value~\cite{Sarantakos:1982bp,Czarnecki:1995fw,Czarnecki:1998xc,Czarnecki:2000ic,Ferroglia:2003wa,Kumar:2013yoa}
\begin{equation}
\label{eq:sIIW_qfunc}
  \sIIW(q^2) = \hat{\kappa}(q^2, \mu) \sIIWh(\mu) \,,
\end{equation}
such that the Thomson limit results in the process-independent
physical observable $\sIIW\equiv\sIIW(0)$, the electroweak analog of
the fine-structure constant $\alpha$. The value $\hat{\kappa}(0,
M_Z)\approx 1.03$ results in $\sIIWh=\num{0.23857(5)}$, quoted by
ref.~\cite{Zyla:2020zbs} as the average of different
results~\cite{Czarnecki:2000ic,Erler:2004in,Kumar:2013yoa,Erler:2017knj},
which is \SI{3}{\percent} larger than the $Z$-pole value used as
input. Excluding uncertainties from experimental input, the error on
the theory prediction of $\sIIWh$ at $q^2=0$ is dominated by the
non-perturbative hadronic contributions.

Experimental determinations of $\sIIW$ from current low-energy
experiments are much less
precise~\cite{Anthony:2005pm,Androic:2013rhu,Wang:2014bba} compared to
$\alpha$, with the current most precise value resulting from the determination of the weak charge of the proton $Q^p_W$ by the Q\textsubscript{weak} experiment at JLab~\cite{Qweak:2018tjf}, obtained at $Q^2=\SI{0.0248}{\GeV\squared}$.
However, future new and upgraded experiments have the
potential of changing the situation. The P2 experiment at
MESA~\cite{Becker:2018ggl}, which is expected to start data taking in
2025, targets \SI{0.15}{\percent} precision on $\sIIW$ at a momentum
transfer of \SI{4.5e-3}{\GeV\squared}~\cite{Becker:2018ggl}, and the
MOLLER and SoLID experiments at JLab have comparable precision
goals~\cite{Benesch:2014bas,Chen:2014psa,Souder:2016xcn}.

Following refs.~\cite{Jegerlehner:1985gq, Jegerlehner:2011mw,
Jegerlehner:2017zsb, Jegerlehner:2019lxt}, the relation between
$\sIIW(-Q^2)$ and its value in the Thomson limit can be written as
\begin{equation}
  \sIIW(-Q^2) = \left( \frac{1-\Delta\alpha_2(-Q^2)}{1-\Delta\alpha(-Q^2)} + \Delta\kappa_b(Q^2) - \Delta\kappa_b(0) \right) \sIIW(0) \,,
\end{equation}
where the bosonic contribution $\Delta\kappa_b$ is given in ref.~\cite{Czarnecki:2000ic}, $\Delta\alpha$ is the contribution to the running of $\alpha$ in eq.~\eqref{eq:Dalpha} and $\Delta\alpha_2$ is the contribution to the running of the $\SU(2)_L$ gauge coupling $\alpha_2\equiv g^2/(4\cpi)$, defined as
\begin{equation}
  \alpha_2(q^2) = \frac{\alpha_2}{1-\Delta\alpha_2(q^2)} \,.
\end{equation}
Similarly to $\Delta\alpha$, $\Delta\alpha_2$ receives the leading hadronic contribution from the \ac{HVP} mixing function
\begin{equation}
\label{eq:Dhad_alph2}
  \DalphaIIhad(q^2) = \frac{4\cpi\alpha}{\sIIW} \SPi^{T_3\gamma}(q^2)
\end{equation}
of the electromagnetic current $j_\mu^\gamma$ with the vector part of
the weak isospin third component $T_3$ current, \ie
\begin{equation}
\label{eq:weakcurrent}
  j_\mu^{T_3} \big|_\mathrm{vector}  = \frac{1}{4}\bar{u}\gamma_\mu u
- \frac{1}{4}\bar{d}\gamma_\mu d - \frac{1}{4}\bar{s}\gamma_\mu s +
\frac{1}{4}\bar{c}\gamma_\mu c +\ldots \,.
\end{equation}
At leading order,
the hadronic contribution to the running of $\sIIW$ is given by~\cite{Jegerlehner:1985gq,Jegerlehner:1986vs,Burger:2015lqa}
\begin{equation}
  \Dhad\sin^2 \thetaW(q^2) = \Dalphahad(q^2) - \DalphaIIhad(q^2) = -\frac{4\cpi\alpha}{\sIIW} \SPi^{Z\gamma}(q^2) \,,
\end{equation}
where $\SPi^{Z\gamma}(q^2)$ is the \ac{HVP} mixing of the
electromagnetic current $j_\mu^\gamma$ and the vector part of the
neutral weak current
\begin{equation}
  j_\mu^Z \big|_\mathrm{vector} = j_\mu^{T_3} \big|_\mathrm{vector} - \sIIW j_\mu^\gamma \,.
\end{equation}
As for the running of $\alpha$, the standard approach is based on a
phenomenological estimate of the hadronic contribution from
experimental data~\cite{Erler:2004in,Erler:2017knj}. However, $R(s)$
alone is not sufficient in this case, as the total cross section
couples only to the electromagnetic current $j_\mu^\gamma$. The
process of assigning individual channels in the hadronic
cross section to the different quark flavor contributions, in order
to reweight them according to the weak isospin charge factors, is
called \emph{flavor separation} and a source of systematic
uncertainty.

In the next section we show that $\SPi^{Z\gamma}(-Q^2)$ admits a decomposition into valence flavor components that can all be
determined directly from suitable correlation functions computable in
lattice
QCD~\cite{Burger:2015lqa,Francis:2015grz,Guelpers:2015nfb,Ce:2018ziv}.
This paves the way for \emph{ab initio} estimates that do not rely on
experimental cross-section data and a reweighting of individual
hadronic channels.

\section{Methodology}
\label{sec:method}

\subsection{The \acs{TMR} method}
\label{sec:TMR}

The main primary observable that we compute in our lattice QCD
simulations is the correlation function, $G_{\mu\nu}(x)$, of two
generic vector currents $j_\mu(x)$, defined by
$G_{\mu\nu}(x)=\ev{j_\mu(x)j_\nu(0)}$. By supplying the appropriate
currents, \ie\ $j_\mu^\gamma$ or $j_\mu^Z$, we can compute the
electromagnetic \ac{HVP} function $\SPi^{\gamma\gamma}$ and its
$Z-\gamma$ counterpart $\SPi^{Z\gamma}$ as functions of $Q^2$ in terms
of these correlators. In this work, we employ the \ac{TMR}, defined
in~\cite{Bernecker:2011gh,Francis:2013qna}, which has emerged as the
standard method to compute the \ac{HVP} in lattice QCD  and
which is well suited to open boundary conditions in the time
direction, which are employed on a large subset of our gauge
ensembles (see section~\ref{sec:lattice_setup}). For concreteness, we
consider the correlator of two electromagnetic currents,
$G_{\mu\nu}^{\gamma\gamma}=\ev{j_\mu^\gamma(x)j_\nu^\gamma(0)}$. In
the continuum and infinite-volume limits, the corresponding subtracted
\ac{HVP} function $\SPi^{\gamma\gamma}(-Q^2)$ is given by the integral
over Euclidean time
\begin{equation}
\label{eq:TMRmethod}
  \SPi^{\gamma\gamma}(-Q^2) = \int_0^\infty \dd{t} G^{\gamma\gamma}(t) K(t,Q^2)
\end{equation}
of the product of the zero-momentum-projected correlator
\begin{equation}
  G^{\gamma\gamma}(t) = -\frac{1}{3} \int\dd[3]{x} \sum_{k=1}^3 \ev{
j_k^\gamma(t,\vec{x}) j_k^\gamma(0) }
\end{equation}
multiplied by a $Q^2$-dependent kernel function
\begin{equation}
\label{eq:TMRkernel}
  K(t,Q^2) = \left[ t^2 - \frac{4}{Q^2}\sin[2](\frac{Qt}{2})\right] .
\end{equation}
The corresponding integral representation of $\SPi^{\gamma Z}(-Q^2)$ is
obtained by replacing one of the electromagnetic currents by $j_\mu^Z$.
After inserting the definitions of the currents in
eqs.~\eqref{eq:emcurrent} and~\eqref{eq:weakcurrent} and performing
the Wick contractions of the quark fields, one can perform explicit
flavor decompositions of both $\SPi^{\gamma\gamma}$ and $\SPi^{\gamma
Z}$, as described in section~\ref{sec:flavor}.

\begin{figure}[t]
  \centering
  \scalebox{.55}{\input{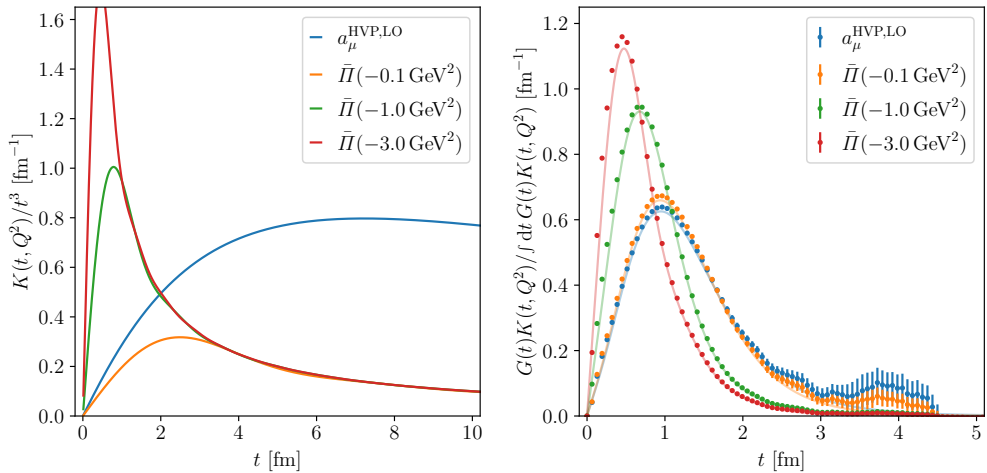}}
  \caption{%
    Left: the kernel $K(t,Q^2)$ of the \acs{TMR} integral in
eq.~\eqref{eq:TMRmethod} for different values of
$Q^2$, compared to the kernel $w(t)$ for
$\aHVP$~\cite{Bernecker:2011gh,DellaMorte:2017dyu} (blue line), as a
function of time $t$.
    All kernels are divided by $t^3$ in the plot, such that they tend to zero at $t\to\infty$ while being still zero at $t=0$.
    Right: contribution of $G(t)K(t,Q^2)$ to the \acs{TMR} integral
normalized to the value of the integral, comparing different kernels.
The light colored lines are drawn using a model for the
Euclidean-time correlator $G(t)$~\cite{Bernecker:2011gh}, that is also
used for the integral, while the data points with error bars are
obtained using actual lattice correlator data at the physical pion
mass.}\label{fig:kernel}
\end{figure}

The properties of the kernel significantly influence the integral and
its systematics: In the left panel of figure~\ref{fig:kernel}, we plot
the kernel function $K(t,Q^2)$ for several different values of $Q^2$
versus Euclidean time. Also shown is the kernel $w(t)$ that appears
in the \ac{TMR} expression for $\aHVP$, which is given explicitly in eq.~(84)
of ref.~\cite{Bernecker:2011gh}. Despite the fact that both $K(t,Q^2)$
and $w(t)$ behave like $t^2$ at long distances, it is evident that
$w(t)$ gives a much higher weight to long distances compared to
$K(t,Q^2)$.
This has several consequences for our calculation:
On the one hand, since the signal-to-noise ratio of the vector correlator
lattice data degrades severely with time, the stronger suppression of
the long-distance contribution by the kernel $K(t,Q^2)$ makes it
easier to achieve good statistical precision for $\SPi(-Q^2)$ in our
range of interest for $Q^2$, compared to $\aHVP$. Moreover,
finite-size effects that affect the correlator mostly at long
distances are more strongly suppressed by $K(t,Q^2)$ relative to
$w(t)$, even though they are still relevant at our target precision,
as explained in section~\ref{sec:FSE}.
On the other hand, the peak of the kernel $K(t,Q^2)$ occurs at
increasingly short distances $t$ for larger values of $Q^2$, which
results in larger discretization effects, both from the correlator
itself and from the approximation of the integral in
eq.~\eqref{eq:TMRmethod} as a discrete sum.\footnote{%
  In this work, we employ the trapezoidal rule to approximate the \ac{TMR} intregral, which has a $\order*{a^2}$ error that is consistent with the use of $\order*{a}$-improved action and operators, see section~\ref{sec:lattice_setup}.
}
Therefore, lattice discretization effects and our ability to estimate
and control the associated systematic error ultimately limit the upper
end of the range of $Q^2$ values.

\subsection{Flavor decomposition}
\label{sec:flavor}

As already mentioned, the \ac{HVP} functions $\SPi^{\gamma\gamma}$ and
$\SPi^{Z\gamma}$ differ only in their flavor content. For the
following discussion, we assume exact strong-isospin symmetry and neglect charm disconnected contributions.
It is convenient to introduce a strong isospin and
$\SU(3)$-flavor basis for the quark triplet $q=(u,d,s)^\tran$, starting from the vector currents $j_\mu^a=\bar{q}\gamma_\mu(\lambda_a/2)q$ where $\lambda_3$, $\lambda_8$ are Gell-Mann matrices and $\lambda_0=\mathrm{Id}_3$,
\begin{subequations}
\begin{equationarray}{rc}
  \text{$I=1$:}                    & j_\mu^3 = \frac{1}{2}         \left(\bar{u}\gamma_\mu u - \bar{d}\gamma_\mu d\right) , \\
  \multirow{2.6}{*}{\text{$I=0$:}} & j_\mu^8 = \frac{1}{2\sqrt{3}} \left(\bar{u}\gamma_\mu u + \bar{d}\gamma_\mu d - 2\bar{s}\gamma_\mu s\right) , \\
                                   & j_\mu^0 = \frac{1}{2}         \left(\bar{u}\gamma_\mu u + \bar{d}\gamma_\mu d + \bar{s}\gamma_\mu s\right) ,
\end{equationarray}
\end{subequations}
such that, with the addition of the charm current
$j_\mu^c=\bar{c}\gamma_\mu c$, the currents $j_\mu^\gamma$ and
$j_\mu^Z$ are represented by
\begin{subequations}
\begin{gather}
  j_\mu^\gamma = j_\mu^3 + \frac{1}{\sqrt{3}} j_\mu^8 + \frac{2}{3} j_\mu^c \,, \qquad
  j_\mu^{T_3}\big|_\mathrm{vector} = \frac{1}{2}\left( j_\mu^\gamma - \frac{1}{3} j_\mu^0 - \frac{1}{6} j_\mu^c \right) , \\
  j_\mu^Z\big|_\mathrm{vector}= j_\mu^{T_3}\big|_\mathrm{vector} - \sin^2\theta_{\mathrm{W}} j_\mu^\gamma = \left(\frac{1}{2}-\sin^2\theta_{\mathrm{W}}\right) j_\mu^\gamma - \frac{1}{6} j_\mu^0 - \frac{1}{12} j_\mu^c \,.
\end{gather}
\end{subequations}
The correlators of interest are
\begin{subequations}
\label{eq:corr_flavor}
\begin{gather}
  G_{\mu\nu}^{\gamma\gamma}(x) = G_{\mu\nu}^{33}(x) + \frac{1}{3}G_{\mu\nu}^{88}(x) + \frac{4}{9} G_{\mu\nu}^{cc}(x) \,, \\
  G_{\mu\nu}^{Z\gamma}(x) =  \left(\frac{1}{2}-\sIIW\right)
G_{\mu\nu}^{\gamma\gamma}(x) - \frac{1}{6\sqrt{3}} G_{\mu\nu}^{08}(x)
- \frac{1}{18} G_{\mu\nu}^{cc}(x) \,,
\end{gather}
\end{subequations}
which can be obtained by computing the building blocks
$G^{33}_{\mu\nu}$, $G^{88}_{\mu\nu}$, $G^{08}_{\mu\nu}$ and
$G^{cc}_{\mu\nu}$. In this paper we will present as intermediate results,
extrapolated to the physical point, the $I=1$ \ac{HVP} function
$\SPi^{33}$, the $I=0$ ones $\SPi^{88}$ and $\SPi^{08}$, with the
latter being relevant for the running of $\sIIW$ case only, and the
charm \ac{HVP} function of $\SPi^{cc}$.

Up to lattice renormalization and O($a$) improvement, the flavor
$\SU(3)$ contributions are defined as\footnote{%
  In the usual lattice notation, $G_{\con}^\ell=2G^{33}$ and $G_{\con}^s=3G_{\con}^{88}-G^{33}$.
  Moreover, $G^{08}_{\con}=\sqrt{3}(G^{33}-G_{\con}^{88})/2$.
}
\begin{subequations}\label{eq:corr_flavor_2}
\begin{gather}
  G_{\mu\nu}^{33}(x) = \frac{1}{2}C^{\ell,\ell}_{\mu\nu}(x) \,, \\
  G_{\mu\nu}^{88}(x) = \frac{1}{6}\left[ C^{\ell,\ell}_{\mu\nu}(x) + 2C^{s,s}_{\mu\nu}(x) + 2D^{\ell-s,\ell-s}_{\mu\nu}(x) \right] , \\
  G_{\mu\nu}^{08}(x) = \frac{1}{2\sqrt{3}}\left[ C^{\ell,\ell}_{\mu\nu}(x) - C^{s,s}_{\mu\nu}(x) + D^{2\ell+s,\ell-s}_{\mu\nu}(x) \right] ,
\end{gather}
\end{subequations}
where the flavor labels $\ell$ and $s$ denote the (isospin averaged)
light and strange quarks, respectively, while $C^{f_1,f_2}_{\mu\nu}$
and $D^{f_1,f_2}_{\mu\nu}$ are, respectively, the connected and
disconnected Wick contractions, schematically given by
\begin{equation}\label{eq:wick_sketch}\contourlength{1.5pt}
  C^{f_1,f_2}_{\mu\nu} = -\ev{\tikz[baseline=-0.5ex]{%
      \draw[thick,->-] (2,0)                                                   to [bend left] node[near start] {\footnotesize\contour{white}{$f_2$}} (0,0);
      \draw[thick,->-] (0,0) node {\footnotesize\contour{white}{$\gamma_\mu$}} to [bend left] node[near start] {\footnotesize\contour{white}{$f_1$}} (2,0) node {\footnotesize\contour{white}{$\gamma_\nu$}};
    }}, \qquad D^{f_1,f_2}_{\mu\nu} = \ev{\tikz[baseline=-0.5ex]{%
      \draw[thick,->-] (0,0) node {\footnotesize\contour{white}{$\gamma_\mu$}} arc[start angle=-180,delta angle=-360,radius=0.4] node[near start] {\footnotesize\contour{white}{$f_1$}};
      \draw[thick,->-] (2,0) node {\footnotesize\contour{white}{$\gamma_\nu$}} arc[start angle=   0,delta angle=-360,radius=0.4] node[near start] {\footnotesize\contour{white}{$f_2$}};
    }}.
\end{equation}
Section~\ref{sec:lat_res} explains the lattice computation of the
connected and disconnected contractions in more detail.

\subsection{Renormalization and \texorpdfstring{$\order{a}$}{O(a)} improvement}
\label{sec:renorm}

We use the vector correlators computed in
ref.~\cite{Gerardin:2019rua}, with updated statistics and ensemble
coverage as listed in table~\ref{tab:ensemble}.
At the sink, we employ both the local (labeled by the superscript
labeled ``$\local$'') and conserved discretizations (labelled ``$\cons$'') of the vector
current, \ie
\begin{subequations}
\begin{gather}
  j_\mu^{a,\local}(x) = \bar{q}(x) \gamma_\mu \frac{\lambda_a}{2} q(x) \,, \\
  j_\mu^{a,\cons}(x) = \frac{1}{2}\left[ \bar{q}(x+a\hat{\mu})(1+\gamma_\mu)U^\dagger_\mu(x) \frac{\lambda_a}{2} q(x) - \bar{q}(x)(1-\gamma_\mu)U_\mu(x) \frac{\lambda_a}{2} q(x+a\hat{\mu}) \right] ,
\end{gather}
\end{subequations}
while only the local current is used at the source.
The $\order{a}$-improvement and renormalization of the vector currents in the flavor basis is complicated by the fact that flavor singlet and non-singlet contributions renormalize differently: it is more convenient to work in the basis introduced in section~\ref{sec:flavor}.
For the local
discretization, the renormalized (``$\ren$'') currents
read~\cite{Bhattacharya:2005rb}
\begin{subequations}
\begin{gather}
  j_{\mu,\ren}^{3,\local} = Z_V \left(1 +
3\bar{b}_Vam_q^{\mathrm{av}} + b_Vam_{q,\ell}\right)
j_\mu^{3,\impr,\local} \,, \\
  \begin{pmatrix}
    j_\mu^8 \\
    j_\mu^0
  \end{pmatrix}_{\ren}^{\local} = Z_V \begin{pmatrix}
    1 + 3\bar{b}_Vam_q^{\mathrm{av}} + b_V\frac{a(m_{q,\ell}+2m_{q,s})}{3} & \left(\frac{b_V}{3}+f_V\right)\frac{2a(m_{q,\ell}-m_{q,s})}{\sqrt{3}} \\
    r_V d_V\frac{a(m_{q,\ell}-m_{q,s})}{\sqrt{3}} & r_V + r_V (3\bar{d}_V+d_V)am_q^{\mathrm{av}}
  \end{pmatrix} \begin{pmatrix}
    j_\mu^8 \\
    j_\mu^0
  \end{pmatrix}^{\impr,\local} , 
\end{gather}
\end{subequations}
where the improved (indicated by the label ``$\impr$'') non-singlet and
singlet local currents are
\begin{equation}
  j_\mu^{a,\impr,\local} = j_\mu^{a,\local} +
ac_V^{\local}\tilde{\partial}_\nu \Sigma_{\nu\mu}^a \,, \qquad
j_\mu^{0,\impr,\local} = j_\mu^{0,\local} +
a\bar{c}_V^{\local}\tilde{\partial}_\nu \Sigma_{\nu\mu}^0 \,, \\
\end{equation}
with the antisymmetric tensor current $\Sigma^a_{\mu\nu}=-(1/2)\bar{q}\comm{\gamma_\mu}{\gamma_\nu}(\lambda_a/2)q$, and the breaking of flavor $\SU(3)$ symmetry introduces a mixing
between the singlet and non-singlet $I=0$ components. Here,
$m_{q,\ell}$ and $m_{q,s}$ are the bare subtracted light and strange
quark masses, with $m_q^\mathrm{av}\equiv (2m_{q,\ell}+m_{q,s})/3$
denoting their average, and $\tilde\partial_\mu$ is the symmetric
lattice derivative. The conserved current is automatically
renormalized, and its $\order{a}$-improved version reads
\begin{equation}
  j_{\mu,\ren}^{a,\cons} = j_\mu^{a,\cons} +
ac_V^{\cons}\tilde{\partial}_\nu \Sigma_{\nu\mu}^a \,, \qquad
j_{\mu,\ren}^{0,\cons} = j_\mu^{0,\cons} +
a\bar{c}_V^{\cons}\tilde{\partial}_\nu \Sigma_{\nu\mu}^0 \,.
\end{equation}
We have used the non-perturbative determination of the renormalization
and improvement coefficients $Z_V$, $b_V$, $\bar{b}_V$ and $c_V$ from
ref.~\cite{Gerardin:2018kpy}. Although a non-perturbative determination
of the renormalization coefficient $r_V$ is not available, one can
avoid relying on the renormalized singlet local current
$j_{\mu,\ren}^{0,L}$ by inserting the conserved singlet current (and thus
$j_\mu^Z$) at the sink. Moreover, $f_V$ and
$\bar{c}_V^{\cons,\local}$ are also not known. We set $f_V=0$ and
$\bar{c}_V^{\cons,\local}=c_V^{\cons,\local}$,
which is valid up to $\order{g_0^6}$ corrections and introduces a negligible
error.\footnote{
  Both $f_V$ and $\bar{c}_V^{\cons,\local}-c_V^{\cons,\local}$ arise
from disconnected diagrams in which at least three gluons are
exchanged. Thus, in perturbation theory this contribution is of $\order{g_0^6}$, see the discusssion after eq.~(27) in ref.~\cite{Gerardin:2018kpy}.
}
We propagate the error on the renormalization coefficients
$Z_V$, $b_V$, $\bar{b}_V$ quoted in ref.~\cite{Gerardin:2018kpy} to
our estimate of the renormalized vector correlator.
The values of the improvement coefficients are taken as a definition of the $\order{a}$-improved theory and no error on $c_V$ and $\bar{c}_V$ is propagated.
In our continuum extrapolations, described in section~\ref{sec:extrapolation}, we have not found any evidence for residual $\order{a}$ discretization effect.

Since our gauge ensembles do not include a dynamical charm quark, the
charm contribution to the vector correlator is computed in the
quenched approximation, with the charm-quark mass tuned using the
experimental $D_s$ meson mass and the local current renormalized
using the mass-dependent $Z_V^c$, as explained in
ref.~\cite{Gerardin:2019rua}.

\subsection{Lattice setup}
\label{sec:lattice_setup}

\begin{table}[tb]
  \centering
  \begin{adjustbox}{center}
  \begin{tabular}{lS[table-format=3]S[table-format=2]S[table-format=1.3]S[table-format=1.3]S[table-format=1.1]S[table-format=4]S[table-format=4]S[table-format=1.1]S[table-format=4]S[table-format=4]}
    \toprule
    & {$T/a$} & {$L/a$} & {$\tnotsym/a^2$} & {$a$ [\si{\fm}]} & {$L$ [\si{\fm}]} & \multicolumn{2}{c}{$m_\pi$, $m_K$ [\si{\MeV}]} & {$m_\pi L$} & \multicolumn{2}{c}{ncfg (con., dis.)} \\
    \midrule
    H101  &  96 & 32 & 2.860 & 0.086 & 2.8 & \multicolumn{2}{c}{$415$} & 5.8 & 2000 &  {-} \\
    H102  &  96 & 32 &       &       & 2.8 & 355 & 440                 & 5.0 & 1900 & 1900 \\
    H105  &  96 & 32 &       &       & 2.8 & 280 & 460                 & 3.9 & 1000 & 1000 \\
    N101  & 128 & 48 &       &       & 4.1 & 280 & 460                 & 5.8 & 1155 & 1155 \\
    C101  &  96 & 48 &       &       & 4.1 & 220 & 470                 & 4.6 & 2000 & 2000 \\
    \midrule
    B450  &  64 & 32 & 3.659 & 0.076 & 2.4 & \multicolumn{2}{c}{$415$} & 5.1 & 1600 &  {-} \\
    S400  & 128 & 32 &       &       & 2.4 & 350 & 440                 & 4.3 & 1720 & 1720 \\
    N451  & 128 & 48 &       &       & 3.7 & 285 & 460                 & 5.3 & 1000 & 1000 \\
    D450  & 128 & 64 &       &       & 4.9 & 215 & 475                 & 5.3 &  500 &  500 \\
    \midrule
    H200  &  96 & 32 & 5.164 & 0.064 & 2.1 & \multicolumn{2}{c}{$420$} & 4.4 & 1980 &  {-} \\
    N202  & 128 & 48 &       &       & 3.1 & \multicolumn{2}{c}{$410$} & 6.4 &  875 &  {-} \\
    N203  & 128 & 48 &       &       & 3.1 & 345 & 440                 & 5.4 & 1500 & 1500 \\
    N200  & 128 & 48 &       &       & 3.1 & 285 & 465                 & 4.4 & 1695 & 1695 \\
    D200  & 128 & 64 &       &       & 4.1 & 200 & 480                 & 4.2 & 2000 & 1000 \\
    E250  & 192 & 96 &       &       & 6.2 & 130 & 490                 & 4.1 &  485 &  485 \\
    \midrule
    N300  & 128 & 48 & 8.595 & 0.050 & 2.4 & \multicolumn{2}{c}{$420$} & 5.1 & 1680 &  {-} \\
    N302  & 128 & 48 &       &       & 2.4 & 345 & 460                 & 4.2 & 2190 & 2190 \\
    J303  & 192 & 64 &       &       & 3.2 & 260 & 475                 & 4.2 & 1040 & 1040 \\
    E300  & 192 & 96 &       &       & 4.8 & 175 & 490                 & 4.3 &  600 &  600 \\
    \bottomrule
  \end{tabular}
  \end{adjustbox}
  \caption{
    List of CLS ensembles employed in this work, with approximate lattice spacings, spatial volume and pion and kaon masses.
    All ensembles realize open boundary conditions in time, except for B450, D450 and E250 on which the temporal boundary conditions are periodic.
    Values of $\tnotsym$ and $a$ are taken from ref.~\cite{Bruno:2016plf}.
    The number of configurations used for connected and disconnected vector correlator measurements is listed in the last two columns.
  }\label{tab:ensemble}
\end{table}

\begin{figure}[t]
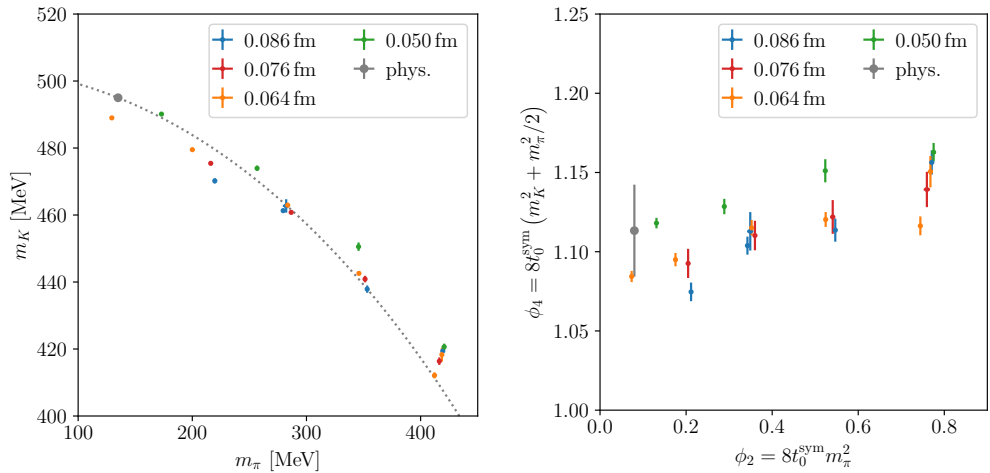

  \centering
  \scalebox{.55}{\input{figures/CLS_landscape_Mpi_MK.pgf}\input{figures/CLS_landscape_phi2_phi4.pgf}}
  \caption{%
    Landscape of the ensembles from the CLS initiative employed in this work.
  }\label{fig:landscape}
\end{figure}

Our calculations are performed on a set of $\Nf=2+1$ ensembles from
the \ac{CLS} initiative~\cite{Bruno:2014jqa}, with tree-level
$\order{a^2}$-improved Lüscher-Weisz gauge action and
non-perturbatively $\order{a}$-improved Wilson
fermions~\cite{Bulava:2013cta}. A list of ensembles is shown in
table~\ref{tab:ensemble}. While \acp{BC} in the spatial directions are
always periodic, most ensembles are characterized by open \acp{BC} in
the time direction, which alleviates the issue of topological charge
freezing at small lattice spacings~\cite{Luscher:2011kk}. Only ensembles B450, D450, and
E250 are characterized by periodic (anti-periodic for fermions)
\acp{BC} in time.
We use four lattice spacings, ranging from $a\approx\SI{0.086}{\fm}$
to $\approx\SI{0.050}{\fm}$. The masses of the $u$ and $d$ quarks are
taken to be degenerate in the calculation, and the pseudoscalar meson
masses span the interval from $m_\pi=m_K\approx\SI{415}{\MeV}$ at the
$\SU(3)$-symmetric point to the physical ones along a trajectory on
which the sum of the bare $u$, $d$ and $s$ quark masses is kept
constant.
We set the scale using the value of the gradient flow scale
$t_0$~\cite{Luscher:2010iy}, which has been determined as
$(8t_0^{\mathrm{phys}})^{1/2}=\SI[parse-numbers=false]{0.415(4)(2)}{\fm}$
in ref.~\cite{Bruno:2016plf}, using the pion and kaon decay constants.

The ensembles have been generated with a small twisted mass applied to
the light quark doublet for algorithmic stability. The correct
$\Nf=2+1$ QCD expectation values are obtained after including the
reweighting factors for the twisted mass and for the RHMC algorithm
used to simulate the strange quark, inclusive of the sign of the
latter~\cite{Mohler:2020txx}. A negative reweighting factor associated
with the simulation of the strange quark is found on less than
\SI{0.5}{\percent} of the total gauge field configurations employed in this work, and on \SI{3.6}{\percent} of the configurations of C101, the most affected ensemble.
The numerical impact of including the sign of the reweighting factor on the \ac{HVP} function $\SPi$ and on the meson masses is negligible with respect to the statistical error.
On the most affected ensembles, a few percent increase in the statistical error is observed, compatible with the loss in statistics due to the negative weight of some configurations.
A few ensembles that had a larger fraction of configurations with
negative weight were excluded entirely from this work.

In this work we use the connected Wick contractions of the
vector-current two-point function that has been computed in
ref.~\cite{Gerardin:2019rua}, albeit with significantly increased
statistics, especially on the ensembles closer to the physical point.
For more details on the connected correlator computation, we refer to ref.~\cite{Gerardin:2019rua}.

\subsection{Quark-disconnected diagrams}
\label{sec:disconnected_diagrams}

The determination of quark-disconnected contributions (see
eq.~\eqref{eq:wick_sketch}) requires the evaluation of quark loops
\begin{equation}
 L_{\mathcal{O}_f}(\vec{p}, t) = \sum_{\vec{x}} e^{i\vec{p}\cdot\vec{x}} \ev{ \mathcal{O}_f(\vec{x}, t) }_F ,
 \label{eq:1pt}
\end{equation}
for some operator $\mathcal{O}_f(\vec{x},t)$ involving a single quark
flavor $f$, where $\ev{\cdots}_F$ denotes the fermionic expectation
value (in a given gauge-field background).
Our computation of quark-disconnected loops has been performed using a
variant of the method introduced in ref.~\cite{Giusti:2019kff}
combining the one-end trick (OET) \cite{McNeile:2006bz} which is
commonly used with twisted-mass fermions
\cite{McNeile:2006bz,Jansen:2008wv,Boucaud:2008xu} with a combination
of the generalized hopping parameter expansion (gHPE)
\cite{Gulpers:2013uca} and hierarchical probing
(HP)~\cite{Stathopoulos:2013aci}.
The difference of two quark-disconnected loops can be written as a product
\begin{equation}
  \tr[\Gamma (D_1^{-1} - D_2^{-1})] =  (m_2-m_1) \tr[\Gamma D_1^{-1} D_2^{-1}] \,,
  \label{eq:oet}
\end{equation}
where $D_f^{-1}$ denotes the inverse of the Dirac operator for a
given quark flavor labeled $f=1,2$ with masses $m_1 \neq m_2$ and
$\Gamma$ is the desired combination of Dirac matrices.
The OET yields a very efficient estimator of the r.h.s.\ of eq.~\eqref{eq:oet} by inserting all-volume stochastic noise at the \enquote{one end} of the trace of the product $\Gamma D_1^{-1}D_2^{-1}$, where the identity (one) matrix in Dirac space is inserted.
This estimator has a lower variance than the standard one that inserts the noise at the \enquote{$\Gamma$ end} of either side of eq.~\eqref{eq:oet}~\cite{Giusti:2019kff}, see appendix~\ref{sec:disconnected} for more details.
In order to derive estimators for loops of a single, individual quark
flavor, an efficient scheme has been proposed in
ref.~\cite{Giusti:2019kff} that relies on computing the OET estimator
for a chain of $f=1,\ldots,N$ different quark flavors with $m_1 < m_2
<\ldots < m_N$ and evaluating the single flavor trace for the heaviest
flavor explicitly, from which it is possible to recursively
reconstruct single-flavor traces for all other quark flavors. To this
end, the hopping parameter expansion is used, which is known to be
very efficient at large quark masses. It is based on a decomposition
of $D_N^{-1}$ into two terms~\cite{Giusti:2019kff}
\begin{equation}
 D_N^{-1} = M_{2n,m} + D_N^{-1} H_m^{2n} \,,
 \label{eq:gHPE}
\end{equation}
where
\begin{equation}
 M_{2n,m} = \frac{1}{D_{ee} + D_{oo}} \sum^{2n-1}_{i=0} H^i_m  \,, \quad H_m = -\left( D_{eo} D_{oo}^{-1} + D_{oe} D_{ee}^{-1} \right) ,
 \label{eq:HPE_M_and_H}
\end{equation}
and $D_{ee}$, $D_{eo}$, $D_{oe}$, $D_{oo}$ denote the blocks of the
even-odd decomposition of the Dirac operator. In
ref.~\cite{Giusti:2019kff} a probing scheme has been introduced that
yields an exact result for the (sparse) first term in
eq.~(\ref{eq:gHPE}) for disconnected loops involving local operators.
However, since we are also interested in computing observables
involving point-split currents, a more general method is required.
Therefore, we evaluate the first term, $M_{2n,m}$, using hierarchical
probing on spin and color diluted stochastic volume sources. For the
second term it is sufficient to use naive stochastic volume sources,
and the required inversion can be reused in the evaluation of
$\tr[\Gamma (D_{N-1}^{-1} - D_{N}^{-1})]$, \ie\ the
last term of the chain of OET estimators.

We find that this method is significantly more efficient than \eg\
plain hierarchical probing, which we have applied in previous studies
in refs.~\cite{Djukanovic:2019jtp,Gerardin:2019rua}. For the local and
conserved vector currents, which are of interest for the present
study, a minor reduction in the resulting errors is already observed
for the case of a single light quark, while a much more significant
improvement is observed when the OET is applied to the $l-s$ combination.
In the case of the conserved vector
current, the errors from the plain hierarchical probing with 512 Hadamard vectors on two
stochastic volume sources exceeds the one from the (OET+gHPE+HP)-based
method by a factor of $\approx 2$, indicating that even with 512 Hadamard
vectors the gauge noise had not nearly been reached for this
observable.
Using OET estimators, we reach the gauge noise for all the disconnected quark loops relevant to this work.
However, even more striking is the difference in
computational cost which is improved by at least a factor five.

Within this study we observe a large gain in precision on the
disconnected contribution to $\SPi^{88}$, since it is the product of
two $\lambda_8$ currents that requires the estimation only of the
first loop difference, proportional to $m_s-m_\ell$. The disconnected
contribution to $\SPi^{08}$ instead has only one factor of
$\lambda_8$, and another factor of the $\SU(3)$-singlet current that
requires the evaluation of the full telescopic sum and is inherently
more noisy. This is clearly visible in the different size of the error
band of the two disconnected contributions in
figure~\ref{fig:running_lattice}.

Finally, we remark that the disconnected contribution to both \ac{HVP}
functions considered here vanishes for $m_s=m_\ell$, \ie\ at the
SU(3)-symmetric point. 

\subsection{Signal-to-noise ratio and bounding method}
\label{sec:bounding_method}

The vector correlator is affected by the well-known exponential
deterioration of the \ac{SN} with Euclidean time
$t$~\cite{Parisi:1983ae,Lepage:1989hd}. In the case of the connected
contribution, the \ac{SN} deteriorates roughly like $\exp{-(E_0-m_\pi)t}$, where
$E_0$ is the lowest energy level in the vector channel.
The problem worsens at lower pion masses. Moreover, for the
quark-disconnected contribution, the statistical error is independent
of the source-sink separation. This is significant, since the kernel
$K(t,Q^2)$ behaves like $t^2$ at long distances. In order to have a
bounded error on the disconnected contribution to the vector
correlator it is necessary to truncate the \ac{TMR} integration, if
one wants to avoid having to increase the \ac{MC} sampling statistics
exponentially with time.
Solving the \ac{SN} problem is an active field of research. One
promising direction is \emph{multi-level} \ac{MC}
sampling~\cite{Luscher:2001up,Meyer:2002cd,Ce:2016idq,Ce:2016ajy,Ce:2019yds}
which has recently been applied to the closely related problem of
computing the \ac{HVP} contribution to $(g-2)_\mu$~\cite{DallaBrida:2020cik}.
To use multi-level MC sampling efficiently, it is crucial that the estimators of connected and disconnected diagrams are in a regime dominated by the gauge noise.
As explained in section~\ref{sec:disconnected_diagrams}, this is indeed the case for disconnected diagrams calculated using the OET estimator.
In turn, this strengthens the case for improving the estimator of connected diagrams to reach the gauge noise, as a first step towards a future application of multi-level MC methods.

\begin{figure}[t]
  \centering
  \scalebox{.55}{\input{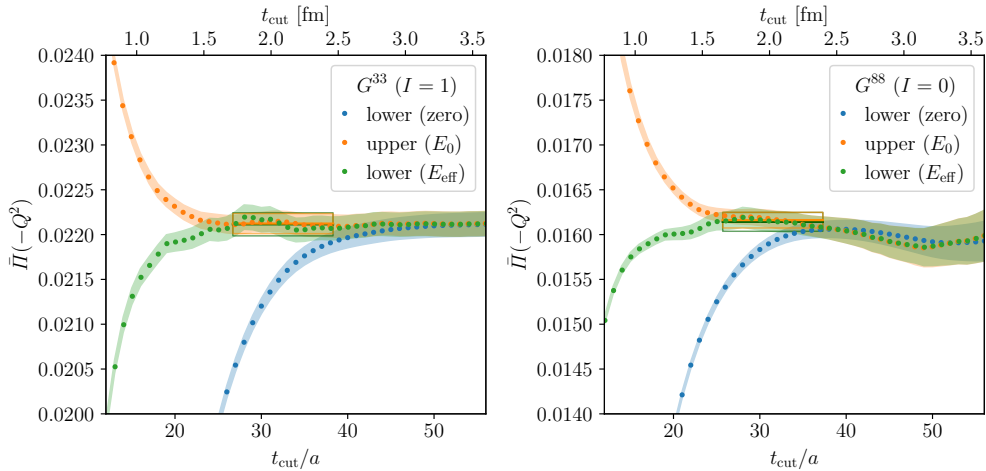}}
  \caption{%
    Bounding method on the $I=1$ (left) and $I=0$ (right) components for ensemble D200 at $Q^2=\SI{0.5}{\GeV\squared}$.
    The upper bound (orange points and band) is chosen as the ground state obtained from the finite volume analysis (left) or as the $\rho$ meson mass (right).
    The lower bound (green points and band) is computed using the effective mass at every time slice.
    The time slices on which the upper and lower bound are averaged are indicated by the limits of the average lines and error boxes.
    The less stringent lower bound given by the integral truncated up to $t$ is also given (blue points and band).
  }
  \label{fig:bounding_method}
\end{figure}

The \emph{bounding method} has established itself as the primary
method to alleviate the \ac{SN} problem in \ac{HVP} computations using
the \ac{TMR}~\cite{Lehner:talkLGT16,Borsanyi:2017zdw,Blum:2018mom}.
The method consists of substituting the correlator $G(t)$ at
$t>\tcut$ with $G(\tcut)$ multiplied by an exponential function that decays with the time
distance. By giving the appropriate exponents to this product, we can
obtain either a lower or an upper bound of the correlator,
\begin{equation}
\label{eq:bounding-method}
	0	\leq G(\tcut) e^{-E_{\mathrm{eff}}(\tcut) (t-\tcut)} \leq G(t) \leq G(\tcut) e^{-E_0 (t-\tcut)} , \qquad t \geq \tcut \,,
\end{equation}
with the effective mass $aE_{\mathrm{eff}}(t)=\log(G(t)/G(t+a))$ and the ground state in a given channel $E_0$.
Once both bounds are saturated within errors, the corresponding estimate \ac{HVP}
contribution can be computed as a function of $\tcut$.
An improved estimate of the \ac{HVP} function is obtained by averaging
both bounds over an interval of about \SI{0.8}{\fm} in $\tcut$,
starting from a timeslice where the two innermost bounds coincide at
least within half the combined uncertainty.
An example is given in figure~\ref{fig:bounding_method} for the $I=1$
and $I=0$ components. We select the bounding method interval for fixed
$Q^2=\SI{0.5}{\GeV\squared}$, where, as noted in
section~\ref{sec:TMR}, the weight of the correlator tail is relatively
high, and use the same interval at all $Q^2$ values, since the results
depend only weakly on $Q^2$.

A dedicated spectroscopy analysis that yields the energy levels in
finite volume is not available for all ensembles used in this work. In
the absence of a precise estimate for $E_0$, we note that any energy
level $\leq E_0$ provides a valid, albeit less stringent upper bound.
Thus, when applying the bounding method, we may supply any realistic
estimate for $E_0$, as long as it does not exceed the true ground
state energy.
Our specific values of $E_0$ depend on the isospin and ensemble
studied. In the $I=1$ channel $\SPi^{33}$, we substitute either the
$\rho$ meson mass $m_\rho$ or the two-pion state $E_{\pi\pi}$ for
$E_0$, depending on the pion mass and box size of the ensemble. Their
respective estimates are obtained from our finite-size effects
analysis. For some ensembles we can employ the spectroscopy
computation of ref.~\cite{Andersen:2018mau} to obtain a precise
estimate of $m_\rho$ while keeping our own bootstrap distribution to
propagate the errors correctly.
The ensembles where $E_{\pi\pi}$ is the ground state are C101, D200,
E250 and E300. We could confirm that the two-pion state used is
lighter than its non-interacting counterpart
$2\sqrt{m_\pi^2+(2\cpi/L)^2}$.

When applying the bounding method to the $I=0$ contribution
$\SPi^{88}$, we have identified $E_0$ with $m_\rho$, which is
motivated by several observations. First, since $m_\rho\lesssim
m_\omega$, this is a more conservative choice. Second, while we have
computed the $I=0$ correlator including quark-disconnected diagrams on
some ensembles, the results are too noisy for applying the
finite-volume analysis to determine the spectrum. Thirdly, while we
could also consider the lightest three-pion state with vector
isoscalar quantum numbers in the non-interacting
case~\cite{Hansen:2020zhy}, \ie
\begin{equation}
\label{eq:three-pion-state}
	E_{3\pi} = 2\sqrt{m_\pi^2 + (2\cpi/L)^2}	+ \sqrt{m_\pi^2 + 2(2\cpi/L)^2} \,.
\end{equation}
we find that $m_\rho<E_{3\pi}$ on our ensembles, mainly due to the
extra energy coming from the momenta needed to get the correct quantum
numbers.

While the effective mass $E_{\mathrm{eff}}$ that provides the lower
bound can be obtained from the asymptotic behavior of the correlator,
its determination is hampered by the \ac{SN} problem at long
distances. In these cases we substitute it by the effective mass
computed on a earlier timeslice, which is in fact larger and therefore
a more conservative choice.

Besides the $\SPi^{33}$ and $\SPi^{88}$ contributions, we apply the
method to the quark-connected $\SPi^{88}_{\con}$, assuming that
asymptotically it behaves like the $\SPi^{33}$ contribution. This
allows us to obtain a more precise estimate of the quark-disconnected
contribution $\SPi^{88}_{\dis}$, by subtracting the bounding method
estimate of $\SPi^{88}_{\con}$ from the one of $\SPi^{88}$.

\begin{table}[tb]
  \centering
  \begin{tabular}{rS[table-format=1.4(2)]S[table-format=1.4(2)]S[table-format=1.4(2)]r@{}S[table-format=1.3(2)]}
    \toprule
    & {$am_\pi$} & {$aE_{2\pi}$} & {$aE_{3\pi}$} & \multicolumn{2}{c}{$am_\rho$} \\
    \midrule
    \begin{filecontents*}{tables/bounding_method_energies.csv}
label,m_pi,E_2pi,E_3pi,m_rho_fit,m_rho_ABHM,m_rho
H101,0.1836( 5),0.5376( 7),0.8704( 9),0.375( 2), {{{-}}} ,0.375( 2)
H102,0.1546( 6),0.4998( 7),0.8176(10),0.358( 3), {{{-}}} ,0.358( 3)
H105,0.1235(13),0.4639(14),0.7679(19),0.338(11), {{{-}}} ,0.338(11)
N101,0.1224( 5),0.3584( 6),0.5803( 9),0.340( 4), {{{-}}} ,0.340( 4)
C101,0.0962( 6),0.3248( 8),0.5335(11),0.334( 4),0.326( 3),0.326( 3)
B450,0.1611( 4),0.5079( 6),0.8290( 8),0.337( 1), {{{-}}} ,0.337( 1)
S400,0.1359( 4),0.4776( 5),0.7868( 7),0.312( 4), {{{-}}} ,0.312( 4)
N401,0.1100( 6),0.3419( 8),0.5572(11),0.303( 6),0.299( 2),0.299( 2)
N451,0.1109( 3),0.3431( 4),0.5589( 5),0.302( 4), {{{-}}} ,0.302( 4)
D450,0.0836( 4),0.2579( 5),0.4200( 7),0.303( 8), {{{-}}} ,0.303( 8)
H200,0.1363( 5),0.4781( 5),0.7874( 8),0.286( 3), {{{-}}} ,0.286( 3)
N202,0.1342( 3),0.3750( 4),0.6036( 6),0.280( 3), {{{-}}} ,0.280( 3)
N203,0.1127( 2),0.3454( 3),0.5621( 4),0.269( 2),0.268( 1),0.268( 1)
N200,0.0923( 3),0.3204( 3),0.5273( 4),0.260( 4),0.252( 2),0.252( 2)
D200,0.0651( 3),0.2357( 3),0.3890( 4),0.256( 3),0.250( 2),0.250( 2)
E250,0.0422( 3),0.1557( 3),0.2574( 4),0.242( 5),0.251( 4),0.251( 4)
N300,0.1062( 2),0.3371( 3),0.5505( 4),0.222( 3), {{{-}}} ,0.222( 3)
N302,0.0872( 3),0.3146( 4),0.5193( 5),0.216( 3), {{{-}}} ,0.216( 3)
J303,0.0648( 2),0.2353( 2),0.3885( 3),0.194( 3),0.200( 2),0.200( 2)
E300,0.0437( 2),0.1574( 2),0.2597( 2),0.198( 2), {{{-}}} ,0.198( 2)
\end{filecontents*}
\csvreader[
      late after line=\ifthenelse{\equal{\CLSlabel}{B450}\or\equal{\CLSlabel}{H200}\or\equal{\CLSlabel}{N300}}{\\\midrule}{\\},
      filter not strcmp={\CLSlabel}{N401},
    ]{tables/bounding_method_energies.csv}%
      {label=\CLSlabel,m_pi=\aMpi,E_2pi=\aEtwopi,E_3pi=\aEthreepi,m_rho=\aMrho,m_rho_ABHM=\aMrhoABHM}%
      {\CLSlabel & \aMpi & \aEtwopi & \aEthreepi & \ifthenelse{\equal{\aMrhoABHM}{{{-}}}}{}{*} & \aMrho}
    \bottomrule
  \end{tabular}%
  \begin{tabular}{r@{}S[table-format=1.2(2)]S[table-format=1.2(2)]}
    \toprule
    & {$m_\rho/m_\pi$} & {$g_{\rho\pi\pi}$} \\
    \midrule
    \begin{filecontents*}{tables/GSmodel_parms.csv}
label,m_pi,mr_rho_cl,g_rhopipi_cl,mr_rho_ll,g_rhopipi_ll
H101,0.1836( 5),2.04( 1),4.80( 2),2.04( 1),4.81( 2)
H102,0.1546( 6),2.32( 2),4.84( 4),2.32( 2),4.85( 4)
H105,0.1235(13),2.74( 9),4.98(19),2.74( 9),5.00(19)
N101,0.1224( 5),2.77( 4),4.92( 6),2.78( 4),4.91( 6)
C101,0.0962( 6),3.48( 4),4.80( 4),3.48( 4),4.81( 4)
B450,0.1611( 4),2.09( 1),4.82( 1),2.09( 1),4.82( 1)
S400,0.1359( 4),2.30( 3),5.00( 5),2.29( 3),5.02( 5)
N401,0.1100( 6),2.75( 6),4.90(12),2.76( 6),4.86(13)
N451,0.1109( 3),2.73( 4),4.98( 7),2.73( 4),4.97( 7)
D450,0.0836( 4),3.67(10),4.64(19),3.63(10),4.72(19)
H200,0.1363( 5),2.10( 2),4.86( 4),2.10( 2),4.86( 4)
N202,0.1342( 3),2.08( 2),4.88( 6),2.08( 2),4.87( 6)
N203,0.1127( 2),2.39( 2),4.90( 4),2.39( 2),4.91( 4)
N200,0.0923( 3),2.81( 5),4.94(10),2.82( 5),4.92( 9)
D200,0.0651( 3),3.92( 5),4.86( 5),3.92( 5),4.86( 5)
E250,0.0422( 3),5.71(13),5.01( 9),5.74(13),4.99( 9)
N300,0.1062( 2),2.09( 3),4.98( 6),2.09( 3),4.98( 6)
N302,0.0872( 3),2.47( 4),4.98( 9),2.47( 4),4.96( 9)
J303,0.0648( 2),3.01( 5),5.09( 8),2.99( 4),5.12( 7)
E300,0.0437( 2),4.53( 5),4.77( 2),4.54( 5),4.77( 2)
\end{filecontents*}
\csvreader[
      late after line=\ifthenelse{\equal{\CLSlabel}{B450}\or\equal{\CLSlabel}{H200}\or\equal{\CLSlabel}{N300}}{\\\midrule}{\\},
      filter not strcmp={\CLSlabel}{N401},
    ]{tables/GSmodel_parms.csv}%
      {label=\CLSlabel,m_pi=\aMpi,mr_rho_cl=\mrcl,g_rhopipi_cl=\gcl,mr_rho_ll=\mrll,g_rhopipi_ll=\gll}%
      { & \mrll & \gll}
    \bottomrule
  \end{tabular}
  \caption{%
    From left to right, label of the CLS ensemble, pion mass, energy
    of the two- and three-pion non-interacting finite-volume states, and
    rho meson mass used in the bounding method. $E_{3\pi}$ is obtained
    employing eq.~\eqref{eq:three-pion-state}. The estimate of $m_\rho$ is
    obtained from a fit to the local-local discretization of the
    correlator $G^{33}$ as described in section~\ref{sec:FSE_MLLGS},
    except when a value is available from a dedicated
    study~\cite{Andersen:2018mau}. In this case, the entry is marked by an
    asterisk (see also table VII in ref.~\cite{Gerardin:2019rua}).
    In the last two coloumns we list $m_\rho/m_\pi$ and
    $g_{\rho\pi\pi}$ which serve as input parameters for the
    Gounaris-Sakurai model used in the \acs{MLLGS} method to correct for
    finite-size effect. For this purpose, we always use the parameters
    obtained from the fit to $G^{33}$.
  }\label{tab:spectroscopy}\label{tab:GSmodel_parms}
\end{table}

It is possible to apply the bounding method also to the $\SPi^{08}$
contribution, with some additional caveats. Indeed, the $G^{08}$
correlator does not have the positive-definite spectral representation
that is needed for eq.~\eqref{eq:bounding-method} to be valid in
general. We know, however, that $G^{08}$ has the same $E_0$ as
$G^{88}$ and that the corresponding amplitude $a_0$ is positive.
Furthermore, the correlator $G^{08}$ approaches its asymptotic
behavior $\sim a_0\exp{-E_0 t}$ from below. Likewise,
$E_{\mathrm{eff}}$ approaches $E_0$ from below.
It follows that, for any $t\geq\tcut$, the correlator $G^{08}(t)$ is
bounded by  $G^{08}(\tcut) e^{-E_{\mathrm{eff}}(\tcut) (t-\tcut)}$ and
$G^{08}(\tcut) e^{-E_0 (t-\tcut)}$ from above and below, respectively,
which is opposite to eq.~\eqref{eq:bounding-method}. We exploit this
fact to apply the bounding method to the $\SPi^{08}$ contribution,
choosing to average the bounds in the same interval of $\tcut$ values
used for $\SPi^{88}$, which, as direct inspection shows, is a
conservative choice. Similarly to the case of $\SPi^{88}$, we apply
the bounding method also to the connected contribution
$\SPi^{08}_{\con}$ and, after taking the difference with $\SPi^{08}$,
obtain a more precise estimate of $\SPi^{08}_{\dis}$.

Finally, we note that the charm correlator does not require any
specific treatment of the tail since it has a very fast exponential
decay and higher precision. The pion masses and energy levels that
enter the bounding method are listed for each ensemble in
table~\ref{tab:spectroscopy}.

\subsection{Correction for finite-size effects}
\label{sec:FSE}

Lattice QCD simulations are performed in a periodic box of finite volume $L^3$
and finite Euclidean time extent $T$. In order to obtain reliable
estimates for $\Dalphahad$, the results must be corrected for finite-size
effects. The leading effect is a shift of the vector correlator that,
for the volumes and pion masses considered here, is of order
$\exp{-m_\pi L}$ and dominated by the $\pi\pi$ channel. It follows
that the $I=1$ contribution $\SPi^{33}$ is mostly affected by
finite-size effects. To correct for this, we follow a strategy similar
to ref.~\cite{Francis:2013qna,DellaMorte:2017dyu,Gerardin:2019rua}. To
this end, we compute the difference between the $I=1$ vector
correlator in infinite and finite volume as a function of Euclidean
time $t$. Depending on the value of $t$ in physical units, different
methods are considered to determine the finite-size correction
reliably.

The $\pi\pi$ contribution to the $I=1$ vector correlator can be
computed in \ac{chiPT}, both in finite and infinite volume. 
In our earlier works~\cite{DellaMorte:2017dyu,Gerardin:2019rua,Ce:2019imp}, we used \ac{chiPT} at \ac{NLO} to correct the correlator at short Euclidean times for finite-size effects, applying the formula given in eq.~(C.4) of
ref.~\cite{DellaMorte:2017dyu} (see also
ref.~\cite{Francis:2013qna}).
This very simple model corresponds to the correction from
noninteracting pions and is known to only account for a fraction of
the finite-volume correction to $\SPi(-Q^2)$ at $Q^2$ values of
$\order*{\SI{1}{\GeV\squared}}$~\cite{Aubin:2015rzx}.
A better estimate of the correction can be obtained using \ac{chiPT} at \ac{NNLO}~\cite{Bijnens:2017esv,Aubin:2019usy}, or using the \ac{HP} method described in section~\ref{sec:FSE_HP}.
We choose to employ the latter for the finite-size correction on the correlator at short time distances.
For the correction  at long time distances, we
use either the \ac{HP} method or the same method as in
refs.~\cite{DellaMorte:2017dyu,Gerardin:2019rua}, described in the next
section.

\begin{figure}[t]
  \centering
  \scalebox{.55}{\input{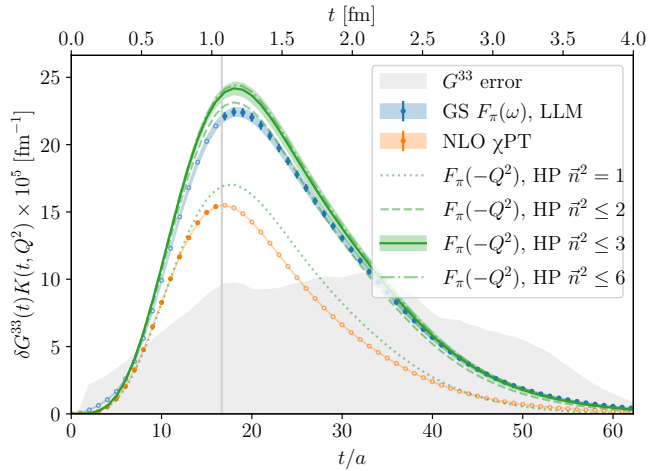}}
  \caption{%
    Comparison of the \ac{TMR} integral at $Q^2=\SI{1}{\GeV\squared}$ of the finite-size correction
    to the $G^{33}$ correlator on the D200 ensemble as a function of the
    \ac{TMR} integration time.
    The \acs{MLLGS} method discussed in section~\ref{sec:FSE_MLLGS}
    (blue points and band) agrees well with the \acs{HP} method discussed
    in section~\ref{sec:FSE_HP} (green line and band),
    while the \acs{NLO} \acs{chiPT} result using eq.~(C.4) from
    ref.~\cite{DellaMorte:2017dyu} (orange points and line) underestimates
    the finite-size correction except at very short time distances.
    The vertical gray line is at $\ti=(m_\pi L/4)^2/m_\pi$.
    For comparison, the gray shaded area indicates the statistical error on the $G^{33}$ correlator.
  }\label{fig:FSE_comparison}
\end{figure}

\subsubsection{Meyer-Lellouch-Lüscher formalism with \acl{GS} parametrization}
\label{sec:FSE_MLLGS}

An accurate description of the finite-size correction on the
correlator tail is obtained making use of a more realistic model for
the time-like pion form factor $F_\pi(\omega)$. In infinite volume the
$\pi\pi$ contribution to the $I=1$ correlator has the spectral
function representation~\cite{OConnell:1995nse,Jegerlehner:2009ry}
\begin{equation}
\label{eq:corr_LLM_infinite}
  G^{33}(t,\infty) = \int_0^\infty \dd{\omega}\omega^2 \rho(\omega^2) \e{-\omega t} , \qquad
  \rho(\omega^2) = \frac{1}{48\cpi^2} \left( 1-\frac{4m_\pi^2}{\omega^2} \right)^{\frac{3}{2}} \abs{F_\pi(\omega)}^2 ,
\end{equation}
while the finite volume correlator is a sum of exponentials
\begin{equation}
\label{eq:corr_LLM_finite}
  G^{33}(t,L) = \sum_n \abs{A_n}^2 \e{-\omega_n t} ,
\end{equation}
with the finite-volume energies $\omega_n$ and amplitudes $A_n$. In
ref.~\cite{Meyer:2011um} it was realized that the amplitudes $A_n$ are
proportional to the timelike pion form factor $F_\pi(\omega)$, with
the proportionality given by a Lellouch-Lüscher
factor~\cite{Luscher:1991cf,Lellouch:2000pv}. Thus, knowledge of the
pion form factor allows one to work out the finite-size correction.
As in our earlier work \cite{DellaMorte:2017dyu,Gerardin:2019rua}, we
use the \ac{GS} parametrization~\cite{Gounaris:1968mw} of
$F_\pi(\omega)$, which depends on only two parameters $m_\rho/m_\pi$
and $g_{\rho\pi\pi}$, and refer to this approach as the \ac{MLLGS}
method. The two parameters are determined empirically from a fit to
our correlator data. In the absence of a better way to isolate its
$\pi\pi$ contribution, we restrict the fit to the tail of the
correlator, and we cut off, using a smoothed step function, the
spectral function representation in eq.~\eqref{eq:corr_LLM_infinite}
at some energy corresponding to the inelastic threshold.
Correspondingly, the sum over Lüscher energies and
Lellouch-Lüscher amplitudes in eq.~\eqref{eq:corr_LLM_finite} is
limited to the same cut-off energy. Non-elastic interactions become
important around the heuristic value
$m_\rho+m_\pi$~\cite{Bernecker:2011gh}, which we use in the smooth
cut-off function. The parameters that we obtain in this way are
tabulated in table~\ref{tab:GSmodel_parms}.

We emphasize that we do not assume that the \ac{GS} parametrization
can be used to accurately model the tail of the correlator.
Instead, we use the model only to correct for the relatively small
finite-volume effect on the correlator.
In the future, we plan to further reduce the model dependence,
employing, where available, a full lattice determination of
$F_\pi(\omega)$~\cite{Andersen:2018mau,Erben:2019nmx} instead of the
\ac{GS} parametrization. The $F_\pi(\omega)$-based model provides a
good spectral representation of the correlator up to the inelastic
threshold, thus we use it for the correlator correction at $t>\ti$,
with $\ti=(m_\pi L/4)^2/m_\pi$, as in
refs.~\cite{DellaMorte:2017dyu,Gerardin:2019rua}.

\subsubsection{\acl{HP} method}
\label{sec:FSE_HP}

An alternative method to correct for the finite-size effect on the correlator has been proposed by Hansen and
Patella~\cite{Hansen:2019rbh,Hansen:2020whp}. Here the leading
finite-volume effects are determined to all orders with respect to the
interactions of a generic, relativistic effective field theory of
pions. Their result is an expansion in the squared momentum vector,
$\abs{\vec{n}}^2=1,2,3,6,\dots$, with each term of order
$\exp\{-\abs{\vec{n}}m_\pi L\}$, and the first neglected effect
arising from a sunset diagram of order $\exp\{-\sqrt{2+\sqrt{3}}m_\pi
L\}\approx\exp\{-\num{1.93}m_\pi L\}$. The coefficient of each term in
the expansion is given by the forward Compton amplitude of the pion,
which is decomposed into a pole and a regular piece. Following
ref.~\cite{Hansen:2020whp}, the dominant contribution is coming from
the pole, and is expressed in terms of the electromagnetic form factor
$F(-Q^2)$ of the pion in the spacelike region $Q^2>0$. In this work,
we model the latter using the monopole representation that describes
the data in ref.~\cite{Brommel:2006ww},
\begin{equation}
  F(-Q^2) = \frac{1}{1+Q^2/M^2(m_\pi^2)} \,, \qquad M^2(m_\pi^2) = \SI{0.517(23)}{\GeV\squared} + \num{0.647(30)} \, m_\pi^2 \,,
\end{equation}
albeit in $\Nf=2$
QCD. The remaining regular piece, which is independent of the pion
form factor, is at most \SI{1}{\percent} of the pole contribution and
can be safely neglected.
As in the case of the \ac{MLLGS} method, this relatively crude
modelling is sufficient, given that it is only used to estimate the
finite-size correction.

While the expansion converges to the leading-order finite-size
correction to the correlator at any Euclidean time $t$, the
convergence is faster at short time distances. Therefore, we use the
sum of the first three terms, $\vec{n}^2=1$, $2$ and $3$, to estimate
the finite-size correction to the \ac{TMR} integrand for either the
whole $t$-range, or for $t<\ti$.
The $\vec{n}^2=3$ term is the last term that we can compute that is parametrically larger than the unknown sunset diagram contribution, and its size is thus taken as a conservative systematic error from the series truncation.
This has a comparable size to the statistical error from our measurement of $m_\pi$.

\begin{table}[t]
  \centering
  \begin{tabular}{rS[table-format=2.2]S[table-format=2.1(1.1)]S[table-format=2.1(1.1)]S[table-format=1.2]S[table-format=2.2(2)]S[table-format=2.1(1.1)]S[table-format=2.1(2.1)]}
    \toprule
                  & \multicolumn{3}{c}{whole $t$ range}     & {$t_i$}        & {$t<t_i$} & {$t>t_i$}    & \\
    $\times 10^5$ & {\ac{chiPT}} & {\ac{HP}} & {\ac{MLLGS}} & {$[\si{\fm}]$} & {\ac{HP}} & {\ac{MLLGS}} & {combined} \\
    \midrule
    \begin{filecontents*}{tables/FSEcorr_contribs.csv}
label,ti,ti_fm,sQED_Pi,sQED_K,HPmethod,GSmodel,sQED_short,HPmethod_short,HPmethod_long,GSmodel_long,sQED_GSmodel,HPmethod_GSmodel,diff_HPmethod_GSmodel_long,HPmethod_HPmethod_GSsyst,HPmethod_HPmethod_GSmodel_avg,Mpi,MK,Fpi,FK
H101,11.7,1.01, 3.37, 1.68,12.0( 5),10.5( 2), 2.31(  0), 4.91( 10), 7.1( 4), 5.7( 1), 8.0( 1),10.6( 2), 1.4( 4),12.0(15),11.3(16),+0.012,+0.012,-0.11,-0.11
H102, 9.9,0.85, 8.23, 1.31,26.9(14),23.9( 5), 3.68(  1), 6.44( 18),20.4(12),17.5( 4),21.2( 4),23.9( 4), 3.0(13),26.9(33),25.4(34),+0.051,+0.000,-0.23,-0.12
H105, 7.9,0.68,21.30, 0.94,61.6(50),57.3(32), 4.51(  1), 6.74( 36),54.9(47),50.7(31),55.2(31),57.4(31), 4.2(55),61.6(66),59.5(56),+0.145,+0.000,-0.60,-0.25
N101,17.6,1.52, 4.27, 0.06, 8.6( 3), 8.2( 2), 3.23(  0), 5.38( 15), 3.3( 2), 2.9( 1), 6.2( 1), 8.3( 2), 0.3( 2), 8.6( 5), 8.5( 5),+0.010,+0.000,-0.04,-0.02
C101,13.8,1.20,13.78, 0.05,25.2(14),23.9( 7), 6.37(  1), 9.29( 34),16.0(11),15.1( 5),21.5( 5),24.4( 6), 0.8(12),25.2(17),24.8(16),+0.038,+0.000,-0.15,-0.06
B450,10.3,0.79, 6.10, 3.05,24.3(11),21.0( 4), 2.62(  2), 4.72( 11),19.6(10),16.1( 3),18.7( 3),20.8( 3), 3.5(10),24.3(36),22.5(38),+0.031,+0.031,-0.28,-0.28
S400, 8.7,0.66,13.37, 2.27,49.8(30),45.6(12), 3.87(  2), 6.22( 21),43.6(29),39.1(11),43.0(11),45.3(11), 4.5(31),49.8(54),47.6(52),+0.114,+0.000,-0.53,-0.29
N401,15.8,1.21, 7.08, 0.16,15.9(11),14.6( 5), 4.26(  1), 7.04( 40), 8.8( 7), 7.9( 3),12.2( 3),14.9( 5), 0.9( 8),15.9(15),15.4(15),+0.022,+0.000,-0.09,-0.04
N451,16.0,1.22, 6.79, 0.16,15.2( 6),14.1( 3), 4.11(  1), 6.77( 16), 8.5( 5), 7.6( 2),11.7( 2),14.3( 2), 0.9( 5),15.2(11),14.8(12),+0.021,+0.000,-0.09,-0.03
D450,21.4,1.63, 7.34, 0.01,12.2( 5),11.6( 3), 4.67(  1), 6.66( 21), 5.5( 3), 5.0( 2), 9.7( 2),11.7( 3), 0.5( 4),12.2( 7),11.9( 7),+0.013,+0.000,-0.05,-0.02
H200, 8.7,0.56,11.36, 5.70,59.9(37),49.3(11), 2.91(  1), 4.81( 17),55.1(35),44.2(10),47.1(10),49.0(10),10.9(37),59.9(115),54.4(122),+0.096,+0.096,-0.86,-0.86
N202,19.3,1.24, 2.11, 1.06, 6.6( 2), 5.8( 1), 1.66(  0), 3.57(  8), 3.0( 2), 2.3( 1), 4.0( 1), 5.9( 1), 0.7( 2), 6.6( 7), 6.2( 8),+0.005,+0.005,-0.05,-0.05
N203,16.2,1.04, 5.70, 0.67,16.2( 7),14.3( 2), 3.25(  1), 5.78( 13),10.5( 6), 8.7( 2),11.9( 2),14.5( 2), 1.8( 6),16.2(19),15.3(21),+0.024,+0.000,-0.11,-0.05
N200,13.3,0.85,14.40, 0.49,37.1(22),33.5( 8), 4.56(  1), 6.94( 23),30.1(20),26.7( 7),31.3( 7),33.7( 7), 3.4(21),37.1(41),35.4(41),+0.070,+0.000,-0.29,-0.12
D200,16.7,1.07,20.73, 0.04,36.1(24),34.1( 6), 7.32(  1),10.26( 39),25.8(20),24.6( 4),31.9( 4),34.8( 6), 1.2(21),36.1(27),35.5(23),+0.056,+0.000,-0.22,-0.08
E250,24.3,1.56,25.26, 0.00,31.7(25),34.1( 8), 9.04(  0),11.11( 50),20.6(20),22.3( 6),31.4( 6),33.4( 8),-1.7(21),31.7(30),32.6(27),+0.031,+0.000,-0.12,-0.05
N300,15.3,0.76, 6.34, 3.17,26.4(12),22.5( 6), 2.66(  0), 4.75( 11),21.7(12),17.6( 5),20.3( 5),22.3( 5), 4.1(13),26.4(43),24.4(45),+0.033,+0.033,-0.29,-0.29
N302,12.5,0.62,15.37, 2.22,58.2(38),48.2(14), 3.81(  1), 5.96( 22),52.3(36),42.2(13),46.0(13),48.2(13),10.1(39),58.2(108),53.2(114),+0.131,+0.000,-0.60,-0.31
J303,16.6,0.83,19.16, 0.34,45.3(31),42.6(10), 5.62(  1), 8.38( 32),36.9(28),34.5( 9),40.2( 9),42.9(10), 2.4(29),45.3(39),44.1(33),+0.086,+0.000,-0.35,-0.14
E300,25.1,1.25,21.17, 0.01,32.3(22),31.3( 4), 7.77(  1),10.42( 40),21.9(18),21.4( 3),29.2( 3),31.9( 5), 0.5(18),32.3(22),32.1(16),+0.041,+0.000,-0.16,-0.06
\end{filecontents*}
\csvreader[
      late after line=\ifthenelse{\equal{\CLSlabel}{B450}\or\equal{\CLSlabel}{H200}\or\equal{\CLSlabel}{N300}}{\\\midrule}{\\},
      filter not strcmp={\CLSlabel}{N401},
    ]{tables/FSEcorr_contribs.csv}%
      {label=\CLSlabel,ti=\ti,ti_fm=\tifm,sQED_Pi=\sQED,HPmethod=\HPm,GSmodel=\GSm,sQED_short=\sQEDs,HPmethod_short=\HPms,GSmodel_long=\GSml,sQED_GSmodel=\sQEDGSm,HPmethod_GSmodel=\HPmGSm,HPmethod_HPmethod_GSmodel_avg=\HPmHPmGSmavg}%
      {\CLSlabel & \sQED & \HPm & \GSm & \tifm & \HPms & \GSml & \HPmHPmGSmavg}
    \bottomrule
  \end{tabular}
  \caption{%
    Finite-size corrections to the $I=1$ \ac{HVP} function $\SPi^{33}(-Q^2)\times 10^5$ at $Q^2=\SI{1}{\GeV\squared}$ for each individual gauge ensemble.
    Columns \num{2}--\num{4} show the finite-size effects estimated using \ac{NLO} \ac{chiPT}, the \acf{HP} method and the \acf{MLLGS} formalism, respectively.
    In the following columns we list the corrections for time distances shorter than $t_i$, estimates using the \ac{HP} method, as well as for distances greater $t_i$ obtained via the \ac{MLLGS} formalism.
    In the last column we specify the chosen combination to correct for finite-size effects on $\SPi^{33}$,
    obtained from the \ac{HP} method at $t<t_i$, and the average of the \ac{HP} and \ac{MLLGS} values, with the difference added to the error as an additional systematic, at $t>t_i$.
  }\label{tab:FSE_HVP}
\end{table}

The finite-size correction to the \ac{TMR} integrand of
$\SPi^{33}(\SI{1}{\GeV\squared})$ on the D200 ensemble as a function
of $t$ is shown in figure~\ref{fig:FSE_comparison} for the three
different methods considered here, including the \ac{HP} partial
series for different values of $\abs{\vec{n}}$, and compared to the
statistical error on the $I=1$ correlator multiplied by the \ac{TMR}
kernel. It is important to recall that the \ac{MLLGS} and \ac{HP}
methods rely on very different input for the pion form factor in the
time-like and space-like regimes, respectively. Thus, the good
agreement between the \ac{MLLGS} approach and the \ac{HP} method for
$\vec{n}^2\leq 3$, especially for $t\gtrsim2$\,fm demonstrates the
robustness of the evaluation of finite-volume corrections based on
these two procedures. By contrast, the correction obtained from
\ac{chiPT} is significantly smaller.
The integral of the correction, at
$Q^2=\SI{1}{\GeV\squared}$, computed using \ac{chiPT}, the \ac{HP}
method and the \ac{MLLGS} approach is listed for each ensemble in table~\ref{tab:FSE_HVP}.
Regarding the finite-size correction over the whole range of
$t$, one observes that the \ac{MLLGS} and the \ac{HP}
methods produce similar results, while the \ac{chiPT} estimates are between
\num{25}--\SI{75}{\percent} of the other two.
With this in mind, we define our best estimate for the correction of the finite-volume effects of $\SPi^{33}$ in the following way: for short time distances ($t<\ti$) we use the \ac{HP}-method estimate, for long time distances ($t>\ti$) we take the average between the \ac{HP}-method and \ac{MLLGS} values including the difference between the two as an additional source of systematic error, added in quadrature.
The \ac{chiPT} estimate is not used at all.
These short- and long-distance corrections are given for each ensemble in table~\ref{tab:FSE_HVP}, with our best estimate in the last column.

We can directly test the reliability of the finite-size corrections,
by comparing the predictions of the \ac{MLLGS} and \ac{HP} models to
results obtained for two different volumes at otherwise identical
simulation parameters. The corresponding pairs of ensembles are H105
and N101 (at $m_\pi\approx\SI{280}{\MeV}$), as well as H200 and N202
(at the $\SU(3)$-symmetric point). For both sets, we confirmed that
the \ac{TMR} integral contribution, which clearly differs before
correcting for finite-size effects, agrees within errors after the
correction is applied.

We do not correct for subleading finite-size effects in the $I=0$
contributions $\SPi^{88}$ and $\SPi^{08}$, except for the case of
$\SU(3)$-symmetric ensembles, where $\SPi^{88}$ and $\SPi^{33}$, and
thus the respective finite-size-effect corrections, coincide. On these
ensembles, the $\SPi^{33}$ and $\SPi^{88}$ finite-size effects are
further enhanced by a factor of $1.5$, due to the contribution from
kaon loops. Away from the $\SU(3)$-symmetric point, the long-distance
behavior of the partially-quenched $G_{\con}^{88}$ and
$G_{\con}^{08}$ correlators is expected to be dominated by the $I=1$
contribution, with a prefactor of $1/3$ and $1/\sqrt{3}$
respectively. Therefore, we include a finite-size correction for
$\SPi_{\con}^{88}$, which is equal to $1/3$ of that of $\SPi^{33}$ and
which cancels the opposite-sign correction on $\SPi_{\dis}^{88}$. The
same procedure is applied to $\SPi_{\con}^{08}$ and
$\SPi_{\dis}^{08}$.

\subsection{Lattice results}
\label{sec:lat_res}

\begin{table}[p]
  \centering
  \begin{tabular}{rS[table-format=4(2)]S[table-format=4(2)]S[table-format=+3(2)]S[table-format=4(2)]S[table-format=3(2)]S[table-format=+3(2)]S[table-format=3(2)]}
    \toprule
    $\times 10^5$ & {$\SPi^{33}$} & {$\SPi^{88}_{\con}$} & {$\SPi^{88}_{\dis}$} & {$\SPi^{88}$} & {$\SPi^{08}_{\con}$} & {$\SPi^{08}_{\dis}$} & {$\SPi^{08}$} \\
    \midrule
    \begin{filecontents*}{tables/contribs.csv}
label,cl33,cl88c,cl88d,cl88,cl08c,cl08d,cl08,ll33,ll88c,ll88d,ll88
H101,2855( 6),2855( 6),   0    ,2855( 6),   0    ,   0    ,   0    ,2769( 7),2769( 7),   0    ,2769( 7)
H102,2985(10),2749( 7), -12( 2),2737( 6), 209( 5), -39( 7), 170( 8),2899(10),2663( 7), -12( 2),2651( 7)
H105,3156(21),2667(11), -13( 8),2654(13), 429(12), -50(20), 379(21),3069(21),2581(11), -15( 8),2566(13)
N101,3170(10),2683( 5), -31( 4),2652( 6), 430( 6), -70(15), 360(18),3086(10),2597( 5), -33( 4),2564( 6)
C101,3349(13),2682( 7), -60( 7),2622( 9), 588( 8),-134(23), 454(24),3264(14),2597( 7), -63( 7),2534(10)
B450,2764( 9),2764( 9),   0    ,2764( 9),   0    ,   0    ,   0    ,2696( 9),2696( 9),   0    ,2696( 9)
S400,2903(13),2659( 8), -14( 2),2645( 8), 216( 6), -36( 9), 180(10),2836(14),2593( 9), -15( 2),2578( 8)
N401,3132(14),2647( 7), -40( 6),2607( 9), 427( 8), nan(nan), nan(nan),3069(15),2582( 8), -45( 6),2537( 9)
N451,3096( 7),2628( 3), -22( 3),2606( 4), 412( 4), -50(11), 363(11),3030( 7),2562( 3), -23( 3),2539( 4)
D450,3279(10),2605( 4), -37( 6),2568( 6), 591( 6), -76(19), 516(18),3214(10),2539( 4), -41( 6),2498( 6)
H200,2697(21),2697(21),   0    ,2697(21),   0    ,   0    ,   0    ,2651(21),2651(21),   0    ,2651(21)
N202,2736(12),2736(12),   0    ,2736(12),   0    ,   0    ,   0    ,2689(13),2689(13),   0    ,2689(13)
N203,2878( 9),2620( 7), -10( 2),2610( 7), 225( 5), -26( 9), 199(11),2830(10),2573( 7), -10( 2),2563( 7)
N200,3023(11),2549( 5), -22( 5),2527( 6), 414( 7), -56(19), 359(20),2977(11),2502( 5), -24( 5),2478( 6)
D200,3248(12),2535( 5), -53( 8),2481( 9), 621( 8),-131(22), 490(23),3200(12),2487( 5), -56( 8),2432( 9)
E250,3530(21),2586( 7),-132(14),2454(14), 826(14),-257(31), 569(31),3482(21),2540( 7),-136(14),2404(14)
N300,2596(13),2596(13),   0    ,2596(13),   0    ,   0    ,   0    ,2569(13),2569(13),   0    ,2569(13)
N302,2737(16),2470( 8), -12( 5),2458( 8), 234( 8),  -9(17), 225(16),2710(16),2442( 8), -12( 5),2430( 8)
J303,3028(18),2455( 8), -35( 5),2420(10), 498(10), -96(14), 402(18),3002(18),2429( 8), -37( 5),2392(10)
E300,3268(27),2462( 9), -66(16),2396(16), 702(16),-171(48), 530(52),3242(27),2434( 9), -67(16),2368(16)
\end{filecontents*}
\csvreader[
      late after line=\ifthenelse{\equal{\CLSlabel}{B450}\or\equal{\CLSlabel}{H200}\or\equal{\CLSlabel}{N300}}{\\\midrule}{\\},
      filter not strcmp={\CLSlabel}{N401},
    ]{tables/contribs.csv}%
      {label=\CLSlabel, cl33=\cltt, cl88c=\cleec, cl88d=\cleed, cl88=\clee, cl08c=\clzec, cl08d=\clzed, cl08=\clze, ll33=\lltt, ll88c=\lleec, ll88d=\lleed, ll88=\llee}%
      {\CLSlabel & \cltt & \cleec & \cleed & \clee & \clzec & \clzed & \clze \\
                 & \lltt & \lleec & \lleed & \llee}
    \bottomrule
  \end{tabular}
  \caption{%
    Estimate of connected and disconnected contribution to $\SPi(-Q^2)\times 10^5$ at $Q^2=\SI{1}{\GeV\squared}$ for the conserved-local (first line) and, when available, local-local (second line) discretization.
    Contributions tabulated as $0$ vanish exactly due to SU(3)-symmetry.
    The contributions are estimated applying the bounding method as explained in section~\ref{sec:bounding_method} and the correction for finite-size effects as of section~\ref{sec:FSE}.
  }\label{tab:results_lattice}
\end{table}

\begin{figure}[t]
  \centering
  \scalebox{.55}{\input{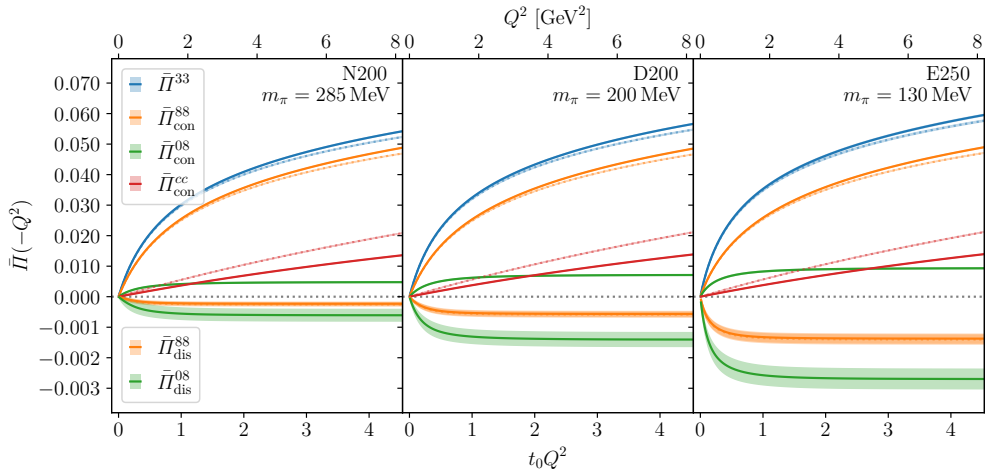}}
  \caption{%
    Running with energy $Q^2$ of different contributions to $\SPi(-Q^2)$ on three different ensembles at $a\approx\SI{0.064}{\fm}$.
    The conserved-local discretization is shown and, when available, the local-local discretization in a lighter color shade.
    The negative side of the vertical axis of the plot is inflated by a factor \num{10} with respect to the positive side.
  }\label{fig:running_lattice}
\end{figure}

We are now in a position to present our finite-volume corrected
results on all our ensembles. Figure~\ref{fig:running_lattice} shows
the running of different contributions to $\SPi(-Q^2)$, defined
through the correlators in eq.~\eqref{eq:corr_flavor}, as a function
of $Q^2$ on three different lattices at the same lattice spacing with
increasingly lighter pions.
While $\SPi(-Q^2)$ is dimensionless, the \ac{TMR} kernel $K(t,Q^2)$ in eq.~\eqref{eq:TMRkernel} and $Q^2$ itself are dimensionful quantities, thus scale setting is needed to translate $Q^2$-values in \si{\GeV\squared} to lattice units.
For this purpose, we insert into the kernel the dimensionless product $t_0Q^2$,\footnote{With the exception of the charm contribution and its extrapolation for which $\tnotsym Q^2$ is used, see section~\ref{sec:extrap_charm}.} where the gradient flow scale $t_0$ introduced in section~\ref{sec:lattice_setup} has been computed on each ensemble, see appendix~\ref{sec:meson} and table~\ref{tab:meson}.
Results on each ensemble at
$Q^2=\SI{1}{\GeV\squared}$ are given in
table~\ref{tab:results_lattice}.
For $\SPi^{08}$, both connected and disconnected, only the results with the conserved-local discretization are available, for the reasons discussed in section~\ref{sec:renorm}.
At the $\SU(3)$-symmetric point $\SPi^{88}_{\con}=\SPi^{33}$, and
$\SPi^{88}_{\dis}$ as well as both components of $\SPi^{08}$ vanish
exactly. The corresponding entries in the table are set to zero.
As one moves away from the
$\SU(3)$-symmetric point, the $\SPi^{33}$ contribution increases,
while the $\SPi_{\mathrm{conn}}^{88}$ contribution becomes smaller.
The quenched charm contribution turns out to be relatively independent
of the pion mass and increases linearly in the range of $Q^2$ values.
The (negative) quark-disconnected contributions are also shown, on a
scale enlarged by a factor \num{10}. It is worth noting that
$\SPi_{\mathrm{disc}}^{88}(-Q^2)$ is constant for
$Q^2\gtrsim\SI{0.5}{\GeV\squared}$, as predicted by perturbation
theory.

\section{Results at the physical point}
\label{sec:extrapolation}

Thanks to the availability of ensembles with four different lattice
spacings and several quark masses, we can reliably extrapolate our
results in section~\ref{sec:lat_res} to vanishing lattice spacing and
physical values of the pseudoscalar meson masses. 
We define the target \enquote{physical} point in the isospin limit
fixing $m_\pi=m_{\pi^0}$ and
$m_K^2-m_\pi^2/2=(m_{K^+}^2+m_{K^0}^2)/2-m_{\pi^+}^2/2$~\cite{Lellouch:discussion2021workshop,Urech:1994hd,Neufeld:1995mu},
which results in $m_\pi=\SI{134.9768(5)}{\MeV}$ and
$m_K=\SI{495.011(10)}{\MeV}$~\cite{Zyla:2020zbs}.
The pion mass of one of our ensembles is slightly below the physical
value, which allows us to interpolate the results.

For $\SPi^{33}(-Q^2)$, $\SPi^{88}(-Q^2)$ and $\SPi^{08}(-Q^2)$, we
perform a combined extrapolation, including both discretizations of
the vector current for $\SPi^{33}(-Q^2)$ and $\SPi^{88}(-Q^2)$. In the
combined fit, we employ an ansatz that implies the same continuum limit for both discretizations, and that encodes the constraints
$\SPi^{33}(-Q^2)=\SPi^{88}(-Q^2)$ and $\SPi^{08}(-Q^2)=0$ at the
$\SU(3)$-symmetric point. The charm contribution $\SPi^{cc}(-Q^2)$ is
treated independently as described in section~\ref{sec:extrap_charm}.

\subsection{Extrapolation strategy}
\label{sec:model}

At any fixed value of $Q^2$ we extrapolate each \ac{HVP} function in
the lattice spacing and the pseudoscalar meson masses to the physical
point. We parametrize the lattice spacing dependence in terms of the
gradient flow time at the SU(3)-symmetric point, $\tnotsym/a^2$, as
determined in ref.~\cite{Bruno:2016plf} and listed in
table~\ref{tab:ensemble}.
As discussed in section~\ref{sec:renorm}, the on-shell quantities
considered in this work are $\order{a}$-improved, and hence we expect
leading discretization effects $\order{a^2}$, up to logarithmic
corrections~\cite{Husung:2019ytz,Ce:2021xgd}. While our favored ansatz
includes only an $\order{a^2}$ term, we also investigate the influence
of higher powers in the lattice spacing, as well as a term
proportional to $a^2\log a$. More details are given in
section~\ref{sec:fit_syst}.

The dependence of the \ac{HVP} function contributions on the meson
masses $m_\pi$ and $m_K$ is modelled using the proxy quantities
$\phi_2=8t_0m_\pi^2$ and $\phi_4=8t_0(m_K^2+m_\pi^2/2)$.
At the isospin-symmetric reference point defined above, the target
values of our extrapolation are
$\phi_2^{\mathrm{phys}}=\num{0.0806(17)}$ and
$\phi_4^{\mathrm{phys}}=\num{1.124(24)}$, where
the conversion to physical units is performed using
$(8t_0^{\mathrm{phys}})^{1/2}=0.415(4)(2)\,\mathrm{fm}$ from
ref.~\cite{Bruno:2016plf}.\footnote{%
  We note that the authors of ref.~\cite{Bruno:2016plf} obtain
  $(8t_0^{\mathrm{phys}})^{1/2}$ from an extrapolation to a slightly
  different reference point, defined by
  $\bar{m}_\pi=\SI{134.8(3)}{\MeV}$ and 
  $\bar{m}_K=\SI{494.2(3)}{\MeV}$~\cite{Aoki:2016frl}, which
  corresponds to $\bar{\phi}_2=\num{0.0804(18)}$ and
  $\bar{\phi}_4=\num{1.120(24)}$.
  Using the data in table~II of ref.~\cite{Bruno:2016plf}, we have
  translated the published value of $(8t_0^{\mathrm{phys}})^{1/2}$ to
  the reference point used in this paper, which results in an
  increase by \SI{0.2}{\percent} in $\phi_2^{\mathrm{phys}}$ and
  $\phi_4^{\mathrm{phys}}$. At our level of precision, the effect on
  the final results can safely be neglected.
}

For the CLS ensembles considered in this work $\phi_4$ is
approximately constant, with values between \SI{-3.5}{\percent} and
\SI[retain-explicit-plus]{+5.5}{\percent} of the target value
$\phi_4^{\mathrm{phys}}$. Therefore, we only employ a linear term in
$\phi_4$ to model small deviations from the line of constant physics
$m_K^2+m_\pi^2/2=\mathrm{const}$.
The interpolation of the pion-mass dependence across a larger range to the
target value $\phi_2^{\mathrm{phys}}$ is more
complex and quantity-dependent. While it is possible to describe
the \ac{HVP} function in \ac{chiPT} including vector mesons as
resonances in the effective theory~\cite{Ecker:1988te,Aubin:2006xv},
this applies only for $Q^2\lesssim m_\pi^2$ and is thus of limited
relevance in our case. Therefore, we choose to model the dependence by
a polynomial in $\phi_2$. However, understanding the behavior of the
various contribution towards the $\SU(2)$ chiral limit and the
$\SU(3)$-symmetric point helps constrain the model choice.

The isovector ($I=1$) contribution $\SPi^{33}$ dominates the \ac{HVP}
function, especially on ensembles that are close to the physical
masses. Indeed, $\SPi^{33}(-Q^2)$ diverges logarithmically in $m_\pi$
in the limit $m_\pi\to 0$~\cite{Golterman:2017njs}. Therefore, we model the $I=1$ contribution
for the conserved-local discretization with an additional non-polynomial divergent term proportional to $\log\phi_2$,
\begin{multline}\label{eq:model_33}
  \SPi^{33,\consloc}(a^2/\tnotsym,\phi_2,\phi_4) = \SPi^{\mathrm{sym}} + \delta_2^{\consloc} a^2/\tnotsym \\
                                                    + \gamma_1^{33} (\phi_2-\phi_2^{\mathrm{sym}}) + \gamma_{\log}^{33} \log\phi_2/\phi_2^{\mathrm{sym}} + \eta_1 (\phi_4-\phi_4^{\mathrm{sym}}) \,,
\end{multline}
and similarly for the local-local discretization, with
$\delta_2^{\consloc}$ replaced by $\delta_2^{\locloc}$.
We also considered other possibilities for the divergent $m_\pi\to 0$
limit, such as including a $1/\phi_2\sim 1/m_\pi^2$ term in addition
to or instead of the $\log\phi_2$ one~\cite{Colangelo:2021moe}.
However, we observed that on our range of pion masses including only the $\log\phi_2$ term results in the best fit to the data.

The isoscalar ($I=0$) contribution, $\SPi^{88}$, has a finite limit
for $m_\pi\to 0$~\cite{Golterman:2017njs}. Therefore, we do not include any divergent term and use instead a
polynomial quadratic in $\phi_2$, since we observe that a simple linear
scaling does not describe the data. This results in
\begin{multline}\label{eq:model_88}
  \SPi^{88,\consloc}(a^2/\tnotsym,\phi_2,\phi_4) =
\SPi^{\mathrm{sym}} + \delta_2^{\consloc} a^2/\tnotsym \\
                                                    + \gamma_1^{88}
(\phi_2-\phi_2^{\mathrm{sym}}) + \gamma_2^{88} (\phi_2-\phi_2^{\mathrm{sym}})^2 + \eta_1
(\phi_4-\phi_4^{\mathrm{sym}}) \,,
\end{multline}
for the conserved-local discretization, and analogously for the
local-local case.

When eqs.~\eqref{eq:model_33} and~\eqref{eq:model_88} are considered in
isolation, $\SPi^{\mathrm{sym}}$, $\phi_2^{\mathrm{sym}}$ and
$\phi_4^{\mathrm{sym}}$ define an arbitrary subtraction point, for
which only one of the three parameters can be fixed by each fit.
As the label \enquote{sym} suggests, we identify this point with the
$\SU(3)$-symmetric point in the continuum limit, which implies the
constraint $2\phi_4^{\mathrm{sym}}=3\phi_2^{\mathrm{sym}}$. Moreover,
$\SPi^{33,\mathrm{sym}}\equiv\SPi^{88,\mathrm{sym}}$, such that all
three parameters $\SPi^{\mathrm{sym}}$, $\phi_2^{\mathrm{sym}}$ and
$\phi_4^{\mathrm{sym}}$ can be fully determined in a combined fit.

Finally, the $\SPi^{08,\consloc}$ contribution includes one $\SU(3)$-singlet current that vanishes linearly in $m_s-m_\ell$ towards the $\SU(3)$-symmetric point.
To leading order this is proportional to $m_K^2-m_\pi^2$ or equivalently $\phi_4-3/2\phi_2$, thus we model $\SPi^{08,\consloc}$ using a simple linear dependence
\begin{equation}\label{eq:model_08}
  \SPi^{08,\consloc}(a^2/\tnotsym,\phi_2,\phi_4) = \lambda_1 (\phi_4-3/2\phi_2) \,.
\end{equation}
In this case, we fit the only available discretization
(conserved-local) without including a term describing the dependence
on the lattice spacing as no discretization effects are observed
within statistical errors.
Moreover, we do not observe any significant deviation from the linear behavior.

\begin{figure}[t]
  \centering
  \scalebox{.55}{\input{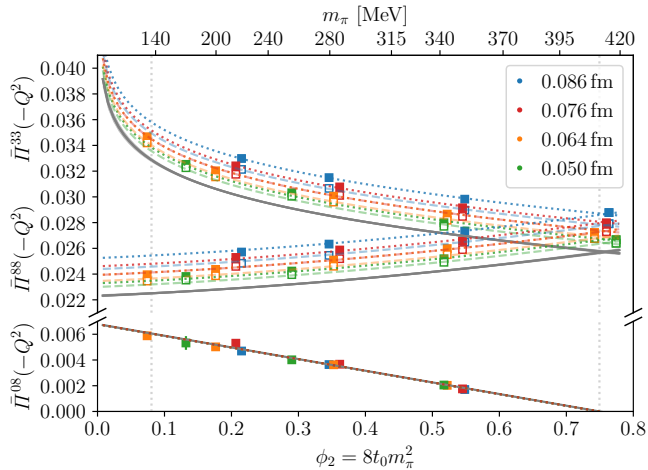}}
  \caption{%
   Plot of $\SPi^{33}(-Q^2)$, $\SPi^{88}(-Q^2)$ (left) and $\SPi^{08}(-Q^2)$ (right) at $Q^2=\SI{1}{\GeV\squared}$ on different ensembles as a function of the pion mass, together with the results of the combined fit.
    With respect to the values in table~\ref{tab:results_lattice}, the data points in the plot include a small shift to the same value of $\phi_4$.
    Filled symbols and dotted lines denote the conserved-local discretization, while open symbols and dashed lines denote the local-local one.
  }\label{fig:extrapolation_results}
\end{figure}

The relative errors on the pion and kaon masses, as well as the scale
$t_0$ that enter $\phi_2$ and $\phi_4$, are of the same order as the
uncertainties of the $\SPi$ contributions. Thus, we fit the quantities
$\phi_2$, $\phi_4$, $\SPi_{\consloc}^{33}$,
$\SPi_{\locloc}^{33}$, $\SPi_{\consloc}^{88}$,
$\SPi_{\locloc}^{88}$ and $\SPi_{\consloc}^{08}$
simultaneously, except for ensembles at the $\SU(3)$-symmetric point
where only the independent quantities $\phi_2=(2/3)\phi_4$,
$\SPi_{\consloc}^{33}=\SPi_{\consloc}^{88}$ and
$\SPi_{\locloc}^{33}=\SPi_{\locloc}^{88}$ are fitted.
We include the correlations between quantities on the same ensemble,
which limits the size of the covariance matrix $C_l$ on ensemble $l$
to either $7\times 7$ or $3\times 3$ in the $\SU(3)$-symmetric case.
Still, we find relatively poor fit quality unless we use a shrunk estimator of the covariance matrix by scaling
the off-diagonal elements of $C_l$ according to~\cite{Ledoit:2004,Touloumis:2015}
\begin{equation}
  \tilde{C}_l(\lambda) = (1-\lambda) C_l + \lambda \diag(C_l) \,.
\end{equation}
We found that the $\chi^2/\mathrm{dof}$ of the fit as a function of the shrinkage parameter $\lambda$ is approximately constant in an interval of small $\lambda$ values, before increasing for $\lambda\to 0$.
Therefore, we select $\lambda=0.05$ as a small value in the constant region.
The errors on the optimal
fit parameters and the extrapolation results are obtained applying the
bootstrap procedure to the fit, and are thus unaffected by this
modification to the covariance matrix.

\subsubsection{Study of the fit model systematics}
\label{sec:fit_syst}

\begin{figure}[t]
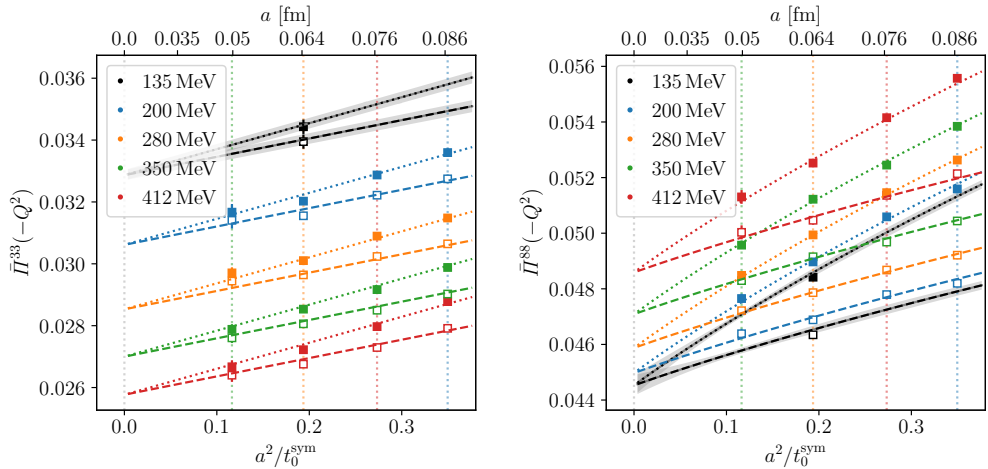

  \centering
  \scalebox{.55}{\input{figures/extrapolation_results_lattice_spacing_phi_33_Q2=1.0.pgf}}%
  \scalebox{.55}{\input{figures/extrapolation_results_lattice_spacing_phi_88_Q2=5.0.pgf}}
  \caption{%
    Plots of $\SPi^{33}(-Q^2)$ at $Q^2=\SI{1}{\GeV\squared}$ (left) and $\SPi^{88}(-Q^2)$ at $Q^2=\SI{5}{\GeV\squared}$ (right) on different ensembles as a function of the lattice spacing, together with the results of the combined fit.
    The left panel shows an example in which the $\order{a^2}$ fit model is used, while a term proportional to $a^3$ has been included in the right plot as well.
    Different colors distinguish different pion masses from a set of five reference values.
    Using the functional form determined by the fit, each data point has been shifted to the closest value of $\phi_2$ matching one of the reference pion masses, and to the same value of $\phi_4$.
    Filled symbols and dotted lines denote the conserved-local
discretization, while open symbols and dashed lines denote the
local-local one.
  }\label{fig:extrapolation_results_lattice_spacing}
\end{figure}

The choice of the fit ansatz introduces a systematic error that we
estimate by considering several variations of the fit model.

The choice of a fit ansatz that constrains both discretizations to have the same continuum limit is motivated by theory.
However, we also check this by performing independent fits including only one of the two discretizations, and observing that the continuum-extrapolated results agree well within errors between different discretizations and with the combined fit ones.
This further supports our choice for the fit ansatz used for the final fit and all the variations in the following.

Our main extrapolation model includes only the leading $\order{a^2}$
discretization effects using two parameters $\delta_2^{\consloc}$
and $\delta_2^{\locloc}$, one for each discretization of the
correlator, common to all flavor contributions in
eqs.~\eqref{eq:model_33} and~\eqref{eq:model_88}. We observe that this
model fits the data well in the energy range below $Q^2$ between
\num{2}--\SI{3}{\GeV\squared}. We also tested a fit model with
independent discretization effects parameters for each flavor
contribution, which resulted in parameters compatible within errors.
Fits to $\SPi^{08}$ were performed without including terms describing
discretization effects (see eq.~\eqref{eq:model_08}), since no
dependence on the lattice spacing could be detected within statistical
errors. Similarly, the fit parameters describing the mass dependence
for both CL and LL discretizations were chosen to be the same.
This is consistent with the choice of not including mass-dependent cut-off effects, and it is supported by the fact that a
fit with independent parameters resulted in compatible results.

For $Q^2$ values larger than \num{2}--\SI{3}{\GeV\squared}, we observe
a rapid deterioration of the quality of the fit. We interpret this as
evidence that discretization effects at larger values of $Q^2$ are not
dominated by $a^2$ effects, but that higher powers in the lattice
spacing are also relevant. Indeed, a modification of
eqs.~\eqref{eq:model_33}, \eqref{eq:model_88} and~\eqref{eq:model_08}
to include both terms proportional to $a^2$ and to $a^3$
fits the data well on an extended range up to
$Q^2\approx\SI{7}{\GeV\squared}$, as was already observed in
ref.~\cite{Ce:2019imp}.
Specifically, we find that an ansatz including
$\delta_2^{\consloc}$ and $\delta_2^{\locloc}$, as well as two
additional parameters $\delta_3^{\consloc}$ and
$\delta_3^{\locloc}$ yields the best fit quality for $\SPi^{33}$,
$\SPi^{88}$ and $\SPi^{08}$. However, including $a^3$ discretization
effects may lead to overfitting the data at lower values of $Q^2$,
where a term proportional to $a^2$ is found to successfully describe
discretization effects.

Therefore, for our final results we switch from the results obtained via a purely $\order{a^2}$ ansatz at low $Q^2$ to those obtained via an ansatz with both $\order{a^2}$ and
$\order{a^3}$ lattice artifacts by applying a smoothed step function
centered around $\SI{2.5}{\GeV\squared}$,
\begin{equation}
  \Theta(Q^2) = \frac{1}{2} + \frac{1}{2}\tanh(\frac{Q^2-\SI{2.5}{\GeV\squared}}{\SI{1.0}{\GeV\squared}}) .
\end{equation}
Both fits agree well within one standard deviation below
$\approx\SI{3}{\GeV\squared}$, and start to disagree above that, in accordance
with the poor quality of the $\order{a^2}$ fit in the high-energy region.
As a consequence, the values of $\SPi$ extrapolated to the physical
point at $Q^2>\SI{2.5}{\GeV\squared}$ are statistically less precise
than values extrapolated at $Q^2<\SI{2.5}{\GeV\squared}$, as it is
clearly visible in figure~\ref{fig:pade_approx}.

To test for possible violations of the leading $\order{a^2}$ scaling due
to the missing $\order{a}$-improvement parameters $f_V$ and
$\bar{c}_V^{\cons,\local}$ we also considered a fit ansatz
with both $a$ and $a^2$ terms. We
observe that this does not describe the data any better than the $\order{a^2}$ fit, so we conclude that residual $\order{a}$ discretization
effects are not significant at our level of precision.

Following ref.~\cite{Ce:2021xgd}, we also considered a
logarithmically-enhanced term of the form
$\tilde{c}_{\SPi}(Q^2)\cdot(a^2/\tnotsym)\log(\tnotsym/a^2)/2$, with
the $Q^2$-dependent coefficient fixed to the free-theory prediction
\begin{equation}
  \tilde{c}_{\SPi}^{\consloc}(Q^2) = \frac{7}{480\cpi^2} \tnotsym Q^2 \,, \qquad \tilde{c}_{\SPi}^{\locloc}(Q^2) = \frac{1}{48\cpi^2} \tnotsym Q^2 \,,
\end{equation}
for both $\SPi^{33}$ and $\SPi^{88}$.
Including this term in eqs.~\eqref{eq:model_33}, \eqref{eq:model_88}
and~\eqref{eq:model_08}, we do not observe any significant change in
the fit quality over the entire range of $Q^2$ values.
The \ac{HVP} functions extrapolated to the physical point are shifted
downwards by less than \SI{0.4}{\percent}, which is always smaller than the
statistical error.

We also tested a variation of the fit model ansatz applying a cut on
the range of pion masses, leaving out those ensembles with
$m_\pi>\SI{400}{\MeV}$. For both the low $Q^2$ and high $Q^2$ fits we
observe a mild deviation with respect to the fit without the mass cut,
always smaller than the statistical error. As was done in
ref.~\cite{Gerardin:2019rua}, we take this as an estimate of the
systematic error due to the chiral and continuum extrapolation and
add it to our error budget.

\subsubsection{Extrapolation of the charm contribution}
\label{sec:extrap_charm}

\begin{figure}[t]
  \centering
  \scalebox{.55}{\input{figures/extrapolation_results_charm.pgf}}
  \caption{%
    Same as figure~\ref{fig:extrapolation_results} for the charm  contribution $\SPi^{cc}(-Q^2)$ at $Q^2=\SI{1}{\GeV\squared}$.
  }\label{fig:extrapolation_results_charm}
\end{figure}

Compared to the isovector and isoscalar channels, the
(quenched) charm contribution $\SPi^{cc}_{\con}$ is smaller and much
more precise. We do not include it in the combined fit and extrapolate
it separately instead, neglecting the small correlation between
$\SPi^{cc}$ and the other channels. For the fit to the conserved-local
discretization we use a linear model in $\phi_2$, \ie
\begin{equation}
  \label{eq:model_charm}
  \SPi^{cc}_{\con}(a^2/\tnotsym,\phi_2) = \SPi^{cc,\mathrm{sym}}_{\con} + \delta_2^{cc,\consloc} a^2/\tnotsym + \gamma_1^{cc} (\phi_2-\phi_2^{\mathrm{sym}}) \,.
\end{equation}
The local-local discretization shows a less favorable extrapolation,
as there is a $\approx\SI{40}{\percent}$ difference between the
coarsest lattice spacings and the continuum limit, while the
conserved-local only shows a $\approx\SI{10}{\percent}$ difference.
As in our previous work \cite{Gerardin:2019rua}, we exclude the
local-local discretization from the subsequent analysis and the final
results.
As in eqs.~\eqref{eq:model_33}, \eqref{eq:model_88} and \eqref{eq:model_08}, we employ $\tnotsym$ from
\cite{Bruno:2016plf} as a proxy for the size of the discrezation effects for each
$\SPi^{cc}_{\con}$ data point.
However, we also use $\tnotsym$ instead of the $t_0$ value computed on each ensemble to set the $Q^2$ scale input in the \ac{TMR} kernel, and to determine the $\phi_2$ of each ensemble entering eq.~\eqref{eq:model_charm}.
While using $\tnotsym$ introduces correlations between ensembles at the same lattice spacing, we found that it significantly reduces the curvature of the data with respect to $\phi_2$ and allows us to use the linear fit model in eq.~\eqref{eq:model_charm}~\cite{SanJose:2022thesis}.
We take into account the additional correlations increasing the size of our covariance matrix
in a straightforward manner. For all other aspects, the method used to
extrapolate the isovector and isoscalar contributions is directly
carried over to the charm contribution. For this component we perform
a cut at $m_\pi<\SI{400}{\MeV}$ and $<\SI{300}{\MeV}$ to estimate any
systematics of the fit. The three extrapolations give compatible
results.

\subsection{The running with energy}
\label{sec:extrapolated_running}

\begin{figure}[t]
  \centering
  \scalebox{.55}{\input{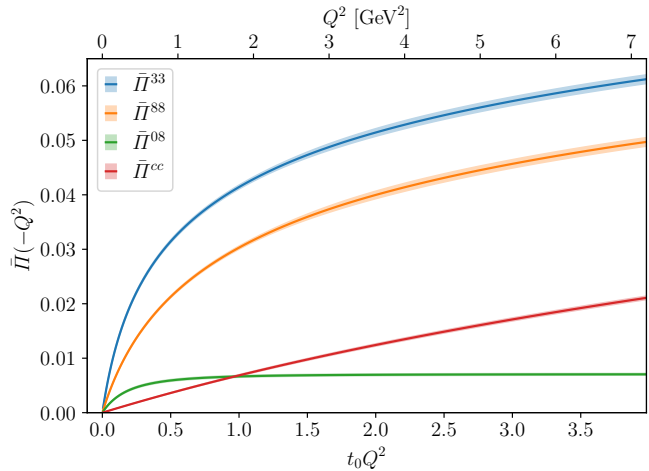}}
  \caption{%
    Contributions to the running extrapolated at the physical point as a function of the momentum transfer squared $Q^2$.
    The contributions are normalized without including the charge factors, according to eqs.~\eqref{eq:corr_flavor}.
  }\label{fig:running_phys}
\end{figure}

\begin{table}[p]
  \centering
\begin{tabular}{
      c@{}S[table-format=1.1]S[table-format=1.4]
      S[table-format=1.5]@{}>{$}r<{$}@{}>{$}r<{$}@{}>{$}r<{$}@{}>{$}r<{$}@{}>{$}r<{$}@{}>{$}r<{$}
      S[table-format=1.5]@{}>{$}r<{$}@{}>{$}r<{$}@{}>{$}r<{$}@{}>{$}r<{$}@{}>{$}r<{$}@{}>{$}r<{$}
    }
    \toprule
    & {$Q^2$ [\si{\GeV\squared}]} & {$t_0Q^2$} & \multicolumn{7}{c}{$\SPi^{33}$} & \multicolumn{7}{c}{$\SPi^{88}$} \\
\end{tabular}\\
  \begin{tabular}{%
      c@{}S[table-format=1.1]S[table-format=1.4]%
      S[table-format=1.5]@{}>{$}r<{$}@{}>{$}r<{$}@{}>{$}r<{$}@{}>{$}r<{$}@{}>{$}r<{$}@{}>{$}r<{$}%
      S[table-format=1.6]@{}>{$}r<{$}@{}>{$}r<{$}@{}>{$}r<{$}@{}>{$}r<{$}@{}>{$}r<{$}@{}>{$}r<{$}%
    }
    \toprule
    & {$Q^2$ [\si{\GeV\squared}]} & {$t_0Q^2$} & \multicolumn{7}{c}{$\SPi^{08}$} & \multicolumn{7}{c}{$\SPi^{cc}$} \\
    \midrule
    \csvreader[%
      late after line=\\,%
      filter=\equal{\QsqGeV}{0.1}\or\equal{\QsqGeV}{0.4}\or\equal{\QsqGeV}{1.0}\or\equal{\QsqGeV}{2.0}\or\equal{\QsqGeV}{3.0}\or\equal{\QsqGeV}{4.0}\or\equal{\QsqGeV}{5.0}\or\equal{\QsqGeV}{6.0}\or\equal{\QsqGeV}{7.0},%
    ]{tables/running_results2.csv}%
      {Q2GeV=\QsqGeV, t0Q2=\tnotQsq,%
       08_mean=\reszem, 08_stat=\reszes, 08_fit=\reszef, 08_scale=\reszesc, 08_ib=\reszeib, 08_cq=\reszecq, 08_comb=\reszecomb,%
       cc_mean=\resccm, cc_stat=\resccs, cc_fit=\resccf, cc_scale=\resccsc, cc_ib=\resccib, cc_cq=\rescccq, cc_comb=\rescccomb}%
      {& \QsqGeV & \tnotQsq & \reszem & \reszes & \reszef & \reszesc & \reszecq & & \reszecomb %
                            & \resccm & \resccs & \resccf & \resccsc & \rescccq & & \rescccomb}
    \bottomrule
  \end{tabular}
  \caption{%
    Contributions to the running extrapolated to the physical point.
    The first quoted uncertainty is the statistical error, the second
is the systematic error from varying the fit model estimated in
section~\ref{sec:fit_syst}, the third is the scale-setting error (see
section~\ref{sec:scale_setting}), and the fourth is the systematic from missing charm sea-quark loops (see section~\ref{sec:charm_quench}).
    The final uncertainty, quoted in square brackets, is the combination of the previous ones.
  }\label{tab:running_phys}
\end{table}
\begin{table}[p]
  \centering
  \begin{adjustbox}{center}
  \begin{tabular}{%
      c@{}S[table-format=1.1]S[table-format=1.4]%
      S[table-format= 1.6]@{}>{$}r<{$}@{}>{$}r<{$}@{}>{$}r<{$}@{}>{$}r<{$}@{}>{$}r<{$}@{}>{$}r<{$}%
      S[table-format=+1.6]@{}>{$}r<{$}@{}>{$}r<{$}@{}>{$}r<{$}@{}>{$}r<{$}@{}>{$}r<{$}@{}>{$}r<{$}%
    }
    \toprule
    & {$Q^2$ [\si{\GeV\squared}]} & {$t_0Q^2$} & \multicolumn{7}{c}{$\Dalphahad$} & \multicolumn{7}{c}{$\Dhad\sIIW$} \\
    \midrule
    \csvreader[%
      late after line=\\,%
      filter=\equal{\QsqGeV}{0.1}\or\equal{\QsqGeV}{0.4}\or\equal{\QsqGeV}{1.0}\or\equal{\QsqGeV}{2.0}\or\equal{\QsqGeV}{3.0}\or\equal{\QsqGeV}{4.0}\or\equal{\QsqGeV}{5.0}\or\equal{\QsqGeV}{6.0}\or\equal{\QsqGeV}{7.0},%
    ]{tables/running_results2.csv}%
      {Q2GeV=\QsqGeV, t0Q2=\tnotQsq,%
       Dalpha_mean=\resggm, Dalpha_stat=\resggs, Dalpha_fit=\resggf, Dalpha_scale=\resggsc, Dalpha_cq=\resggcq, Dalpha_ib=\resggib, Dalpha_comb=\resggcomb,%
       Ds2thW_mean=\resZgm, Ds2thW_stat=\resZgs, Ds2thW_fit=\resZgf, Ds2thW_scale=\resZgsc, Ds2thW_cq=\resZgcq, Ds2thW_ib=\resZgib, Ds2thW_comb=\resZgcomb}%
      {& \QsqGeV & \tnotQsq & \resggm & \resggs & \resggf & \resggsc & \resggcq & \resggib & \resggcomb %
                            & \resZgm & \resZgs & \resZgf & \resZgsc & \resZgcq & \resZgib & \resZgcomb}
    \bottomrule
  \end{tabular}
  \end{adjustbox}
  \caption{%
    Total \ac{HVP} contribution to the running of $\alpha$ and $\sIIW$.
    After the statistical error and the fit, scale-setting and charm-sea-quark systematic errors propagated from the $\SPi$ results in table~\ref{tab:running_phys}, the fifth uncertainty is the systematic error from missing isospin-breaking effects (see section~\ref{sec:isospin_breaking}).
    The final uncertainty quoted is the combination of the previous ones.
  }\label{tab:running_phys_2}
\end{table}

\begin{figure}[t]
  \centering
  \scalebox{.55}{\input{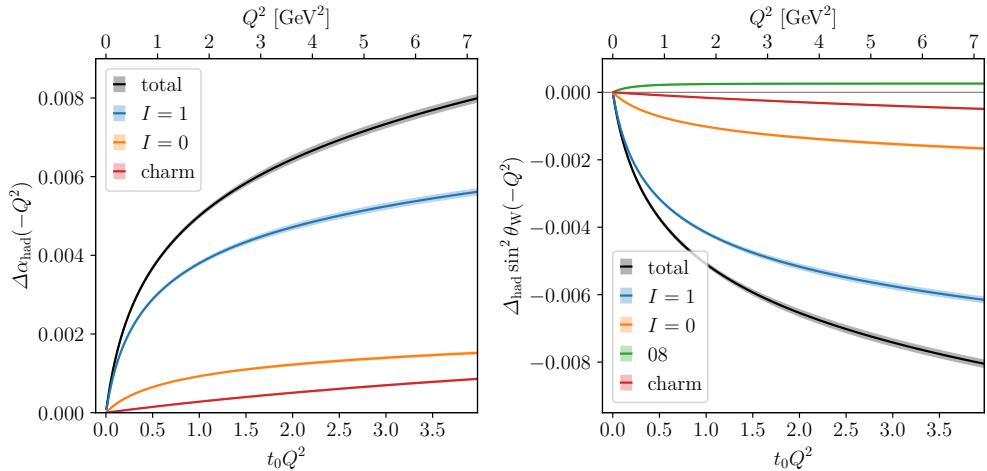}}
  \caption{%
    Total \ac{HVP} contribution to the running of $\alpha$ (left
    panel) and $\sIIW$ (right panel) as a function of $Q^2$, together
    with the $I=1$, $I=0$, charm and, for $\sIIW$, $Z\gamma$-mixing
    contributions.
  }\label{fig:running_phys_2}
\end{figure}

The results of the extrapolation of the \ac{HVP} functions $\SPi^{33},
\SPi^{88}, \SPi^{08}$ and $\SPi^{cc}$ to the physical point are
plotted as a function of $Q^2$ in figure~\ref{fig:running_phys}.
Furthermore, the corresponding numerical estimates are listed in
table~\ref{tab:running_phys} for several values of $Q^2$. These
numbers constitute the main result of this paper. The quoted errors
include all statistical and systematic uncertainties on the result
extrapolated to the continuum limit, provided that exact isospin
symmetry is assumed. According to the flavor decomposition described
in section~\ref{sec:flavor}, one can use these results to construct
the hadronic running of $\alpha$ and $\sIIW$. The corresponding
results for $\Dalphahad(-Q^2)$ and $\Dhad\sIIW(-Q^2)$ are plotted in
figure~\ref{fig:running_phys_2} and listed in
table~\ref{tab:running_phys_2}, including the uncertainty due to
isospin-breaking effects.

The results in tables~\ref{tab:running_phys}
and~\ref{tab:running_phys_2} for an extended set of \num{109} values
of $Q^2$ between \SI{0.01}{\GeV\squared} and \SI{7}{\GeV\squared} are
given in appendix~\ref{sec:supplementary}.

In the following two subsections we discuss the estimation of the
systematic errors due to scale setting, the quenching of the charm
quark and neglecting isospin breaking.

\subsubsection{Scale-setting error}
\label{sec:scale_setting}

To set the relative scale between the ensembles employed in this work,
we use $t_0$~\cite{Luscher:2010iy} which can be computed to
very high precision with a small computing investment, see
appendix~\ref{sec:meson} and table~\ref{tab:meson}.
In order to convert $t_0$ into physical units, we use
$(8t_0^{\mathrm{phys}})^{1/2}=\SI[parse-numbers=false]{0.415(4)(2)}{\fm}$,
which has been determined in ref.~\cite{Bruno:2016plf} on a subset of
ensemble used in this work, using a combination of the pion and kaon
decay constants $f_\pi$ and $f_K$. Therefore, we have a
\SI{1.1}{\percent} error on our absolute scale.

Since the (subtracted) \ac{HVP} function is a dimensionless quantity,
scale setting enters only indirectly, through the value of $Q^2$ in
physical units that appears in the TMR kernel, and in the
extrapolation to the physical point through the definition of the
point in the ($m_\pi$, $m_K$) plane that corresponds to an
isosymmetric version of the physical world.

In analogy to the case of $\aHVP$ considered in section~B.2 of
ref.~\cite{DellaMorte:2017dyu}, the error $\Delta l_0$ on the scale
$l_0=(8t_0)^{1/2}$ propagates to $\SPi$ according to
\begin{gather}
\label{eq:scale_error_linear}
  \frac{\Delta\SPi}{\SPi} \simeq \abs{ \frac{l_0}{\SPi}\dv{\SPi}{l_0} } \frac{\Delta l_0}{l_0}
  = \abs{ \frac{2t_0Q^2}{\SPi}\pdv{\SPi}{(t_0Q^2)} +
\frac{2\phi_2}{\SPi}\pdv{\SPi}{\phi_2} +
\frac{2\phi_4}{\SPi}\pdv{\SPi}{\phi_4} } \frac{\Delta l_0}{l_0} \,.
\end{gather}
The first term in the absolute value on the r.h.s.\ is proportional to
the slope of $\SPi$ as a function of $Q^2$.
For all contributions, it is positive and monotonically decreasing
with $Q^2$, relatively more important at low $Q^2$, where $\SPi$
varies faster, than at high $Q^2$.
For $\SPi^{\gamma\gamma}$, it evaluates to $\approx\num{0.9}$ at
$Q^2=\SI{1}{\GeV\squared}$, decreasing to $\approx\num{0.6}$ at
$Q^2=\SI{7}{\GeV\squared}$ and increasing to $\approx\num{1.7}$ at
$Q^2=\SI{0.1}{\GeV\squared}$.
Empirically, we observe that the third term in the r.h.s.\ of
eq.~\eqref{eq:scale_error_linear} is of the same order and negative,
which has the effect of partially cancelling the $Q^2$ contribution and reducing the scale setting error.
Specifically for the $\SPi^{33}$ contribution, also the second term in
the r.h.s.\ of eq.~\eqref{eq:scale_error_linear} is non-negligible and
negative as the $I=1$ contribution at small $m_\pi$ increases faster
with decreasing $\phi_2$.

To reliably estimate the scale setting error including cases in which
the three terms nearly cancel, we employ bootstrap sampling, which
allows us to go beyond the first-order error propagation in
eq.~\eqref{eq:scale_error_linear}.
Artificial bootstrap samples with a normal distribution are generated
for $(8t_0^{\mathrm{phys}})^{1/2}$ and, in turn,
$\phi_2^{\mathrm{phys}}$ and $\phi_4^{\mathrm{phys}}$, which define the
physical point in the fit model, and $t_0Q^2$.
The induced distribution of $\SPi(Q^2)$ is obtained evaluating the optimized fit model at $(\phi_2^{\mathrm{phys}}, \phi_4^{\mathrm{phys}})$ samples from these distributions, and using numerical derivatives in the case of the $t_0Q^2$ distribution to account for the small deviation of the samples with respect to set of values at which the extrapolation is performed.
The resulting scale setting error is the third error contribution
given for each quantity in tables~\ref{tab:running_phys}
and~\ref{tab:running_phys_2}.
The scale-setting error as a function of energy for both
$\SPi^{\gamma\gamma}$ and $\SPi^{Z\gamma}$ is compared to other
sources of uncertainty in figure~\ref{fig:pade_approx}.
In both cases, the systematic error from scale setting is larger than
the statistical for $\SI{0.5}{\GeV\squared}\lesssim
Q\lesssim\SI{2.5}{\GeV\squared}$, while the statistical error
dominates at $Q\gtrsim\SI{2.5}{\GeV\squared}$.

\subsubsection{Charm quark loop effects}
\label{sec:charm_quench}

Our computation is performed using gauge ensembles with $\Nf=2+1$
flavors of dynamical quarks, such that the light and strange quarks
are present in the \enquote{sea}, while for the charm quark only the
connected Wick contraction of the valence contribution is included in
the result.
We include the missing contributions from charm sea quarks, as well as
disconnected diagrams involving charm valence quarks as a systematic
uncertainty in our error budget.

As explained in detail in appendix~\ref{sec:CharmQuenching}
we quantify the charm quenching effect phenomenologically, by
estimating the contributions from $D$-meson loops to the connected
vector correlator involving $(u,\,d,\,s)$ quarks. In particular, we
determine the contributions of $D^+D^-$, $D^0\bar{D}^0$ and
$D_s^+D_s^-$ loops to the $R$-ratio and, in turn, the subtracted
\ac{HVP} function, by treating the $D$-meson form factors in scalar
QED. Other non-perturbative effects, such as changes in the $\omega$
and $\phi$ masses in QCD with $\Nf=2+1$, due to  mixing with the
$J/\psi$ and higher charmonium vector resonances are found to be
negligible. For $Q^2=\SI{5}{\GeV\squared}$ we estimate that the size of the
charm sea contribution is only about $3$ permil of the corresponding
$(u,\,d,\,s)$ quark contribution.

Regarding the charm disconnected valence quark contribution, we note
that the BMW collaboration has reported it to be less than one percent
of the light and strange disconnected contributions to
$\aHVP$~\cite{Borsanyi:2017zdw}. We assume that the effect is of
similar size for the hadronic running of the electromagnetic and weak
couplings. Since the light and strange disconnected correlators
already contribute at most one percent to the total hadronic running,
the contribution from disconnected charm loops is expected to be
\SI{0.01}{\percent}. This is subleading with respect to the quenched
charm systematic error already included in
table~\ref{tab:running_phys_2}.

\subsubsection{Isospin-breaking effects}
\label{sec:isospin_breaking}

As discussed in section~\ref{sec:lattice_setup}, our simulations are
performed in the limit of strong isospin symmetry, \ie\ we
work with degenerate up and down quark masses ($m_u=m_d=m_\ell$) and
neglect effects caused by \ac{QED}. To estimate the systematic effect
due to this assumption, we have evaluated the \ac{HVP} functions in
\ac{QCD}+\ac{QED} on a subset of our isospin-symmetric ensembles based
on the techniques described in~\cite{Ferrenberg:1988yz,Duncan:2004ys,Hasenfratz:2008fg,Finkenrath:2013soa,deDivitiis:2011eh,deDivitiis:2013xla}.
We use the QED\textsubscript{L} prescription~\cite{Hayakawa:2008an} to
regularize the IR divergence of non-compact lattice \ac{QED}.
Furthermore, we choose the same boundary conditions for the photon
field as for the \ac{QCD} gauge field. In a next step, \ac{QCD}+\ac{QED}
obtained from reweighted isosymmetric \ac{QCD} is expanded
up to leading order around isosymmetric \ac{QCD} in terms of the
electromagnetic coupling $e^{2}$ as well as the shifts in the bare quark
masses $\Delta m_u$, $\Delta m_d$ and $\Delta m_s$,
as applied by the RM123 collaboration~\cite{deDivitiis:2011eh,deDivitiis:2013xla}. This procedure results in Feynman
diagrams which represent perturbative quark mass shifts and the
interaction between quarks and photons~\cite{Risch:2021hty,Risch:2019xio,Risch:2018ozp,Risch:2017xxe}.

To match both theories we utilize a scheme based on leading-order
\ac{chiPT}, including leading order strong and
electromagnetic isospin breaking
corrections~\cite{Neufeld:1995mu,Risch:2021hty}. On each ensemble, we
match the results for $m_{\pi^{0}}^{2}$ and $m_{K^{+}}^{2} + m_{K^{0}}^{2} - m_{\pi^{+}}^{2}$
in both theories, which serve as proxies for the average light and
strange quark masses, respectively. These conditions are compatible
with the definition of the \enquote{physical} point of isosymmetric
QCD in section~\ref{sec:extrapolation}. We extend this scheme by the
corresponding proxy for the light quark mass splitting
$m_{K^{+}}^{2}-m_{K^{0}}^{2}-m_{\pi^{+}}^{2}+m_{\pi^{0}}^{2}$~\cite{Risch:2021hty}
and set it to its physical value. As we consider
leading-order effects, the electromagnetic coupling does not
renormalize and, hence, is fixed via the fine-structure constant
$e^{2}=4\cpi\alpha$~\cite{deDivitiis:2013xla}. Isospin-breaking effects
in the determination of the scale are neglected.

We have computed the leading-order \ac{QCD}+\ac{QED} quark-connected
contribution to $\SPi^{\gamma\gamma}$ and $\SPi^{Z\gamma}$ for
the three ensembles D450, N200 and H102
as well as the pseudo-scalar meson masses required for the above hadronic renormalization
scheme. The considered Feynman diagrams are evaluated by means of stochastic
$U(1)$ quark sources with support on a single time slice and $Z_{2}$
photon sources that are used to stochastically estimate the all-to-all photon
propagator in Coulomb gauge. To reduce the stochastic noise, covariant
approximation averaging~\cite{Shintani:2014vja} in combination with the
truncated solver method~\cite{Bali:2009hu} is applied. The noise problem
of the vector-vector correlation function at large time separations is
treated via a reconstruction using a single exponential function.
A more detailed description of the computation is given in refs.~\cite{Risch:2021nfs,Risch:2021hty,Risch:2019xio}.
Since the renormalization procedure of the local vector current in our
\ac{QCD}+\ac{QED} computation is based on a comparison of the local-local and the conserved-local discretisations of the vector-vector correlation function and hence differs
from the purely isosymmetric \ac{QCD} calculation~\cite{Gerardin:2018kpy}
we determine the relative correction by isospin breaking in
the \ac{QCD}+\ac{QED} setup. We observe that the size of the relative
first-order corrections for $\SPi^{\gamma\gamma}$~\cite{Risch:2021hty}
and $\SPi^{Z\gamma}$ is largest on D450. To rate the systematic error
of disregarding isospin-breaking corrections, which is added to the
error budget of the final result, we multiply
the obtained relative correction on D450
by the final results obtained from the isosymmetric \ac{QCD}
calculation. In figure~\ref{fig:pade_approx}, we compare this error
for both $\SPi^{\gamma\gamma}$ and $\SPi^{Z\gamma}$ to other
sources of uncertainty as a function of energy. We find that
isospin-breaking effects make a larger contribution to
the running of $\alpha$ compared to $\sIIW$.
However, this systematic uncertainty makes only a small contribution to the
total error. It is comparable to the statistical error of $\Dalphahad$
for $Q^2\lesssim\SI{2.5}{\GeV\squared}$ but the scale setting
uncertainty presently dominates in this regime.

\subsubsection{Rational approximation of the running}
\label{sec:rat_approx}

In addition to sampling the \ac{HVP} function at the $Q^2$ values in
table~\ref{tab:running_phys_2}, we provide an analytic function of
$Q^2$ that can be used to interpolate the \ac{HVP} function to any
value of $Q^2$ in the range up to $\SI{7}{\GeV\squared}$.
For this purpose, we use a rational function
\begin{equation}
\label{eq:approx_func}
  \SPi(-Q^2) \approx R_M^N(Q^2) = \frac{\sum_{j=0}^M a_j Q^{2j}}{1 + \sum_{k=1}^N b_k Q^{2k}} \,,
\end{equation}
where the numerator and the denominator are polynomials of degree $M$
and $N$ respectively, with $b_0=1$ in the denominator.
This choice is motivated by the fact that the \ac{HVP} function
$\Pi(-Q^2)$ can be expressed as a Stieltjes series with a finite
radius of convergence through a once-subtracted dispersion
relation~\cite{Aubin:2012me}.
This guarantees the existence of a convergent series of multi-point
Padé approximants with rigorous error
bounds~\cite{Baker:1969sw,Barnsley:1973zi}, which is particularly
useful when the sampling of the \ac{HVP} function is constrained to
the lattice discrete momenta, see ref.~\cite{Aubin:2012me}.
The \ac{TMR} method used in this work gives us more flexibility in the
choice of the momenta to sample, and allows for a very straightforward way
to obtain the rational approximation, by solving the over-constrained
system~\cite{Press:2007}
\begin{equation}
\label{eq:approx_system}
  \sum_i \frac{1}{\delta\SPi(-Q_i^2)} \left[ \sum_{j=0}^M a_j Q^{2j} - \left(1 + \sum_{k=1}^N b_k Q_i^{2k}\right) \SPi(-Q_i^2) \right] = 0 \,,
\end{equation}
via a least-squared fit, weighted by the inverse of the total error
$\delta\SPi$ at each $Q_i$ and in the range of energies
$0<Q^2\leq\SI{7}{\GeV\squared}$.
The minimization uses the constraint that $R_M^N(Q^2)$ has poles at
$Q_i^2<0$. Since the subtracted \ac{HVP} function $\SPi$ vanishes by
definition at $Q^2=0$, we set $a_0=0$.

We observe that a rational function of degree $M=3$ and $N=3$
describes the data very well. Using the set of \num{109} values of $Q^2$ between \SI{0.01}{\GeV\squared} and \SI{7}{\GeV\squared} that we sampled, we find that
higher-order coefficients are small and poorly determined by
eq.~\eqref{eq:approx_system}.
The resulting rational approximation for $\SPi^{\gamma\gamma}$ is
\begin{equation}
\label{eq:approx_gg}
  \SPi^{\gamma\gamma}(-Q^2) \approx \frac{\num{0.1094(23)}\, x + \num{0.093(15)}\, x^2 + \num{0.0039(6)}\, x^3}{1 + \num{2.85(22)}\, x + \num{1.03(19)}\, x^2 + \num{0.0166(12)}\, x^3} \,, \qquad x = \frac{Q^2}{\si{\GeV\squared}} \,,
\end{equation}
where the errors assigned to the coefficients in the numerator and
denominator, together with the correlation matrix
\begin{equation}
\label{eq:approx_gg_corr}
  \mathrm{corr}^{\gamma\gamma}\begin{pmatrix}
    a_1 \\ a_2 \\ a_3 \\ b_1 \\ b_2 \\ b_3
  \end{pmatrix} = \begin{pmatrix}
    1     \\
    0.455 & 1     \\
    0.17  & 0.823 & 1     \\
    0.641 & 0.946 & 0.642 & 1     \\
    0.351 & 0.977 & 0.915 & 0.869 & 1     \\
    0.0489 & -0.0934 & 0.0667 & -0.044 & -0.115 & 1     \\
  \end{pmatrix}
\end{equation}
reproduce the error band very accurately.\footnote{
  For both eqs.~\eqref{eq:approx_gg} and~\eqref{eq:approx_Zg}, we observe that a rational approximation with the same coefficients and errors except for $b_3=0$ approximates the data equally well.
  We choose to include the $b_3$ since this makes the extrapolation to higher $Q^2$ better behaved.
  However, we stress that the rational approximations in eqs.~\eqref{eq:approx_gg} and~\eqref{eq:approx_Zg} are valid only in the range of $Q^2\leq\SI{7}{\GeV\squared}$ and are not suitable for an extrapolation outside this range.
}
For $\SPi^{Z\gamma}$, the rational approximation is
\begin{equation}
\label{eq:approx_Zg}
  \SPi^{Z\gamma}(-Q^2) \approx \frac{\num{0.0263(6)}\, x + \num{0.025(5)}\, x^2 + \num{0.00089(34)}\, x^3}{1 + \num{2.94(29)}\, x + \num{1.12(27)}\, x^2 + \num{0.015(8)}\, x^3} \,, \qquad x = \frac{Q^2}{\si{\GeV\squared}} \,,
\end{equation}
with the correlation matrix
\begin{equation}
\label{eq:approx_Zg_corr}
  \mathrm{corr}^{Z\gamma}\begin{pmatrix}
    a_1 \\ a_2 \\ a_3 \\ b_1 \\ b_2 \\ b_3
  \end{pmatrix} = \begin{pmatrix}
    1     \\
    0.48  & 1     \\
    0.278 & 0.734 & 1     \\
    0.619 & 0.964 & 0.644 & 1     \\
    0.402 & 0.983 & 0.815 & 0.91  & 1     \\
    0.236 & 0.416 & 0.882 & 0.389 & 0.486 & 1     \\
  \end{pmatrix} .
\end{equation}

\begin{figure}[t]
  \centering
  \scalebox{.55}{\input{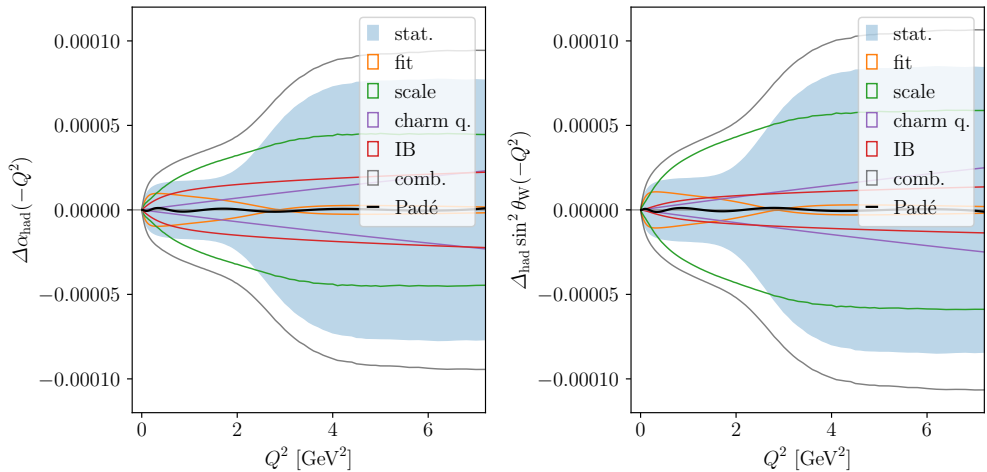}}
  \caption{%
    The deviation of the rational approximation of $\Dalphahad$
    (left) and $\Dhad\sIIW$ (right) from the data, plotted as a
    function of $Q^2$ and compared to the statistical error
    (blue-shaded area) as well as different
    sources of systematic uncertainty: fit model (orange-bordered
    area), scale setting (green-bordered area) and isospin breaking
    (red-bordered area). The plots show that statistical errors
    increase when a term of $\order{a^3}$ is added to the leading
    discretization effect of $\order{a^2}$ in the fit model for
    $Q\gtrsim\SI{2.5}{\GeV\squared}$. The gray lines represent the
    total error.
  }\label{fig:pade_approx}
\end{figure}

The deviation of the approximation from our measured values is
compared to the different sources of uncertainty in
figure~\ref{fig:pade_approx}. We find that the deviation is always
much smaller than the combined error: For instance, for
$Q^2>\SI{1.5}{\GeV\squared}$ it is less than $1/5$ of the combined
error, and less than \SI{0.3}{\percent} of that of the actual data.

\subsubsection{Dependence on the definition of the physical point}

As discussed in section~\ref{sec:extrapolation}, the results quoted in tables~\ref{tab:running_phys}
and~\ref{tab:running_phys_2} have been obtained by extrapolation to
a reference point in the isospin-symmetric limit.
The shift in $\SPi^{\gamma\gamma}(Q^2)$ and $\SPi^{Z\gamma}(Q^2)$ 
corresponding to a small change in the choice of convention for the
physical point can be estimated from the derivatives of the
extrapolated values with respect to $\phi_2$ and $\phi_4$.
An effective description of the derivatives as a function of $Q^2$ is given by the rational approximations
\begin{gather}
  \pdv{\SPi^{\gamma\gamma}(-Q^2)}{\phi_2} = -\frac{\num{0.2676}\,  x + \num{0.3960}\,  x^2}{1 + \num{6.944}\, x + \num{12.06}\, x^2} \,, \quad
  \pdv{\SPi^{\gamma\gamma}(-Q^2)}{\phi_4} = -\frac{\num{0.06393}\, x}{1 + \num{1.569}\, x} \,, \\
  \pdv{\SPi^{Z\gamma}(-Q^2)}{\phi_2}      = -\frac{\num{0.06388}\, x + \num{0.08935}\, x^2}{1 + \num{6.880}\, x + \num{11.83}\, x^2} \,, \quad
  \pdv{\SPi^{Z\gamma}(-Q^2)}{\phi_4}      = -\frac{\num{0.01887}\, x}{1 + \num{1.663}\, x} \,,
\end{gather}
where $x=Q^2/\si{\GeV\squared}$.

Combined with the derivatives with respect to the momentum transfer
variable $t_0Q^2$, which can be easily obtained from
eqs.~\eqref{eq:approx_gg} and~\eqref{eq:approx_Zg}, the given rational
approximations can also be used to account for a small variation of
the global scale according to eq.~\eqref{eq:scale_error_linear}.

\section{Comparison and discussion}
\label{sec:comparison}

The main results of this paper are the contributions from
$u$, $d$, $s$ and $c$ quarks to the hadronic running of the QED coupling
$\alpha$ and the electroweak mixing angle $\sIIW$, as a function of
the space-like momentum $Q^2>0$, computed in lattice QCD. In this
section, we present a detailed comparison of our results to those from
other lattice calculations, and to phenomenological analyses based on
dispersion theory and hadronic cross section data.

\subsection{Hadronic running of the electromagnetic coupling}
\label{sec:alpha_comp}

\begin{figure}[t]
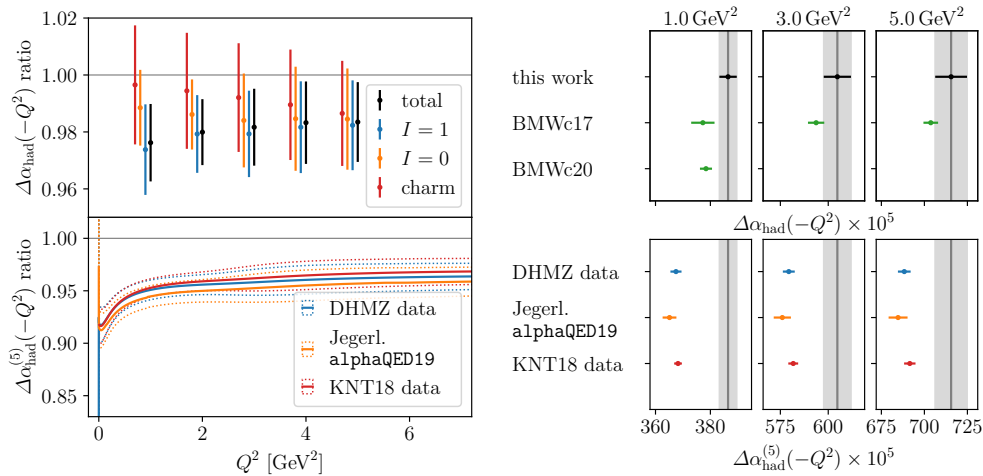

  \centering
  \scalebox{.55}{\input{figures/running_comparison_ratio.pgf}\input{figures/running_comparison_detail.pgf}}
  \caption{%
    Left, upper panel: ratio of the hadronic running $\Dalphahad$
    computed by \ac{BMWc}~\cite{Borsanyi:2017zdw} divided by our results,
    for five different momenta. In addition to the total
    contribution, we show the isovector ($I=1$), isoscalar ($I=0$) and
    charm quark components.
    Left, lower panel: the total hadronic running $\Dalphahad^{(5)}$ from
    various phenomenological
    estimates~\cite{Davier:2019can,Jegerlehner:alphaQEDc19,Keshavarzi:2018mgv}
    and the lattice result of ref.~\cite{Borsanyi:2017zdw}, normalized
    by the result of this work.
    Right: Compilation of results for the four-flavor $\Dalphahad$ lattice computations~\cite{Borsanyi:2017zdw,Borsanyi:2020mff} (above) and the five-flavor $\Dalphahad^{(5)}$ phenomenological estimates (below) at selected values of $Q^2$.
    The gray vertical error band for the result of this work includes the small bottom quark contribution as an additional systematic error, see section~\ref{sec:alpha_comp} for details.
  }\label{fig:running_comparison_alpha}
\end{figure}

Our estimates for $\Dalphahad(-Q^2)$ can be directly compared to the lattice
calculation results by \ac{BMWc}, given in table~S3 in the supplementary material of ref.~\cite{Borsanyi:2017zdw}, after correcting the latter for
finite-size effects determined in that same reference.
Ratios between results obtained by \ac{BMWc} and our estimates are plotted in
the upper left panel in figure~\ref{fig:running_comparison_alpha}, for
the total contribution as well as for its various components.
While there is good agreement for the isoscalar ($I=0$)
component, a slight tension at the level of $1$--$2$ standard deviations
is observed in the isovector ($I=1$) channel that dominates
the total contribution. We note that estimates by \ac{BMWc} are smaller by
\num{2}--\SI{3}{\percent} for $Q^2\lesssim\SI{3}{\GeV\squared}$.
For the charm contribution, our results are up to
\SI{2}{\percent} larger than \ac{BMWc}'s, but they are compatible within the errors, which are dominated by scale setting.
The comparison of the
absolute values of the two lattice results is depicted in the right
panel of figure~\ref{fig:running_comparison_alpha}, which shows the
slightly smaller error of the \ac{BMWc} result.
The most recent result from \ac{BMWc}~\cite{Borsanyi:2020mff}, also shown in the right panel of figure~\ref{fig:running_comparison_alpha}, has a smaller error but it is only available at $Q^2=\SI{1}{\GeV\squared}$.
We also mention that the first lattice calculation of the
quark-connected \ac{HVP} contributions to running of $\alpha$ and
$\sIIW$ up to $Q^2=\SI{10}{\GeV\squared}$ was published by Burger
\emph{et al.}~\cite{Burger:2015lqa}, who reported a \num{2}--\SI{3}{\percent} error
dominated by systematic effects. However, we do not include this
result in our comparison since the disconnected contribution has not
been determined in that reference.

In the lower left panel of figure~\ref{fig:running_comparison_alpha} 
we show the ratios of three recent phenomenological determinations of
$\dalpfive{-Q^2}$ and the rational approximation of our result as
continuous curves.
Our result lattice results for $\Dalphahad(-Q^2)$ includes the
contributions from $u$, $d$, $s$ and $c$ quarks. In order to account
for the contributions from bottom quarks that are needed to complete
the estimate for $\dalpfive{-Q^2}$, we use results by the HPQCD
collaboration for the lowest four time moments of the
\ac{HVP}~\cite{Colquhoun:2014ica}. We determine the contribution from
bottom quarks by constructing Padé approximants from the moments,
which results in a few-permil effect on the total hadronic
running of the coupling (up to $2.6$ permil at the largest $Q^2=\SI{7}{\GeV\squared}$).
This effect is larger than the $0.4$ permil
effect reported for the \ac{HVP} contribution to the muon
$g-2$~\cite{Chakraborty:2016mwy} due to the fact that the
running coupling scale $Q^2$ is not well separated from the bottom
quark mass, in contrast to the muon mass case.\footnote{As a
crosscheck, we have reproduced the bottom quark contribution to the
muon $g-2$ reported by HPQCD~\cite{Chakraborty:2016mwy}.}
However, this effect is a small fraction of the percent-level total error on $\Dalphahad(-Q^2)$ and we include it as an additional source of systematic error.

Results from Davier \emph{et
al.}~\cite{Davier:DHMZdata,Davier:2019can} (labellel ``DHMZ data''),
Keshavarzi \emph{et al.}~\cite{Keshavarzi:KNT18data,Keshavarzi:2018mgv} (KNT18 data),
and based on Jegerlehner's \texttt{alphaQEDc19} software package~\cite{Jegerlehner:alphaQEDc19,Jegerlehner:2019lxt} 
show good agreement among each other, but are between \num{3} and
\SI{6}{\percent} lower than our estimate.\footnote{%
  The estimate of $\dalpfive{-Q^2}$ in the space-like region
  corresponding to ref.~\cite{Davier:2019can}
  was kindly provided to us by Davier, Hoecker, Malaescu, and Zhang.
  We are grateful to Keshavarzi, Nomura and Teubner for providing the
  full covariance matrix of the $R$-ratio, allowing for a
  determination of $\dalpfive{-Q^2}$ consistent with ref.~\cite{Keshavarzi:2018mgv}.
}
After taking the errors into account, we observe a sizeable tension of
up to 3.5 standard deviation between our lattice calculation and
phenomenological estimates for space-like momenta in the range between
$3$ and \SI{7}{\GeV\squared}. For smaller space-like momenta,
the tension is even larger, due to the fact that the extrapolation to
the continuum limit has been performed with an $a^2$-term only, which
results in a smaller error.

The electromagnetic coupling at the $Z$ pole,
$\dalpfive{M_Z^2}$, puts a limit on the sensitivity of global
electroweak precision fits~\cite{Jegerlehner:2011mw,Baak:2014ora,Jegerlehner:2019lxt}.
This quantity also receives growing interest with respect to searches
for physics beyond the Standard Model (BSM) at a future International
Linear Collider. Our lattice results for $\Dalphahad(-Q^2)$ for
space-like $Q^2$ up to \SI{7}{\GeV\squared} can be combined with
either perturbative QCD or phenomenology to obtain the five-flavor hadronic
running at the $Z$ pole, $\dalpfive{M_Z^2}$ with no or much reduced
reliance on experimental data.

The connection between $\Dalphahad(-Q^2)$ and the hadronic running of
$\alpha$ for five quark flavors at the $Z$ pole in the time-like region
can be established via the so-called Euclidean split
technique (also known as Adler function approach)~\cite{Eidelman:1998vc,Jegerlehner:2008rs}.
As the name suggests, this technique allows for separating the contribution to the running at space-like kinematics, which is accessible to computational frameworks formulated in Euclidean spacetime such as lattice \ac{QCD}, from the small subleading contribution associated with the space-like to time-like rotation at high energies.
The method amounts to rewriting the hadronic contribution to the running at $M_Z$ as
\begin{multline}
\label{eq:Eu_split}
  \dalpfive{M_Z^2} = \dalpfive{-Q_0^2} \\
  + \left[ \dalpfive{-M_Z^2} - \dalpfive{-Q_0^2} \right]
  + \left[ \dalpfive{M_Z^2}  - \dalpfive{-M_Z^2} \right]_{\text{pQCD}} ,
\end{multline}
where the threshold energy $Q_0^2$ is typically around \SI{5}{\GeV\squared}.
The first term on the r.h.s.\ is proportional to the
space-like \ac{HVP} according to eq.~\eqref{eq:Dhad_alpha}.
In the literature~\cite{Eidelman:1998vc,Jegerlehner:2008rs},
$\dalpfive{-Q_0^2}$ has been evaluated by employing the dispersive
approach. Here, we evaluate this quantity using our lattice QCD
results $\Dalphahad(-Q_0^2)$ shown in table~\ref{tab:running_phys_2} as input, with the addition of the small bottom quark contribution as a systematic error.

\begin{table}[tb]
  \centering
  \begin{tabular}{S[table-format=1.1]S[table-format=1.6(3)]S[table-format=1.6(3)]}
    \toprule
    {$Q_0^2$} & {{\pqcdptxt}[Adler]} & {KNT18[data]} \\
    \midrule
    0.1  & {--}          & 0.026798(110) \\
    0.4  & {--}          & 0.025372(107) \\
    0.5  & {--}          & 0.025045(106) \\
    1.0  & 0.023928(223) & 0.023889(103) \\
    2.0  & 0.022492(149) & 0.022578(097) \\
    3.0  & 0.021640(129) & 0.021754(093) \\
    4.0  & 0.021020(116) & 0.021144(089) \\
    5.0  & 0.020528(107) & 0.020656(086) \\
    6.0  & 0.020117(099) & 0.020247(083) \\
    7.0  & 0.019763(093) & 0.019894(080) \\
    \bottomrule
  \end{tabular}
  \caption{%
    The contribution from $[\dalpfive{-M_Z^2} - \dalpfive{-Q_0^2}]$ for various threshold energy $Q_0^2$.
    The second column is based on the {\pqcdptxt}[Adler] approach in eq.~\eqref{eq:alp_adl}.
    The third column is obtained with KNT18[data] approach in eq.~\eqref{eq:alphaMZ-t2}.
    See the text for details.
  }\label{tab:alp_lat_ph}
\end{table}

The second term in eq.~\eqref{eq:Eu_split} is the high-energy contribution
$[\dalpfive{-M_Z^2} - \dalpfive{-Q_0^2}]$.
To estimate it, we follow Jegerlehner's idea~\cite{Jegerlehner:1999hg} of utilizing
the Adler function $D(Q^2)$ defined in eq.~\eqref{eq:adler_def}.
For sufficiently large $Q^2$, the Adler function is calculable within
perturbative QCD (pQCD) plus minor non-perturbative (NP)
corrections~\cite{Eidelman:1998vc,Chetyrkin:1996cf}. Our
implementation of this approach, which we call {\pqcdptxt}[Adler], is
based on the public code \texttt{pQCDAdler} by
Jegerlehner~\cite{Jegerlehner:pQCDAdler}. It takes into account full three-loop QCD with charm and bottom quark mass effects
as well as massless four- and five-loop effects to improve high-energy tails.
The code also accounts for the NP corrections via
the operator product expansion and Padé approximants.
We note that the {\pqcdptxt}[Adler] approach does not rely on the $R$-ratio integral,
and does not suffer from the systematics of the cross-section data.

Once the Adler function $D(Q^2)$ has been determined, we can calculate
\begin{equation}
\label{eq:alp_adl}
  \left[\dalpfive{-M_Z^2} - \dalpfive{-Q_0^2}\right]_{\pqcdpeq} = \frac{\alpha}{3\cpi}\int_{Q_0^2}^{M_Z^2}\frac{\dd{Q^2}}{Q^2}D(Q^2) \,,
\end{equation}
where $\alpha$ is the \ac{QED} coupling in the Thomson limit, and
\pqcdptxt{} indicates that perturbative QCD has been augmented by
small NP corrections when estimating $D(Q^2)$. Our results for
$[ \dalpfive{-M_Z^2} - \dalpfive{-Q_0^2} ]$ obtained in this way are
shown in the second column in table~\ref{tab:alp_lat_ph}.
The quoted errors, which amount to about \SI{0.5}{\percent} at
$Q_0^2=\SI{5}{\GeV\squared}$, originate from uncertainties in the strong coupling
at the $Z$ pole and heavy-quark pole masses, which are used as input
quantities. For smaller $Q_0^2$, the uncertainty associated with NP
corrections to $D(Q_0^2)$ grows significantly, hence we cannot access
very small values of $Q_0^2$ due to the Landau pole appearing in the
strong coupling.

In addition to using {\pqcdptxt}[Adler] for the evaluation of
$[\dalpfive{-M_Z^2} - \dalpfive{-Q_0^2}]$, we also consider the dispersive integral
\begin{equation}
\label{eq:alphaMZ-t2}
  \left[ \dalpfive{-M_Z^2} - \dalpfive{-Q_0^2} \right] =
  \frac{\alpha}{3\cpi}(M_Z^2-Q_0^2) \int_{m_{\pi^0}^2}^\infty \dd{s} \frac{R(s)}{(s+Q_0^2)(s+M_Z^2)} \,.
\end{equation}
This allows for a consistency check of the {\pqcdptxt}[Adler] approach
described above. The appearance of $Q_0^2$ in the denominator of the integrand implies
that contributions from the $R$-ratio at low energies are suppressed
along with any experimental uncertainties in their determination.
To compute the dispersion integral in eq.~(\ref{eq:alphaMZ-t2}), we
use the $R$-ratio data from KNT18~\cite{Keshavarzi:2018mgv}.
Since ref.~\cite{Keshavarzi:2018mgv} does not quote a result for
$[\dalpfive{-M_Z^2} - \dalpfive{-Q_0^2} ]$, we have performed the
integration of the $R$-ratio ourselves, using the full covariance
matrix~\cite{Keshavarzi:KNT18data} in the error estimate.
In the following, we shall refer to this method as ``KNT18[data]''.
The corresponding results are shown in the third column of table~\ref{tab:alp_lat_ph}.
The results are consistent with the \pqcdptxt{} approach (second
column) within the uncertainty.

Finally, we focus on the second combination in square brackets in eq.~\eqref{eq:Eu_split},
which provides the link between the space-like and time-like regions at $M_Z$.
We quote the pQCD estimate by
Jegerlehner~\cite{Jegerlehner:1985gq,Jegerlehner:2019lxt},
\begin{equation}
  \left[ \dalpfive{M_Z^2} - \dalpfive{-M_Z^2} \right]_{\text{pQCD}} = \num{0.000045(2)} \,.
\end{equation}

\begin{figure}[t]
  \centering
  \scalebox{.55}{\input{figures/comparison_Zpole_pQCDp.pgf}}%
  \scalebox{.55}{\input{figures/comparison_Zpole_KNT18.pgf}}
  \caption{%
    The hadronic contribution to the running coupling for five
    flavors at the $Z$ pole mass, $\dalpfive{M_Z^2}$, evaluated according to
    eq.~\eqref{eq:Eu_split} and using our lattice result for
    $\dalpfive{-Q_0^2}$, plotted as a function of the threshold energy $Q_0^2$.
    Left: The higher energy contribution $[\dalpfive{-M_Z^2} -
    \dalpfive{-Q_0^2}]$
    computed via the \pqcdptxt approach in eq.~\eqref{eq:alp_adl}
    using the \texttt{pQCDAdler} software package
    \cite{Jegerlehner:pQCDAdler}.
    Right: Results based on the KNT18[data] approach of
    eq.~\eqref{eq:alphaMZ-t2} 
    using the $R$-ratio data with full covariance
    matrix~\cite{Keshavarzi:2018mgv,Keshavarzi:KNT18data}.
    The red symbols in each panel are taken to produce the final
    estimates for each method, while the maxima and minima of the blue
    bands within the non-shaded region are used to estimate the uncertainty.
  }\label{fig:alp_mz_latph_pqcd}
\end{figure}

With these ingredients in hand, we can provide an estimate for the
phenomenologically relevant quantity $\dalpfive{M_Z^2}$, using our
lattice estimate for $\dalpfive{-Q_0^2}$ as input in
eq.~\eqref{eq:Eu_split}. In figure~\ref{fig:alp_mz_latph_pqcd}, we
show $\dalpfive{M_Z^2}$ as a function of the Euclidean squared
momentum transfer $Q_0^2$. In the left panel the contribution from 
$[\dalpfive{-M_Z^2}-\dalpfive{-Q_0^2}]$ has been
determined in perturbative QCD via the Adler function
({\pqcdptxt}[Adler]), while in the right panel the same quantity has
been evaluated using the $R$-ratio data and correlation matrix from
KNT18 in eq.~\eqref{eq:alphaMZ-t2}. The blue bands represent the total
error obtained by adding in quadrature all uncertainties that enter
eq.~\eqref{eq:Eu_split}. In both cases we find that the
estimates for $\dalpfive{M_Z^2}$ are very stable for
$Q_0^2\gtrsim\SI{3}{\GeV\squared}$. The slight upward trend and the loss of precision observed
for $Q_0^2\lesssim\SI{2}{\GeV\squared}$ when using the
{\pqcdptxt}[Adler] approach is symptomatic of the failure of pQCD at strong couplings.
Alternatively, when employing the dispersive approach of
eq.~\eqref{eq:alphaMZ-t2}, one observes a decreasing trend for
$Q_0^2\lesssim\SI{2}{\GeV\squared}$, which is due to the enhanced
contributions from low-lying resonances ($\rho$, $\omega$, $\phi$) in
eq.~\eqref{eq:alphaMZ-t2} as $Q_0^2$ is lowered. For our final
results in both approaches we choose $Q_0^2=\SI{5}{\GeV\squared}$
and estimate the uncertainty associated with the choice of $Q_0^2$
from the maxima and minima of the blue bands in the region
\num{3}--\SI{7}{\GeV\squared}. In this momentum range, our lattice
results for the hadronic running can be extrapolated reliably to the
continuum limit. Furthermore, this choice of interval guarantees that
our final estimate is not affected by the Landau pole when using the
{\pqcdptxt}[Adler] approach, nor is it dominated by the experimentally
determined $R$-ratio when employing eq.~\eqref{eq:alphaMZ-t2}
instead.

\begin{figure}[t]
  \centering
  \scalebox{.55}{\input{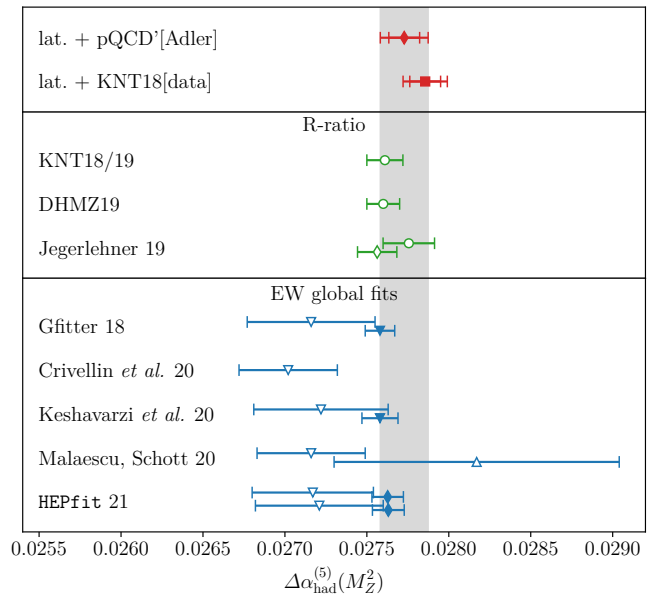}}
  \caption{%
    Compilation of results for $\dalpfive{M_Z^2}$.
    The first two data points (red symbols)
    represent the results from the Euclidean split technique using our
    lattice estimate for $\dalpfive{-Q_0^2}$.
    Green circles denote results based on the standard dispersive
    approach, where the $R$-ratio integration is performed over the entire
    momentum range. From top to bottom, we plot the results from
    refs.~\cite{Keshavarzi:2018mgv,Keshavarzi:2019abf},
    \cite{Davier:2019can}, and \cite{Jegerlehner:2019lxt}.
    The estimate based on the Adler function in
    ref.~\cite{Jegerlehner:2019lxt} is shown as a green diamond.
    Blue symbols represent the results from global EW fits,
    published in refs.~\cite{Haller:2018nnx,Crivellin:2020zul,Keshavarzi:2020bfy,Malaescu:2020zuc,deBlas:2021wap} (see the text for further details).
    The upper triangle point from ref.~\cite{Malaescu:2020zuc} does not use the Higgs mass.
    The gray band represents our final result quoted in eq.~\eqref{eq:alp5_mz_lat_padler}.
  }\label{fig:alp5}
\end{figure}

For our main result of the hadronic running of the QED coupling at the
$Z$ pole we adopt the {\pqcdptxt}[Adler] approach and quote
\begin{equation}
\label{eq:alp5_mz_lat_padler}
  \begin{split}
    \dalpfive{M_Z^2}|_{\text{Lat} + {\pqcdpeq}[\text{Adler}]} &=
  0.02773(9)_{\text{lat}}(2)_{\text{bottom}}(12)_{{\pqcdpeq}[\text{Adler}]} \\
    &= \num[separate-uncertainty]{0.02773 \pm 0.00015} \,.
  \end{split}
\end{equation}
The first error is the total uncertainty of our lattice estimate of
$\Dalphahad(-\SI{5}{\GeV\squared})$ as listed in
table~\ref{tab:alp_lat_ph}, while the second error accounts for the
neglected contribution from bottom quark effects. The error labeled
{\pqcdptxt}[Adler] is associated with the evaluation of
$[\dalpfive{-M_Z^2} - \dalpfive{-\SI{5}{\GeV\squared}} ]$ (see the
second column in table~\ref{tab:alp_lat_ph}), augmented by the maximum
deviations from the central value in the region $Q_0^2\in [3,
7]\,\si{\GeV\squared}$.

For completeness, we also list the result obtained via the KNT18[data]
approach, which yields
\begin{equation}
\label{eq:alp5_mz_lat_knt}
  \begin{split}
    \dalpfive{M_Z^2}|_{\text{Lat} + \text{KNT18[data]}} &=
    0.02786(9)_{\text{lat}}(2)_{\text{bottom}}(10)_{\text{KNT18[data]}} \\
    &= \num[separate-uncertainty]{0.02786 \pm 0.00013} \,,
  \end{split}
\end{equation}
where the meaning of the errors is similar to
eq.~\eqref{eq:alp5_mz_lat_padler}. The relative difference to the
result obtained via the Adler function amounts to less than \SI{0.5}{\percent} and
is indicative of the different treatment of the non-lattice
contribution in eq.~\eqref{eq:Eu_split}.

In figure~\ref{fig:alp5}, we present a compilation of results for
$\dalpfive{M_Z^2}$ obtained using our lattice estimate of the HVP,
the standard dispersive approach, as well as global EW fits.
The first two symbols (red filled diamond/square) show our results
represented by eqs.~\eqref{eq:alp5_mz_lat_padler}
and~\eqref{eq:alp5_mz_lat_knt}.

We shall first focus on our main result --- the red filled diamond
(Lattice + {\pqcdptxt}[Adler]). The inner and outer error bars
represent the total uncertainty and the combination of the first two
errors in eq.~\eqref{eq:alp5_mz_lat_padler}, respectively. The total
error of about \SI{0.5}{\percent} is close to the precision of the
dispersive approach (open green circles/diamond). Our main result is
consistent with the latter and also broadly agrees with the estimates
from global EW fits (blue upper/lower triangles).

It is instructive to compare our main result --- which is represented
by the gray vertical band --- with
Jegerlehner's evaluation~\cite{Jegerlehner:2019lxt}, also based on
the Euclidean split method and represented by the open green diamond
in the figure. The two estimates differ chiefly by the contribution at
the hadronic energy scale $Q_0^2=\SI{5}{\GeV\squared}$: while our
result is based on lattice QCD, the open green diamond has been
obtained from the $R$-ratio. In
figure~\ref{fig:running_comparison_alpha}, we have
observed a clear tension between the two evaluations (black vs.\
orange). However, from eq.~\eqref{eq:alp5_mz_lat_padler} one easily
reads off that $\dalpfive{-Q_0^2}$ contributes at most
\SI{60}{\percent} to the total uncertainty of $\dalpfive{M_Z^2}$,
resulting in smaller tension, due to the additional, albeit correlated, uncertainty from the high-energy contribution.

Next, we compare our results to the estimate from global EW fits.
This category includes results from the Gfitter group~\cite{Haller:2018nnx}, ref.~\cite{Crivellin:2020zul} (obtained using the \texttt{HEPfit} code~\cite{DeBlas:2019ehy}), refs.~\cite{Keshavarzi:2020bfy,Malaescu:2020zuc} (employing the Gfitter library), and ref.~\cite{deBlas:2021wap} (with two different scenarios).
The blue open lower triangles in figure~\ref{fig:alp5} were all
obtained by a fit to EW precision data, treating $\dalpfive{M_Z^2}$ as a free parameter.
This allows for $\dalpfive{M_Z^2}$ to be determined exclusively from the other EW precision observables, and
favors smaller values compared to both lattice and $R$-ratio determinations.
The precision of these estimates of $\dalpfive{M_Z^2}$ is lower
compared to results extracted from the lattice and the $R$-ratio.
With the exception of ref.~\cite{Crivellin:2020zul}, all results are
compatible with our value within $1.3\sigma$.
Therefore, we conclude that lattice calculations of
the HVP contribution to the hadronic running of $\alpha$ and the muon
$g-2$ are not in contradiction with global EW
fits~\cite{Crivellin:2020zul}.
The data point from ref.~\cite{Malaescu:2020zuc} marked by the blue open upper triangle results from treating both the Higgs mass
$M_H$ and $\dalpfive{M_Z^2}$ as fit parameters without
priors. It shows that if the precise experimental input for $M_H$ is
not used, the resulting determination of $\dalpfive{M_Z^2}$ favors a larger value with a significantly larger uncertainty~\cite{Malaescu:2020zuc}.
Fits represented by blue filled triangles employ priors for
$\dalpfive{M_Z^2}$ centered about the
$R$-ratio estimate~\cite{Keshavarzi:2020bfy,Haller:2018nnx}. It is
interesting to note that the output for $\dalpfive{M_Z^2}$ is very
close to the input $R$-ratio estimate.
Finally, the authors of ref.~\cite{deBlas:2021wap} apply the Euclidean
split technique in eq.~\eqref{eq:Eu_split}, by combining the
perturbative running with the lattice result by
\ac{BMWc}~\cite{Borsanyi:2017zdw} at $Q_0^2=\SI{4}{\GeV\squared}$. In
this way they obtain $\dalpfive{M_Z^2}=\num{0.02766(10)}$, which is consistent with our result in eq.~\eqref{eq:alp5_mz_lat_padler}.
Using this value as prior for the fit, the posterior probability is
shown in figure~\ref{fig:alp5} with a blue filled diamond for two
different scenarios.
The observed pull of $1.1$--$1.3\sigma$ further supports the
conclusion that lattice results are not inconsistent with global EW fits~\cite{deBlas:2021wap}.
Alternatively, one could use  
our estimate $\dalpfive{M_Z^2}$ as a prior, which is left to future work.

\subsection{Hadronic running of the electroweak mixing angle}

\begin{figure}[t]
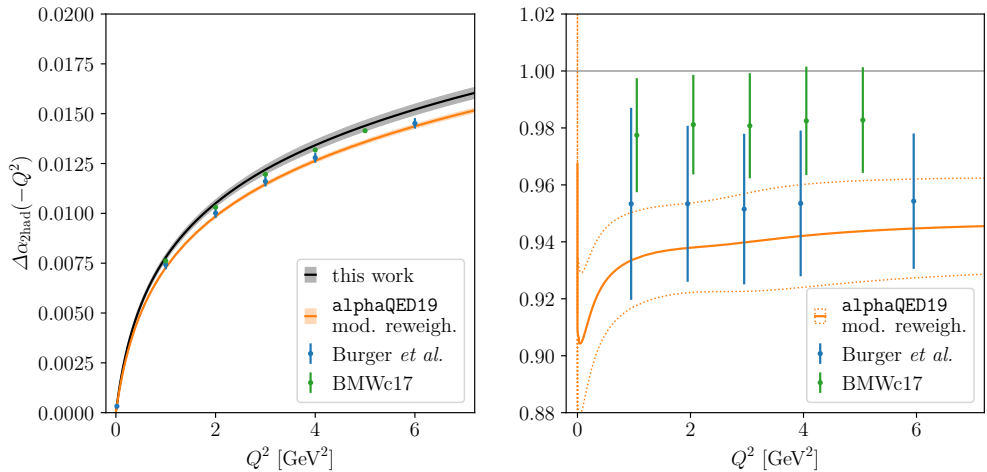

  \centering
  \scalebox{.55}{\input{figures/running_comparison_alph2.pgf}\input{figures/running_comparison_alph2_ratio.pgf}}
  \caption{%
    Left: Hadronic contributions to the running of the weak coupling
    $\alpha_2$ from the our computation, compared to lattice results
    from ref.~\cite{Burger:2015lqa}, as well as the
    phenomenological estimate determined using \texttt{alphaQEDc19}~\cite{Jegerlehner:alphaQEDc19,Jegerlehner:2019lxt}, as a function of the space-like
    momentum transfer $Q^2$.
    Right: Ratios of the phenomenological estimate and of the lattice
    computation in ref.~\cite{Burger:2015lqa} over our result.
  }\label{fig:running_comparison_alph2}
\end{figure}

The lattice formalism gives us full control over the quark flavor
charge factors that are used to construct the quark-level vector
currents in eqs.~\eqref{eq:emcurrent} and~\eqref{eq:weakcurrent}.
The ability to perform an exact separation of the vacuum polarization
function $\SPi^{\gamma Z}$ in terms of individual valence quark flavors (see
section~\ref{sec:method}) is an inherent feature of the lattice
approach. It eliminates the need to perform a reweighting of exclusive
channels in hadronic cross section data, which is a source of
systematic uncertainty in phenomenological determinations of the
hadronic running of $\sIIW$.
In section~\ref{sec:extrapolated_running} we have reported our
results for the hadronic contribution to the running of the
electroweak mixing angle, $\Dhad\sIIW(-Q^2)$, as a function of the
space-like momentum $Q^2>0$. Our result can thus replace  estimates
from the data-driven approach in studies that apply the running with
energy to determine the electroweak mixing angle $\thetaW$ in the
Thomson limit, with the understanding that current lattice QCD results
are limited to $Q^2\approx\SI{7}{\GeV\squared}$.
In the following, we compare our data with previous results in the
literature.

Phenomenological estimates of $\Dhad\sIIW(-Q^2)$ can be obtained by
applying Jegerlehner's \texttt{alphaQEDc19} software
package~\cite{Jegerlehner:alphaQEDc19}, see
ref.~\cite{Jegerlehner:2019lxt}. Since \texttt{alphaQEDc19} provides
$\Dalphahad$ and $\Delta\alpha_{2,\mathrm{had}}$ as primary quantities
with their respective error estimates, we perform the comparison with
our lattice determination for the running of the SU(2) gauge coupling
$\alpha_2$.
As explained in ref.~\cite{Jegerlehner:2019lxt}, the software package
implements a modified flavor separation scheme assuming SU(3) flavor
symmetry \cite{Jegerlehner:1986vs}, which differs from the
``perturbative'' separation scheme advocated in
refs.~\cite{Czarnecki:1995fw, Czarnecki:2000ic, Erler:2004in}.
This modification is motivated by the observation that it brings the
phenomenological estimate into better agreement with previous lattice
results, see for instance the comparison plot in figure~9 of
ref.~\cite{Burger:2015lqa} and the discussion in
ref.~\cite{Jegerlehner:2019lxt}.

A comparison between the \texttt{alphaQEDc19} estimate of
$\Dhad\sIIW(-Q^2)$ and our lattice results is shown in
figure~\ref{fig:running_comparison_alph2}.
In performing this comparison, one has to take into account that the
result of ref.~\cite{Jegerlehner:2019lxt} has been obtained using
$\sin^2\theta^\ell_{\mathrm{eff}}=\num{0.23153}$~\cite{ALEPH:2005ab}
as a reference value in the normalization of eq.~\eqref{eq:Dhad_alph2}.
This amounts to a \SI{3}{\percent} difference which is accounted for in
the plot. We observe that, even though the modified reweighting brings
the phenomenological estimate closer to our lattice results, it still
falls short by about \SI{8}{\percent} at smaller $Q^2$
and \SI{4}{\percent} at higher values.

In figure~\ref{fig:running_comparison_alph2} we include points from
previous lattice calculations. While the \ac{BMWc}
paper~\cite{Borsanyi:2017zdw} does not quote results for the running
of $\alpha_2$, or for the $\SPi^{08}$ component, we can still estimate
the $\alpha_2$ contribution using the fact that the Wick-connected
component is $\SPi^{08,\con}=\sqrt{3}/2(\SPi^{33}-\SPi_{\con}^{88})$,
together with the observation from our data that the same component
varies between \SI{130}{\percent} at $Q^2=\SI{1}{\GeV\squared}$ and
\SI{125}{\percent} at $Q^2\geq\SI{5}{\GeV\squared}$ of $\SPi^{08}$.
For the data points for $\Delta\alpha_{2,\mathrm{had}}$ plotted in
figure~\ref{fig:running_comparison_alph2}, we correct for this by
adding a $\order{\SI{0.5}{\percent}}$ positive contribution with
negligible errors to the \ac{BMWc} data points, which brings them into
good agreement with our result, especially at larger $Q^2$.
The result by Burger \emph{et al.}~\cite{Burger:2015lqa} is missing the
quark-disconnected contribution, as in the $\Dalphahad$ case. Thus, it
is plausible that their estimate is lower than our result. 
Since the SU(3)-symmetric flavor-separation scheme implemented in
\texttt{alphaQEDc19} was motivated by the findings of earlier lattice
calculations~\cite{Bernecker:2011gh,Francis:2013jfa,Burger:2015lqa},
the missing quark-disconnected contributions in those references might
explain the observed shortfall between the phenomenological estimate
on the one hand, and the more comprehensive results of our work and
ref.~\cite{Borsanyi:2017zdw} on the other.

\begin{figure}[t]
  \centering
  \scalebox{.55}{\input{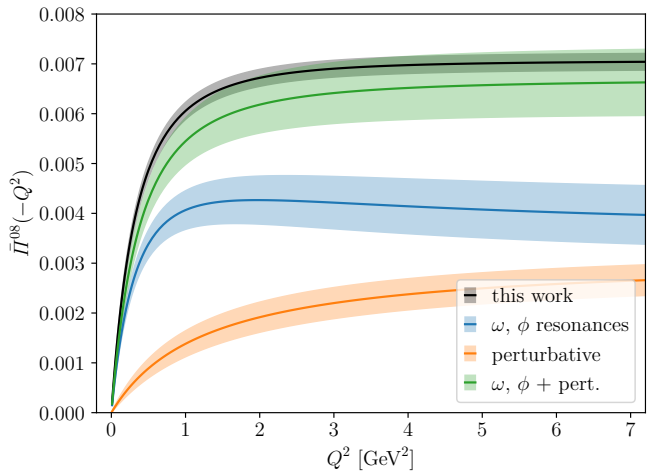}}
  \caption{%
    \ac{HVP} mixing function $\SPi^{08}$ as a function of $Q^2$ from our lattice computations, compared to the phenomenological model described in appendix~\ref{sec:model_08}.
    Two contributions to the model are also shown in isolation: the finite perturbative \ac{QCD} one, and the narrow resonances one resulting from the $\omega$ and $\phi$ mesons with opposite signs.
  }\label{fig:running_comparison_08}
\end{figure}

Another possible strategy to estimate the hadronic contribution to the
running of $\thetaW$ at low energies is to combine the
phenomenological evaluation of $\Dalphahad(-Q^2)$ from $R$-ratio
experimental data with extra input providing the exact flavor
separation and reweighting obtained from precise lattice data.
From eq.~\eqref{eq:corr_flavor} we observe that the difference between
$\SPi^{\gamma\gamma}$ and $\SPi^{Z\gamma}$ is proportional to the
$I=0$ mixing \ac{HVP} function $\SPi^{08}$ given in
table~\ref{tab:running_phys}, plus a small charm quark contribution.
In figure~\ref{fig:running_comparison_08}, we plot our results for
$\SPi^{08}$ as a function of $Q^2$ up to \SI{7}{\GeV\squared}.
To our knowledge, this is the most precise ab-initio computation of
this quantity and the first to include both connected and disconnected
contributions. In the same plot we also show a phenomenological model
estimate, which is discussed in detail in appendix~\ref{sec:model_08}.
The error band represents the statistical uncertainty on the model,
which is dominated by the experimental error on the partial decay
widths into $e^+e^-$ of the $\omega$ and $\phi$ vector resonances. The
model is obtained assuming that the quark-disconnected diagrams of
type $(s,s)$ and $(\ell,s)$ are negligible, and this assumption is
shown to give a good estimate in an analogous model for the $\aHVP$
strange and light isoscalar contributions. Even neglecting this
sources of systematic uncertainty, the plot shows clearly that the
lattice results are in good agreement with the model over the whole
range of $Q^2$ values, but significantly more precise.

$\SPi^{08}(Q^2)$ over the whole $Q^2$ range is well approximated
within \SI{15}{\percent} of its total error by a rational
approximation of order $[2/2]$ of the kind employed in
section~\ref{sec:rat_approx}, \ie
\begin{equation}
\label{eq:approx_08}
  \SPi^{08}(-Q^2) = \frac{\num{0.0217(11)}\, x + \num{0.0151(12)}\, x^2}{1 + \num{2.93(8)}\, x + \num{2.15(12)}\, x^2} \,, \qquad x = \frac{Q^2}{\si{\GeV\squared}} \,,
\end{equation}
where the numerator $a_i$ and denominator $b_j$ parameters are
strongly correlated according to
\begin{equation}
\label{eq:approx_08_corr}
  \mathrm{corr}\begin{pmatrix}
    a_1 \\
    a_2 \\
    b_1 \\
    b_2
  \end{pmatrix} = \begin{pmatrix}
    1     \\
    0.97  & 1     \\
    0.97  & 0.984 & 1     \\
    0.944 & 0.994 & 0.98  & 1     \\
  \end{pmatrix} .
\end{equation}
At large $Q^2$, $\SPi^{08}$ varies very slowly with $Q^2$. For
$Q^2\to\infty$, the rational approximation tends to the finite value
$a_2/b_2=\num{0.00704(17)}$ which coincides with the value at our
largest $Q^2=\SI{7}{\GeV\squared}$
\begin{equation}
\label{eq:Pi08_largeQ2}
  \SPi^{08}=\num{0.00704(17)}
\end{equation}
and is only $\SI{2}{\percent}$ larger than the value at
$Q^2=\SI{3}{\GeV\squared}$. We take this value as our main result for
the $I=0$ $Z\gamma$-mixing \ac{HVP} function at large $Q^2$.

\section{Conclusions}

In this paper we have presented a computation of the leading hadronic
contribution to the running of the \ac{QED} coupling $\alpha$ and of
the electroweak mixing angle $\thetaW$ from first principles using
lattice \ac{QCD}.

For the \ac{QED} coupling, our main result is presented in
eqs.~\eqref{eq:approx_gg} and~\eqref{eq:approx_gg_corr} in terms of a
rational approximation of $\Dalphahad(-Q^2)$ as a function of the
space-like $Q^2>0$ up to $Q^2\approx\SI{7}{\GeV\squared}$. Our results
are slightly larger but still compatible with an earlier calculation
by BMWc. However, there is a significant tension with the predictions
based on the data-driven method.
Since the tension is larger at lower $Q^2$, where our result is most sensitive to the scale setting error, a new determination of the $t_0$ scale that is underway~\cite{Strassberger:2021tsu} will help clarify the significance of this tension.

Combining our result obtained in a $Q^2$-range between \num{3} to
\SI{7}{\GeV\squared} range with perturbative \ac{QCD}, we obtain an
estimate for $\Dalphahad^{(5)}(M_Z^2)$ in
eq.~\eqref{eq:alp5_mz_lat_padler} that does not rely on any
experimental data as input, except for the calibration of the lattice
scale. Our estimate, based on the Euclidean split technique and the
perturbative Adler function, is consistent with and of similar
precision (\ie\ \SI{0.55}{\percent}) as estimates employing the
data-driven approach. Thus, the tension in $\Dalphahad(-Q^2)$ observed
between our results and the $R$-ratio is largely washed out when
running the result up to the $Z$ pole.

For the electroweak mixing angle $\thetaW$, we also provide a
description in terms of a rational function of $\Dhad\sIIW(-Q^2)$ for
$Q^2$ up to \SI{7}{\GeV\squared} (see eqs.~\eqref{eq:approx_Zg}
and~\eqref{eq:approx_Zg_corr}). Here we take advantage of the fact that
the different flavor structure of the (vector part of the) weak
neutral current with respect to the electromagnetic current is easily
implemented in the lattice approach. This results in estimates for
$\Dhad\sIIW(-Q^2)$ with a similar error budget as that for
$\Dalphahad(-Q^2)$, which is a vast improvement over what can be
obtained with the data-driven method which relies on a heuristic
flavor separation affected by large systematic uncertainties.
Specifically, in eq.~\eqref{eq:approx_08} we provide a rational
representation for the flavor-singlet mixing contribution
$\SPi^{08}(Q^2)$, which for a large $Q^2$ tends to the constant value
in eq.~\eqref{eq:Pi08_largeQ2}. This result can be used in
comprehensive studies of the running of the electroweak mixing angle
in combination with the experimental $R$-ratio data to reduce the
systematics from the flavor separation.

\addsec{Acknowledgements}
We thank Gilberto Colangelo, Jens Erler, Rodolfo Ferro-Hernández, Tim
Harris, Martin Hoferichter, Fred Jegerlehner, Daniel Mohler, Massimo
Passera, Thomas Teubner and Arianna Toniato for valuable discussions.
Further thanks go to Simon Kuberski for the determination of auxiliary
data that were used in our analysis.
We are grateful to Bogdan Malaescu and the authors of
refs.~\cite{Davier:DHMZdata,Davier:2019can} for sharing their data for
the running of $\alpha$ at $Q^2>0$.
We thank Alex Keshavarzi and the authors of
refs.~\cite{Keshavarzi:KNT18data,Keshavarzi:2018mgv} for sharing the
$R$-ratio data with covariance matrix, which were used to calculate
tabulated data for the running of $\alpha$ at $Q^2>0$.
Calculations for this project have been performed on the HPC clusters
Clover and HIMster-II at Helmholtz Institute Mainz and Mogon-II at
Johannes Gutenberg-Universität (JGU) Mainz, on the HPC systems JUQUEEN
and JUWELS at Jülich Supercomputing Centre (JSC), and on Hazel Hen at
Höchstleistungsrechenzentrum Stuttgart (HLRS)~\cite{Ce:2021hlrs}.
The authors gratefully acknowledge the support of the Gauss Centre for
Supercomputing (GCS) and the John von Neumann-Institut für Computing
(NIC) for project HMZ21 and HMZ23 at JSC and project GCS-HQCD at HLRS.
Our programs use the deflated SAP + GCR solver from the
\textsc{openQCD} package~\cite{Luscher:2012av,Luscher:openQCD}, as
well as the QMP and QDP++ library~\cite{Edwards:2004sx}.
Our data analysis has been performed using Python and the libraries
NumPy~\cite{Harris:2020xlr}, SciPy~\cite{Virtanen:2019joe},
pandas~\cite{McKinney:2010,Reback:pandas},
mpmath~\cite{Johansson:mpmath} and
uncertainties~\cite{Lebigot:uncertainties}, and using GNU
Parallel~\cite{Tange:parallel}.
Plots have been produced using Matplotlib~\cite{Hunter:2007ouj}.
We are grateful to our colleagues in the \ac{CLS} initiative for sharing ensembles.
This work has been supported by Deutsche Forschungsgemeinschaft
(German Research Foundation, DFG) through project HI 2048/1-2 (project
No.\ 399400745) and through the Cluster of Excellence ``Precision Physics,
Fundamental Interactions and Structure of Matter'' (PRISMA+ EXC
2118/1), funded within the German Excellence strategy (Project ID 39083149).
The work of M.C.\ has been supported by the European Union's Horizon 2020
research and innovation program under the Marie Skłodowska-Curie Grant
Agreement No.\ 843134.
A.G. received funding from the Excellence Initiative of Aix-Marseille University - A*MIDEX, a French \enquote{Investissements d'Avenir} programme, AMX-18-ACE-005 and from the French National Research Agency under the contract ANR-20-CE31-0016.

\appendix
\section{Pseudoscalar meson and gradient flow observables}
\label{sec:meson}

\begin{table}[b!]
  \centering
  \begin{adjustbox}{center}
  \begin{tabular}{rS[table-format=1.5(2)]S[table-format=1.5(2)]S[table-format=1.5(2)]S[table-format=1.5(2)]}
    \toprule
    & {$am_\pi$} & {$am_K$} & {$af_\pi$} & {$af_K$} \\
    \midrule
    \begin{filecontents*}{tables/mesons.csv}
label,aMpi,aMK,amll,amls,afpi,afK
H101,0.18356(48),0.18356(48),0.009184(42),0.009184(42),0.06387(19),0.06387(19)
H102,0.15457(55),0.19164(48),0.006456(57),0.010184(54),0.06035(29),0.06392(21)
H105,0.12353(128),0.20251(87),0.004033(73),0.011355(62),0.05824(48),0.06446(33)
N101,0.12236(47),0.20189(28),0.004009(29),0.011330(24),0.05748(23),0.06420(19)
C101,0.09616(65),0.20578(32),0.002422(26),0.011806(23),0.05459(27),0.06342(15)
B450,0.16108(43),0.16108(43),0.008100(31),0.008100(31),0.05643(16),0.05643(16)
S400,0.13592(43),0.17056(38),0.005679(32),0.009143(30),0.05381(35),0.05701(26)
N401,0.10997(59),0.17786(32),0.003769(34),0.010092(27),0.05218(33),0.05730(31)
N451,0.11089(29),0.17827(18),0.003783(18),0.010096(14),0.05216(16),0.05767( 9)
D450,0.08362(39),0.18393(18),0.002139(19),0.010795(18),0.04967(25),0.05741(11)
H200,0.13633(47),0.13622(67),0.006881(26),0.006881(26),0.04750(26),0.04752(27)
N202,0.13423(30),0.13421(29),0.006859(24),0.006859(24),0.04832(16),0.04833(16)
N203,0.11266(23),0.14413(19),0.004761(21),0.007899(12),0.04634(12),0.04907(11)
N200,0.09234(28),0.15076(21),0.003140(15),0.008633(13),0.04414(14),0.04907(14)
D200,0.06515(28),0.15615(16),0.001551( 8),0.009379( 6),0.04217(12),0.04910(11)
E250,0.04217(28),0.15924( 8),0.000645( 8),0.009763( 5),0.04008(22),0.04852(10)
N300,0.10618(24),0.10618(24),0.005509(13),0.005509(13),0.03803(13),0.03803(13)
N302,0.08725(34),0.11373(32),0.003720(10),0.006400( 9),0.03632(15),0.03854(15)
J303,0.06481(23),0.11964(20),0.002043( 8),0.007181(20),0.03428(11),0.03866(14)
E300,0.04367(16),0.12372(13),0.000917( 5),0.007724( 4),0.03237(12),0.03817(23)
\end{filecontents*}
\csvreader[
      late after line=\ifthenelse{\equal{\CLSlabel}{B450}\or\equal{\CLSlabel}{H200}\or\equal{\CLSlabel}{N300}}{\\\midrule}{\\},
      filter not strcmp={\CLSlabel}{N401},
    ]{tables/mesons.csv}%
      {label=\CLSlabel,aMpi=\aMpi,aMK=\aMK,amll=\amll,amls=\amls,afpi=\afpi,afK=\afK}%
      {\CLSlabel & \aMpi & \aMK & \afpi & \afK}
    \bottomrule
  \end{tabular}%
  \begin{tabular}{r@{}S[table-format=1.3(2)]S[table-format=2.3(2)]}
    \toprule
    & {$t_0/a^2$} & {$(w_0/a)^2$} \\
    \midrule
    \begin{filecontents*}{tables/t0w0.csv}
label,t0,w0sq
H101, 2.847( 5), 3.694( 9)
H102, 2.881( 6), 3.759(11)
H105, 2.889( 7), 3.779(14)
N101, 2.890( 2), 3.789( 5)
C101, 2.913( 3), 3.836( 6)
B450, 3.663( 8), 4.845(18)
S400, 3.686( 7), 4.895(15)
N401, 3.674( 5), 4.874(10)
N451, 3.682( 2), 4.896( 5)
D450, 3.696( 3), 4.933( 6)
H200, 5.151(16), 7.003(42)
N202, 5.165(12), 7.020(29)
N203, 5.146( 6), 6.982(15)
N200, 5.164( 6), 7.040(14)
D200, 5.179( 3), 7.099( 7)
E250, 5.203( 2), 7.176( 6)
N300, 8.544(19),11.821(48)
N302, 8.526(19),11.784(49)
J303, 8.621(10),12.101(31)
E300, 8.622( 6),12.163(15)
\end{filecontents*}
\csvreader[
      late after line=\ifthenelse{\equal{\CLSlabel}{B450}\or\equal{\CLSlabel}{H200}\or\equal{\CLSlabel}{N300}}{\\\midrule}{\\},
      filter not strcmp={\CLSlabel}{N401},
    ]{tables/t0w0.csv}%
      {label=\CLSlabel,t0=\tnot,w0sq=\wnotsq}%
      { & \tnot & \wnotsq}
    \bottomrule
  \end{tabular}
  \end{adjustbox}
  \caption{%
    Meson masses and decay constants, and reference scales $t_0$ and $w_0^2$.
  }\label{tab:meson}
\end{table}

Thanks to the high statistics on the vector correlator, the results presented in section~\ref{sec:lat_res} have a better than \SI{1}{\percent} precision on $\SPi(-Q^2)$ for all ensembles.
This is comparable to the statistical error of the meson masses and decay constants on the ensembles in table~\ref{tab:ensemble}, therefore it is sensible to include this source of uncertainty in the extrapolation to the physical point.
To implement the strategy presented in section~\ref{sec:extrapolation}, we performed a dedicated computation of pseudoscalar density and axial current two-point functions, that we used to obtain $m_\pi$, $m_K$, $f_\pi$ and $f_K$ to subpercent precision.
Crucially, this allows us to include the correlation between these observables and the values of $\SPi(-Q^2)$ on the same ensemble into the analysis.
The axial current is non-perturbatively $\order{a}$-improved using $c_A$ from ref.~\cite{Bulava:2016ktf} and renormalized using $Z_A$ and $b_A$ from refs.~\cite{DallaBrida:2018tpn,Korcyl:2016ugy}.
The values of the meson masses and decay constants are tabulated in table~\ref{tab:meson}.

For a similar reason, we computed the gradient-flow quantities $t_0$~\cite{Luscher:2010iy} and $w_0$~\cite{Borsanyi:2012zs} that set the scale, entering in the $Q^2$ scale that is input in the kernel and in positioning of the ensemble in the $(m_\pi,m_K)$ plane, see also figure~\ref{fig:landscape}.
We use the same procedure as in ref.~\cite{Bruno:2014jqa}.
Specifically, to minimize the effect of boundary effects and additional discretization effects from the boundary, on open \acp{BC} ensembles we use only the value of the action density with clover-type discretization $E(t,t_{\mathrm{fl}})$ on a single time slice at $t=T/2$.
On periodic \acp{BC} ensembles the four-dimensional volume average is used.
The values of $t_0$ and $w_0^2$ are also tabulated in table~\ref{tab:meson}, and, in the case of $t_0$, they agree within errors with the values quoted in refs.~\cite{Bruno:2014jqa,Bruno:2016plf}.
In the following, we choose to use $t_0$.

The meson masses and meson decay constants in table~\ref{tab:meson} are also affected by finite-size effects.
These effects are reliably computed in \ac{chiPT}~\cite{Colangelo:2005gd} and listed in table~\ref{tab:FSE_meson}.
They amount to a positive, permil-level shift on the masses and a negative sub-percent shift on the decay constants.

\begin{table}[tb]
  \centering
  \begin{tabular}{rS[table-format=1.3]S[table-format=1.3]S[table-format=+1.2]S[table-format=+1.2]}
    \toprule
    {$\delta$ [\si{\percent}]} & {$m_\pi$} & {$m_K$} & {$f_\pi$} & {$f_K$} \\
    \midrule
    \csvreader[
      late after line=\ifthenelse{\equal{\CLSlabel}{B450}\or\equal{\CLSlabel}{H200}\or\equal{\CLSlabel}{N300}}{\\\midrule}{\\},
      filter not strcmp={\CLSlabel}{N401},
    ]{tables/FSEcorr_contribs.csv}%
      {label=\CLSlabel,Mpi=\Mpi,MK=\MK,Fpi=\Fpi,FK=\FK}%
      {\CLSlabel & \Mpi & \MK & \Fpi & \FK}
    \bottomrule
  \end{tabular}
  \caption{%
    Finite-size effects on mesonic observables.
  }\label{tab:FSE_meson}
\end{table}

\section{Autocorrelation study}
\label{sec:autocorrelation}

We employ the $\Gamma$-method \cite{Wolff:2003sm,DePalma:2017lww} to estimate the integrated autocorrelation time, $\tau_\text{int}$ associated to $\Delta\alpha$ for each ensemble.
Then, we use this information to estimate a common bin size. Taking into account all autocorrelations, the expected increase of the uncertainty is estimated as
\begin{equation}
\label{eq:auto:growth-error}
    \Delta\Pi/\Delta\Pi_\text{naive} = \sqrt{2 \cdot \tau_\text{int}} \,,
\end{equation}
where $\Delta\Pi$ is the true standard deviation and $\Delta\Pi_\text{naive}$ is the uncertainty without taking into account autocorrelations.
The next step is to plot the ratio $\Delta\Pi(B)/\Delta\Pi_\text{naive}$ vs. the bin size, $B$.
For a similar analysis see ref.~\cite{Kelly:2019wfj}.
We find perfect agreement between \eqref{eq:auto:growth-error} and the region where $\Delta\Pi(B)/\Delta\Pi_\text{naive}$ plateaus.

The noise of the correlator at long distances could, in principle, hide the autocorrelations, artificially reducing $\tau_\text{int}$.
To avoid this we repeat the analysis several times, each time discarding more time slices at the tail of the correlator.
However, we found that the result is largely independent of the amount of noise removed.
We also observe a clear trend to smaller bin sizes as the physical point is approached.
While this effect is in apparent contrast with the expected $\sim 1/a^2$ scaling with lattice spacings~\cite{Bruno:2014jqa}, it is most likely explained with the noisier observables and shorter \ac{MC} chains of ensembles at finer lattice spacings and lighter pion masses.

\begin{table}[tb]
  \centering
  \begin{tabular}{rS[table-format=2]S[table-format=1.2(2)]}
    \toprule
    & {bin size} & {$\tau_\text{int}$} \\
    \midrule
    \begin{filecontents*}{tables/autocorrelations_bin.csv}
id,binsize,tauint
H101,25,1.70(26)
H102,25,1.73(27)
H105,20,1.32(27)
N101,15,0.79(11)
C101,20,0.79(10)
B450,25,1.45(24)
S400,20,2.17(32)
N451,10,0.73(10)
D450,5,0.55(7)
H200,30,1.20(19)
N202,35,1.86(45)
N203,20,1.15(17)
N200,15,0.77(10)
D200,10,0.58(6)
E250,5,0.47(4)
N300,40,3.36(67)
N302,30,2.07(33)
J303,20,1.41(26)
E300,20,1.07(22)
\end{filecontents*}
\csvreader[
      head to column names,
      late after line=\ifthenelse{\equal{\id}{B450}\or\equal{\id}{H200}\or\equal{\id}{N300}}{\\\midrule}{\\},
      filter ifthen={\equal{\id}{H101}\or\equal{\id}{H102}\or\equal{\id}{H105}\or\equal{\id}{N101}\or\equal{\id}{C101}\or%
                     \equal{\id}{B450}\or\equal{\id}{S400}\or\equal{\id}{N451}\or\equal{\id}{D450}},
    ]{tables/autocorrelations_bin.csv}{}
    {\id & \binsize & \tauint}
    \bottomrule
  \end{tabular}\qquad
  \begin{tabular}{rS[table-format=2]S[table-format=1.2(2)]}
    \toprule
    & {bin size} & {$\tau_\text{int}$} \\
    \midrule
    \csvreader[
      head to column names,
      late after line=\ifthenelse{\equal{\id}{B450}\or\equal{\id}{H200}\or\equal{\id}{N300}}{\\\midrule}{\\},
      filter ifthen={\equal{\id}{H200}\or\equal{\id}{N202}\or\equal{\id}{N203}\or\equal{\id}{N200}\or\equal{\id}{D200}\or\equal{\id}{E250}\or%
                     \equal{\id}{N300}\or\equal{\id}{N302}\or\equal{\id}{J303}\or\equal{\id}{E300}},
    ]{tables/autocorrelations_bin.csv}{}
    {\id & \binsize & \tauint}
    \bottomrule
  \end{tabular}
  \caption{%
    Bin size and integrated autocorrelation time in units of saved configurations for each of the ensembles included in our study.
    Configurations are saved every 4 \acfp{MDU} (2 trajectories of length 2 \acp{MDU}), except for J303, for which configurations are separated by 8 \acp{MDU} (4 trajectories).
  }
  \label{tab:bin-tau}
\end{table}

\section{Treatment of quark-disconnected diagrams}
\label{sec:disconnected}

In this appendix we collect further technical details regarding the evaluation of quark loops
\begin{equation}
 L_{\mathcal{O}_f}(\vec{p}, t) = \sum_{\vec{x}} e^{i\vec{p}\cdot\vec{x}} \ev{ \mathcal{O}_f(\vec{x}, t) }_F ,
 \label{eq:1pt_app}
\end{equation}
for some operator $\mathcal{O}_f(\vec{x},t)$ involving a single quark flavor $f$, which represents the computationally most expensive part of this study. A simple all-to-all estimator can be constructed using stochastic (four-dimensional) volume sources $\xi_i$, $i=1,...,N_s$ which satisfy
\begin{equation}
 \mathbb{E}[\xi_i^\dag \xi_j] = \delta_{ij} \,,
 \label{eq:stoch_volume_sources}
\end{equation}
A simple estimator for the quark loop function in eq.~(\ref{eq:1pt_app}) is then given by the average over $N_s$ noise sources,
\begin{equation}
 L_{\mathcal{O}_f}(\vec{p}, t) = \frac{1}{N_s}\sum_{s=1}^{N_s} \sum_{\vec{x}}
e^{i\vec{p}\cdot\vec{x}} \tr\left[\xi^\dag_s \Gamma_\mathcal{O} (D_f)_{xx}^{-1} \xi_s\right] ,
 \label{eq:1pt_naive_estimator}
\end{equation}
where $D_f$ denotes the Dirac operator for a single quark flavor and $\Gamma_\mathcal{O}$ the desired combination of Dirac matrices corresponding to a local (bilinear) operator $\mathcal{O}_f(x)=\bar{\psi}(x)\Gamma\psi(x)$. The generalization to non-local operators (\ie\ point-split currents or operators involving derivatives) is straightforward and does not require additional inversions. However, the resulting statistical error for this naive estimator behaves as $1/\sqrt{N_s}$, implying an insufficient rate of convergence for many observables (\eg\ vector currents) with respect to the computational cost required to saturate to gauge noise. \par

Therefore, we have computed the quark-disconnected loops using a variant of the
method introduced in ref.~\cite{Giusti:2019kff} combining the one-end trick
(OET) \cite{McNeile:2006bz} with a combination of the generalized hopping
parameter expansion (gHPE) \cite{Gulpers:2013uca} and hierarchical probing
\cite{Stathopoulos:2013aci}.
We write the difference between the quark loop function of operators $\mathcal{O}_1$ and $\mathcal{O}_2$ with different flavors (with different quark masses $m_1$ and $m_2$ respectively) using eq.~\eqref{eq:oet} to get the OET estimator
\begin{equation}
  L_{\mathcal{O}_1}(\vec{p}, t) - L_{\mathcal{O}_2}(\vec{p}, t) = (m_1 - m_2) \frac{1}{N_s}\sum_{s=1}^{N_s} \sum_{\vec{x}} \sum_{y}
e^{i\vec{p}\cdot\vec{x}} \tr\left[\xi^\dag_s (D_2)_{yx}^{-1} \Gamma_\mathcal{O} (D_1)_{xy}^{-1} \xi_s\right] ,
\end{equation}
where in the r.h.s.\ we have used the cyclic property of the trace and inserted the volume sources satistfing eq.~\eqref{eq:stoch_volume_sources} at the other end of the propagator product with respect to the local operator insertion.
Using the OET is it possible to saturate to the gauge noise for all relevant observables using at most $\mathcal{O}(100)$ sources. Note that this does not require any spin or color dilution but is achieved with plain stochastic $4d$ sources, which greatly reduces the computational cost compared to \eg\ plain hierarchical probing.
Using the gHPE, cf.\ eqs.~(\ref{eq:gHPE}-\ref{eq:HPE_M_and_H}), we generalize
the method of ref.~\cite{Giusti:2019kff} to apply also to point-split currents
by evaluating the first term on the r.h.s.\ of eq.~(\ref{eq:gHPE}) using hierarchical probing on spin and color diluted stochastic volume sources. We find that $N_h=512$ probing vectors are sufficient to reach gauge noise for non-local operators, while for local operators the saturation occurs already at $N_h=32$. For the second term it is sufficient to use naive stochastic volume sources, and the required inversion can be reused in the evaluation of $\mathrm{tr}\left[\Gamma (D_{N-1}^{-1} - D_{N}^{-1})\right]$, \ie\ the last term of the chain of OET estimators. Regarding the order of the gHPE, we found a choice of $n=2$ to be most effective. In practice, we use always four quark flavors, \ie\ light, strange, an additional, intermediate valence quark and a charm valence quark. For the (bare) mass of the intermediate quark flavor $i$ we found that the following choice for the value of the intermediate $\kappa_i$
\begin{equation}
 \frac{1}{\kappa_i} = \frac{1-X}{\kappa_s} + \frac{X}{\kappa_c} \,, \quad X=1/4 \,,
 \label{eq:kappa_i}
\end{equation}
works well for our ensembles. The values for $\kappa_{s,c}$ are listed in table~\ref{tab:ensembles_disc} and most of the values for the charm quark have been previously published in ref.~\cite{Gerardin:2019rua}. We use 512 stochastic volume sources for the light quark and double this number for each heavier quark flavor. Moreover, we average over stochastic sources such that effectively four independent blocks remain, which we found to be an acceptable compromise between storage requirements and the resulting loss of effective statistics in (unbiased) estimators for \eg\ the quark-disconnected contribution to meson two-point functions. Since certain other projects require position space data on large lattices such a compromise is needed, \ie\ to keep overall storage consumption at a feasible level when storing lattice-wide objects involving all sixteen local operators. One-link displaced operators have only been stored in momentum space.

\begin{table}[tb]
 \centering
 \begin{adjustbox}{center}
 \begin{tabular}{lS[table-format=1.9]S[table-format=1.16]S[table-format=1.20]S[table-format=1.6]S[table-format=2]}
 \toprule
       & {$\kappa_\ell$} & {$\kappa_s$} & {$\kappa_i$} & {$\kappa_c$} & {$q_\mathrm{max}^2$} \\
 \midrule
  H102 & 0.136865    & 0.136549339        & 0.13290161176800299531 & 0.123041 & 12 \\
  H105 & 0.136970    & 0.13634079         & 0.13281239106145185162 & 0.123244 & 12 \\
  N101 & 0.136970    & 0.13634079         & 0.13281239106145185162 & 0.123244 & 16 \\
  C101 & 0.137030    & 0.136222041        & 0.13276176090690750203 & 0.123361 & 16 \\
 \midrule
  S400 & 0.136984    & 0.136702387        & 0.13364790438310457391 & 0.125252 & 12 \\
  N451 & 0.1370616   & 0.1365480771       & 0.13359033405198803699 & 0.125439 & 25 \\
  D450 & 0.137126    & 0.136420428639937  & 0.13353998320731906973 & 0.125585 & 36 \\
 \midrule
  N203 & 0.137080    & 0.136840284        & 0.13443858045576029663 & 0.127714 & 16 \\
  N200 & 0.137140    & 0.13672086         & 0.13439191552174857156 & 0.127858 & 16 \\
  D200 & 0.137200    & 0.136601748        & 0.13434086201784468360 & 0.127986 & 16 \\
  E250 & 0.137232867 & 0.136536633        & 0.13431178258222831433 & 0.128052 & 36 \\
 \midrule
  N302 & 0.137064    & 0.1368721791358    & 0.13515349063190918961 & 0.130247 & 16 \\
  J303 & 0.137123    & 0.1367546608       & 0.13509842962222663237 & 0.130362 & 16 \\
  E300 & 0.137163    & 0.1366751636177327 & 0.13505900450196004851 & 0.130432 & 36 \\
 \bottomrule
 \end{tabular}
 \end{adjustbox}
  \caption{Values of (valence) $\kappa_{\ell,s,i,c}$ for the ensembles for which disconnected loops have been produced, together with the lattice momentum cutoff $q_\mathrm{max}^2$ in units of $(2\cpi/L)^2$ up to which momentum space data has been saved for both local and one-link displaced operators.}
 \label{tab:ensembles_disc}
\end{table}

\begin{figure}[t]
 \centering
 \includegraphics[width=0.67\columnwidth]{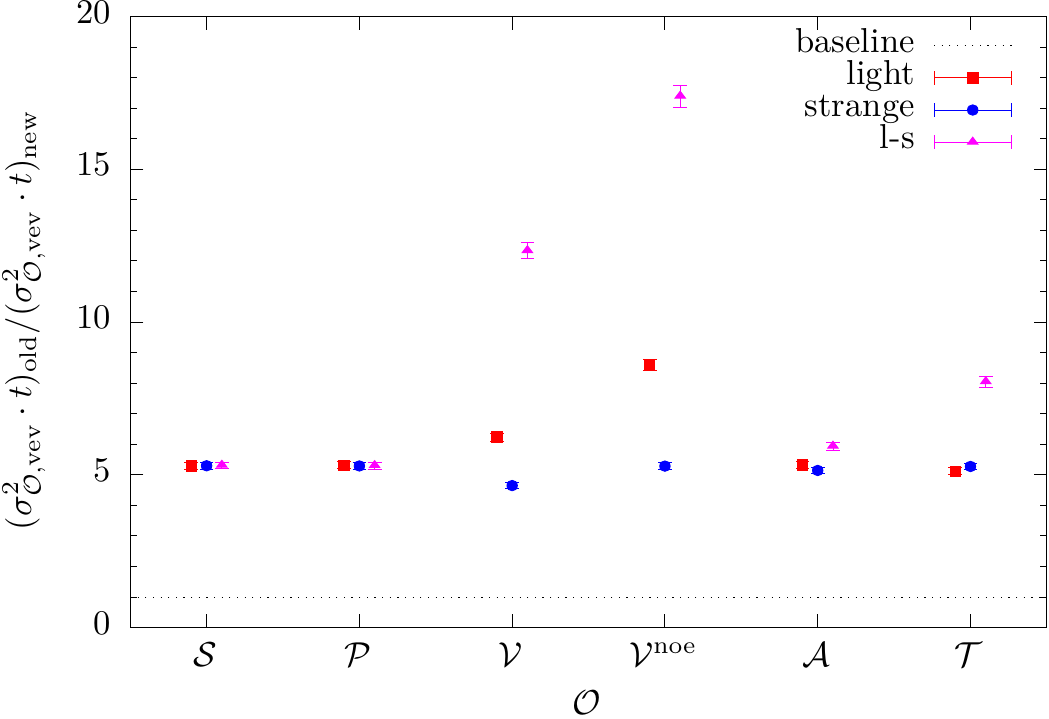}
 \caption{Speedup of the ``new'' (OET + gHPE + HP) method over the ``old'' (plain hierarchical probing) setup  for scalar ($\mathcal{S}$), pseudoscalar ($\mathcal{P}$), vector ($\mathcal{V}$), conserved vector ($\mathcal{V}^\mathrm{noe}$), axialvector ($\mathcal{A}$) and tensor ($\mathcal{T}$) quark bilinear operators. Results are shown for individual light and strange flavors as well as for the $l-s$ combination obtained directly from the OET. For operators that had reached gauge noise already for the old setup a speedup of roughly a factor five is obtained, while for the $l-s$ combination of the vector currents speedups of more than an order of magnitude are observed.}
 \label{fig:speedup}
\end{figure}

The resulting method is significantly more efficient than \eg\ plain hierarchical probing as applied in refs.~\cite{Djukanovic:2019jtp,Gerardin:2019rua}. While the statistical errors for the scalar, pseudoscalar and axial vector currents are very consistent between the two methods, this is not the case for the vector and tensor currents. Considering the products of variance $\sigma_{\mathcal{O}, \mathrm{vev}}^2$ and computational cost $t$, an operator-dependent, effective speedup $s_\mathcal{O}$ can be defined as the ratio of this quantity for the ``old'' (hierarchical probing) and the ``new'' (OET+gHPE+HP)-based method
\begin{equation}
 s_\mathcal{O} = \frac{(\sigma^2_{\mathcal{O}, \mathrm{vev}} t)_\mathrm{old}}{(\sigma^2_{\mathcal{O}, \mathrm{vev}} t)_\mathrm{new}}\,,
 \label{eq:effective_speedup}
\end{equation}
where $\sigma^2_{\mathcal{O}, \mathrm{vev}}$ denotes the variance of the vacuum expectation values at zero-momentum, \ie\ for the observable
\begin{equation}
 \mathrm{vev} = \sum_t \ev{ L_\mathcal{O}(\vec{0}, t) } .
\end{equation}
Assuming independent measurements (\ie\ a sufficiently large set of gauge configurations), this ratio determines the difference in absolute computational cost for the two methods for a given statistical target precision. The results are shown in fig.~\ref{fig:speedup} for the same set of operators and flavors. Depending on the observable this effective speedup reaches an order of magnitude. The largest difference between the two methods is observed for the $l-s$ combination of vector currents with an improvement of more than $17$ in case of the conserved vector current. \par

\section{Phenomenological model of the \texorpdfstring{$\SPi^{08}$}{̅Π⁰⁸} contribution}
\label{sec:model_08}

Modelling the $\SPi^{08}$ contribution using experimental data as input requires some assumptions.
In this appendix we describe the phenomenological model that we compare to the lattice result in figure~\ref{fig:running_comparison_08}.
Using the definitions in eqs.~\eqref{eq:corr_flavor_2},
\begin{equation}
  G^{\gamma\gamma}_{I=0} = \frac{1}{3} G^{88} \simeq \frac{1}{18} G^\omega + \frac{1}{9} G^\phi \,, \qquad
  G^{08} \simeq \frac{1}{2\sqrt{3}} \left[  G^\omega - G^\phi \right] ,
\end{equation}
where $G^\omega=C^{\ell,\ell}+2D^{\ell,\ell}$, $G^\phi=C^{s,s}$, and we neglected the disconnected diagram contributions $D^{\ell,s}$ and $D^{s,s}$.
The $I=0$ light and strange contributions are labeled with the vector meson that contributes the most to that channel, respectively the $\omega$ and the $\phi$.
Indeed, we can model the contribution to the hadronic cross-section in these channels with a narrow vector resonance contribution, plus the appropriate perturbative contribution
\begin{gather}
\label{eq:R_isoscalar_model}
  R^{\ell}_{I=0}(s) = \frac{1}{18} A_\omega m_\omega^2 \delta(s-m_\omega^2) + \theta(s-s_0) \frac{\Nc}{18} (1+\alpha_s/\cpi) \,, \\
  R^s(s)            = \frac{1}{9}  A_\phi m_\phi^2 \delta(s-m_\phi^2)       + \theta(s-s_1) \frac{\Nc}{9}  (1+\alpha_s/\cpi) \,,
\end{gather}
where the masses are $m_\omega=\SI{782.65(12)}{\MeV}$ and $m_\phi=\SI{1019.461(16)}{\MeV}$~\cite{Zyla:2020zbs}.
The amplitudes $A_\omega$ and $A_\phi$ can be estimated observing that the contribution for a narrow resonance is proportional to its electronic decay width $\Gamma(V\to e^+e^-)$, resulting in
\begin{equation}
  \frac{A_\omega}{18} = \frac{9\cpi}{\alpha^2} \frac{\Gamma(\omega\to e^+e^-)}{m_\omega} = \frac{\num{7.33(24)}}{18} \,, \qquad \frac{A_\phi}{9} = \frac{9\cpi}{\alpha^2} \frac{\Gamma(\phi\to e^+e^-)}{m_\phi} = \frac{\num{5.86(10)}}{9} \,,
\end{equation}
with $\Gamma(\omega\to e^+e^-)=\SI{0.60(2)}{\keV}$ and $\Gamma(\phi\to e^+e^-)=\SI{1.251(21)}{\keV}$~\cite{Zyla:2020zbs}.
Setting $\Nc=3$, $\alpha_s=\num{0.30}$ and the perturbative contribution thresholds $\sqrt{s_0}=\SI{1.02}{\GeV}$ and $\sqrt{s_1}=\SI{1.24}{\GeV}$, eqs.~\eqref{eq:R_isoscalar_model} yield a respectively \num{50.2e-10} and \num{53.4e-10} for their contribution to $\aHVP$.
We can add these contributions to get \num{103.6e-10} for the total $uds$ isoscalar contribution to $\aHVP$, where we have neglected the $\ell,s$-type disconnected diagrams, consistent with the findings of ref.~\cite{Bijnens:2016ndo}.
These numbers are in excellent agreement with lattice results~\cite{Gerardin:2019rua}.
We also consider the exact sum rule for the spectral function of $G^{08}$,
\begin{equation}
  \int_0^\infty \dd{s} R^{08}(s) = \frac{1}{2\sqrt{3}} \left[ A_\omega m_\omega^2 - A_\phi m_\phi^2 + \Nc \left(1+\frac{\alpha_s}{\cpi}\right) (s_1 - s_0) \right] = 0 \,,
\end{equation}
where $R^{08}(s)=[18R^{\ell}_{I=0}(s)-9R^{s}(s)]/(2\sqrt{3})$.
This is satisfied within errors by the model, with the resonances contributing \SI{-0.46(5)}{\GeV\squared} and the perturbative piece contributing \SI{0.47}{\GeV\squared}.

This gives us confidence in the model for the $\SPi^{08}$ mixing function
\begin{multline}
  \SPi^{08}_{\mathrm{model}}(-Q^2) = \frac{Q^2}{12\cpi^2} \int_0^\infty \dd{s} \frac{R^{08}(s)}{s(s+Q^2)} = \frac{1}{12\cpi^2} \frac{1}{2\sqrt{3}} \\
  \times \left[ 
      A_\omega \frac{Q^2}{m_\omega^2+Q^2} - A_\phi \frac{Q^2}{m_\phi^2+Q^2}
    + \Nc \left(1+\frac{\alpha_s}{\cpi}\right) \log(\frac{s_1(s_0+Q^2)}{s_0(s_1+Q^2)})
  \right] .
\end{multline}
The perturbative contribution cancels except for differences in the light and strange quark thresholds, that leads to a finite contribution in the $Q^2\to\infty$ limit.
With the inclusion of a $\pm\SI{100}{\MeV}$ correlated error in the $\sqrt{s_0}$ and $\sqrt{s_1}$ threshold that gives the error band in figure~\ref{fig:running_comparison_08}, the value is $\SPi^{08}_{\mathrm{pert.}}\to \Nc(1+\alpha_s/\cpi)\log(s_1/s_0)/(12\cpi^2 2\sqrt{3})=\num{0.00313(28)}$.
Similarly, the $\omega$ and $\phi$ resonances contribution in the $Q^2\to\infty$ limit is $\SPi^{08}_{\omega,\phi}\to (A_\omega-A_\phi)/(12\cpi^2 2\sqrt{3})=\num{0.00357(64)}$, for a total of \num{0.00669(70)}, with the $\order{\SI{10}{\percent}}$ error dominated by the experimental uncertainty on the $\omega$ and $\phi$ partial decay widths into $e^+e^-$.
At $Q^2=\SI{7}{\GeV\squared}$, our lattice result is $\SPi^{08}=\num{0.00704(17)}$ and very slowly varying with $Q^2$, only $\SI{2}{\percent}$ larger than the $Q^2=\SI{3}{\GeV\squared}$ value, that makes it compatible with the model value but significantly more precise.

\section{Phenomenological estimate of the charm quenching effect
\label{sec:CharmQuenching}}

Our goal in this appendix is to estimate the rough size of
 the effect of neglecting sea charm quarks on the running of
 $\alpha_{\mathrm{em}}$ up to $Q_0^2\approx\SI{5}{\GeV\squared}$.
To this end, we may split up the subtracted vacuum polarisation into two terms,
\begin{equation}
\SPi(-Q_0^2) = [\Pi(-Q_0^2) - \Pi(-\SI{1}{\GeV\squared})] + \SPi(-\SI{1}{\GeV\squared}) \,.
\end{equation}
The first term can be estimated using perturbation theory;
the dynamical charm quark effects enter at order $\alpha_s^2$ and is numerically small.
For the second term, we will see that the contribution of the $D \bar
D$ channels to the light-quark correlators amounts to a roughly
one-permil effect if one neglects the virtuality dependence of the $D$-meson form factor.

We assume that the pion and kaon masses ared used to set the $(u,d,s)$
quark masses, and that the same low-energy scale-setting quantity is
used in the $\Nf=2+1+1$ and the $\Nf=2+1$ theory. To be
clear, we do not claim to have a quantitative estimate of the charm
quenching effect on $\SPi(-Q_0^2)$, which represents a
non-perturbative problem. Rather, we have tried to identify some of
the differences between the two theories and assume the size of these
differences to be representative of the total quenching effect.

\subsection{\texorpdfstring{$D$}{D} meson loops \label{sec:Dloop}}
 
In the $\Nf=2+1+1$ theory, $D^+ D^-$, $D^0 \bar D^0$ and $D_s^+ D_s^-$ pairs can contribute to the
connected vector correlator of the $(u,d,s)$ quarks, while these contributions are absent in the $\Nf=2+1$ theory.
The real production of these heavy-light meson pairs makes a positive contribution to the spectral function above the
threshold of $\sqrt{s}=2m_D$. The contribution to the R-ratio is 
\begin{equation}
R_{D^+D^-}(s) = 
\frac{1}{4} \left(1-\frac{4m_D^2}{s}\right)^{3/2} \,|F_{D^+}(s)|^2
\end{equation}
and a similar expression for the $D^0 \bar D^0$ and $D_s^+D_s^-$ channels.
Since the form factors $F_D$ are not known precisely, we will set them to their values at $s=0$,
which amounts to treating these mesons in the scalar QED framework.
With the form factors $F_{D^+}(s)$ and $F_{D_s^+}(s)$ set to $1/3$, since that is
the charge of these mesons with respect to the $(u,d,s)$ electromagnetic current, and similarly
with $F_{D^0}(s)$ set to $-2/3$, we obtain 
for the subtracted HVP the contribution
\begin{equation}
  \delta\SPi^{\gamma\gamma}(-Q^2) = \frac{4}{9} f(Q^2/m_{D^0}^2) + \frac{1}{9} f(Q^2/m_{D^+}^2) + \frac{1}{9} f(Q^2/m_{D_s}^2) \,,
\label{eq:PiDloops}
\end{equation}
with\footnote{For $z\to 0$, we have $f(z)=\frac{z}{480\cpi^2} + \order*{z^2}$.} 
\begin{equation} \label{eq:fDloops}
f(z) = \frac{1}{144\cpi^2} \left[ -8(1+3/z) + 3 (1+4/z)^{3/2} \log\bigg(\frac{2+z + \sqrt{z (4+z )}}{2}\bigg)\right].
\end{equation}
For $Q^2=\SI{1}{\GeV\squared}$, we find
\begin{equation}\label{eq:PifromDloops}
\delta\SPi^{\gamma\gamma}(-\SI{1}{\GeV\squared}) = \num{0.39e-4}
\end{equation}
for the combined contribution of the  $D_s^+  D_s^-$, $D^+D^-$ and $D^0\bar D^0$  channels.
For comparison, the $(u,d,s)$-quark contribution to this quantity is roughly 0.040.
Thus the contribution \eqref{eq:PifromDloops} represents a one-permil effect.
For $Q^2=\SI{5}{\GeV\squared}$, we find
\begin{equation}\label{eq:Pi5fromDloops}
\delta\SPi^{\gamma\gamma}(-\SI{5}{\GeV\squared}) = \num{1.81e-4} \,,
\end{equation}
a $2.6$ permil effect compared to about $0.070$ for the $(u,d,s)$-quark contribution.
In particular,
\begin{equation}
\delta\SPi^{\gamma\gamma}(-\SI{1}{\GeV\squared}) - \delta\SPi^{\gamma\gamma}(-\SI{5}{\GeV\squared}) = \num{-1.42e-4} \,.
\label{eq:dPifromDloops}
\end{equation}
The numerical values provided here can be enhanced by the presence of
$D\bar D$ $1^{--}$ resonances, such as the $\psi(3770)$; however, the
latter is thought to be a `good $\bar c c$ resonance', therefore
having little coupling to the electromagnetic current carried by the
$(u,d,s)$ quarks.
As an analogy, it may also be worth noting that the magnitude of the
effective proton timelike form factor describing the cross-section
$e^+e^-\to \bar p p$ starts at about $0.4$ at
threshold~\cite{BESIII:2021rqk}, and thus well below its value at $s=0$.

The corresponding contribution for the running of the mixing angle is easily estimated considering the charges of the D mesons with respect to the weak isospin current, that results in
\begin{equation}
  \delta\SPi^{T_3\gamma}(-Q^2) = \frac{1}{12} \left[ 2 f(Q^2/m_{D^0}^2) + f(Q^2/m_{D^+}^2) + f(Q^2/m_{D_s}^2) \right] ,
\end{equation}
and using the relation $\delta\SPi^{Z\gamma}=\delta\SPi^{T_3\gamma}-\sIIW\delta\SPi^{\gamma\gamma}$.

\subsection{Perturbative estimate\label{sec:pertq}}

It is interesting to compare eq.~\eqref{eq:dPifromDloops} the difference of
eqs.~\eqref{eq:PifromDloops} and \eqref{eq:Pi5fromDloops} to the purely
perturbative prediction for the effect of unquenching the charm quark.

The required formulae for the contribution of heavy sea quarks to the
spectral function of the $(u,d,s)$ quarks can be found in ref.~\cite{Hoang:1994it}.
The `valence quarks' $(u,d,s)$,
\ie\ those coupling to the electromagnetic current, are treated as
massless.  There is a virtual correction to the spectral function
starting already at $s=0$, and a contribution corresponding to the
`real emission' of a $\bar cc$ pair, which opens at $s=4m^2$.  Using
the spectral representation, and setting $\alpha_s =0.30$, $m_c =
\SI{1.25}{\GeV}$ we find
\begin{equation}
\SPi^{\gamma\gamma}(-\SI{1}{\GeV\squared}) - \SPi^{\gamma\gamma}(-\SI{5}{\GeV\squared}) = \num{-0.19e-4} \,.
\end{equation}
This prediction is about seven times smaller than the prediction
in eq.~\eqref{eq:dPifromDloops} based on the $D$-meson loops treated in the
scalar QED framework.

\subsection{Change in the \texorpdfstring{$\omega$}{ω} and
\texorpdfstring{$\phi$}{φ} masses and decays constants due to mixing with
\texorpdfstring{$J/\psi$}{J/ψ}}

A further non-perturbative effect to be expected in QCD with dynamical
charm quarks is a small shift in the masses and decay constants of the
$\omega$ and $\phi$ mesons, relative to their respective values in
QCD without dynamical charm quarks.  Simply speaking, this may be
viewed as a result of the $\omega$ and $\phi$ mixing with the $J/\psi$
and possibly higher vector charmonium mesons.  Parametrically, this
effect is of order $1/m_{J/\psi}^2$, with a significant additional
suppression associated with the small rate at which $J/\psi$
decays. Even the mixing of the $\omega$ and $\phi$ mesons among
themselves is known to be very small in the single flavor quark basis; see for
instance the recent article~\cite{Volkov:2020jor} on this topic and references
therein.  Since it is difficult to isolate and quantitatively estimate
the effect, we refrain from doing so here.

\subsection{Synthesis}

The largest potential charm-quenching correction we have identified
comes from the $D$-meson loop, assuming conservatively
virtuality-independent form factors. We will therefore use that
correction as the basis of our estimate for the overall systematic
uncertainty associated with the neglect of dynamical charm quark
efects. Specifically, we will use
eqs.~\eqref{eq:PiDloops}--\eqref{eq:fDloops} to estimate the
charm-quenching error.  As for the sign of the effect of quenching the
charm quark, we remark that the specific effects considered in
sections~\ref{sec:Dloop} and~\ref{sec:pertq} both lead to
$\SPi(-Q^2)$ in the $\Nf=2+1$ theory being slightly
underestimated. Nevertheless, we will quote symmetric systematic
errors, since we are unable to perform a complete analysis.

As a final remark, we have not addressed the quark-disconnected
contributions to the electromagnetic-current correlator involving one
charm quark, \ie\ $2\ev{j_\mu^8 j_\nu^c}/(3\sqrt{3})$.  At
least in perturbation theory, such contributions are of order
$\alpha_s^3$, \ie\ of higher order than those considered in section
\ref{sec:pertq}.

\KOMAoption{captions}{tableheading}
\section{Extended table of results at the physical point}
\label{sec:supplementary}

The results for the $\SPi^{33}$, $\SPi^{88}$, $\SPi^{08}$ and $\SPi^{cc}$, and for the total \ac{HVP} contribution to the running of $\alpha$ and $\sIIW$, extrapolated to the physical point as explained in section~\ref{sec:extrapolation}, are given in tables~\ref{tab:supplementary1} and~\ref{tab:supplementary2} respectively for all the values of $Q^2$ sampled.

\begin{landscape}
\setlength{\LTcapwidth}{\columnwidth}
\setlength\LTleft{-5cm plus 5cm minus 5cm}
\setlength\LTright{-5cm plus 5cm minus 5cm}
\begin{longtable}{%
    c@{}S[table-format=1.1]S[table-format=1.6]%
    S[table-format=1.6]@{}>{$}r<{$}@{}>{$}r<{$}@{}>{$}r<{$}@{}>{$}r<{$}@{}>{$}r<{$}@{}>{$}r<{$}%
    S[table-format=1.6]@{}>{$}r<{$}@{}>{$}r<{$}@{}>{$}r<{$}@{}>{$}r<{$}@{}>{$}r<{$}@{}>{$}r<{$}%
    S[table-format=1.6]@{}>{$}r<{$}@{}>{$}r<{$}@{}>{$}r<{$}@{}>{$}r<{$}@{}>{$}r<{$}@{}>{$}r<{$}%
    S[table-format=1.7]@{}>{$}r<{$}@{}>{$}r<{$}@{}>{$}r<{$}@{}>{$}r<{$}@{}>{$}r<{$}@{}>{$}r<{$}%
  }
  \caption{%
    Results for the $\SPi^{33}$, $\SPi^{88}$, $\SPi^{08}$ and $\SPi^{cc}$ \ac{HVP} contribution extrapolated to the physical point for all the values of $Q^2$ sampled.
    The first quoted uncertainty is the statistical error, the second is the systematic error from varying the fit model estimated in section~\ref{sec:fit_syst}, the third is the scale-setting error (see section~\ref{sec:scale_setting}), and the fourth is the systematic from missing charm sea-quark loops (see section~\ref{sec:charm_quench}).
    The final uncertainty, quoted in square brackets, is the combination of the previous ones.
  }\label{tab:supplementary1} \\
  \toprule
  & {$Q^2$ [\si{\GeV\squared}]} & {$t_0Q^2$} %
  & \multicolumn{7}{c}{$\SPi^{33}$}  & \multicolumn{7}{c}{$\SPi^{88}$} %
  & \multicolumn{7}{c}{$\SPi^{08}$}  & \multicolumn{7}{c}{$\SPi^{cc}$} \\
  \midrule\endfirsthead
  \caption[]{(continued)} \\
  \toprule
  & {$Q^2$ [\si{\GeV\squared}]} & {$t_0Q^2$} %
  & \multicolumn{7}{c}{$\SPi^{33}$}  & \multicolumn{7}{c}{$\SPi^{88}$} %
  & \multicolumn{7}{c}{$\SPi^{08}$}  & \multicolumn{7}{c}{$\SPi^{cc}$} \\
  \midrule\endhead
  \midrule
  \multicolumn{31}{c}{continued} \\
  \bottomrule\endfoot
  \bottomrule\endlastfoot
  \csvreader[%
    late after line=\\,%
    filter=\lengthtest{\QsqGeV pt<7.01 pt},%
  ]{tables/running_results2.csv}%
    {Q2GeV=\QsqGeV, t0Q2=\tnotQsq,
     33_mean=\resttm, 33_stat=\restts, 33_fit=\resttf, 33_scale=\resttsc, 33_ib=\resttib, 33_cq=\resttcq, 33_comb=\resttcomb,%
     88_mean=\reseem, 88_stat=\resees, 88_fit=\reseef, 88_scale=\reseesc, 88_ib=\reseeib, 88_cq=\reseecq, 88_comb=\reseecomb,%
     08_mean=\reszem, 08_stat=\reszes, 08_fit=\reszef, 08_scale=\reszesc, 08_ib=\reszeib, 08_cq=\reszecq, 08_comb=\reszecomb,%
     cc_mean=\resccm, cc_stat=\resccs, cc_fit=\resccf, cc_scale=\resccsc, cc_ib=\resccib, cc_cq=\rescccq, cc_comb=\rescccomb,%
     Dalpha_mean=\resggm, Dalpha_stat=\resggs, Dalpha_fit=\resggf, Dalpha_scale=\resggsc, Dalpha_cq=\resggcq, Dalpha_ib=\resggib, Dalpha_comb=\resggcomb,%
     Ds2thW_mean=\resZgm, Ds2thW_stat=\resZgs, Ds2thW_fit=\resZgf, Ds2thW_scale=\resZgsc, Ds2thW_cq=\resZgcq, Ds2thW_ib=\resZgib, Ds2thW_comb=\resZgcomb}%
    {\footnotesize %
     & \QsqGeV & \tnotQsq
     & \resttm & \restts & \resttf & \resttsc & \resttcq &          & \resttcomb %
     & \reseem & \resees & \reseef & \reseesc & \reseecq &          & \reseecomb %
     & \reszem & \reszes & \reszef & \reszesc & \reszecq &          & \reszecomb %
     & \resccm & \resccs & \resccf & \resccsc & \rescccq &          & \rescccomb}
\end{longtable}
\end{landscape}

\begin{center}
\setlength{\LTcapwidth}{\columnwidth}
\setlength\LTleft{-5cm plus 5cm minus 5cm}
\setlength\LTright{-5cm plus 5cm minus 5cm}
\begin{longtable}{%
    c@{}S[table-format=1.1]S[table-format=1.6]%
    S[table-format=1.7]@{}>{$}r<{$}@{}>{$}r<{$}@{}>{$}r<{$}@{}>{$}r<{$}@{}>{$}r<{$}@{}>{$}r<{$}%
    S[table-format=+1.7]@{}>{$}r<{$}@{}>{$}r<{$}@{}>{$}r<{$}@{}>{$}r<{$}@{}>{$}r<{$}@{}>{$}r<{$}%
  }
  \caption{%
      Results for the total \ac{HVP} contribution to the running of $\alpha$ and $\sIIW$ extrapolated to the physical point for all the values of $Q^2$ sampled.
      Following the statistical error, and the systematic errors from varying the fit model estimated, scale-setting and missing charm sea-quark loops, the fifth uncertainty is the systematic error from missing isospin-breaking effects (see section~\ref{sec:isospin_breaking}).
    The final uncertainty, quoted in square brackets, is the combination of the previous ones.
  }\label{tab:supplementary2} \\
  \toprule
  & {$Q^2$ [\si{\GeV\squared}]} & {$t_0Q^2$} %
  & \multicolumn{7}{c}{$\Dalphahad$} & \multicolumn{7}{c}{$\Dhad\sIIW$} \\
  \midrule\endfirsthead
    \caption[]{(continued)} \\
  \toprule
  & {$Q^2$ [\si{\GeV\squared}]} & {$t_0Q^2$} %
  & \multicolumn{7}{c}{$\Dalphahad$} & \multicolumn{7}{c}{$\Dhad\sIIW$} \\
  \midrule\endhead
  \midrule
  \multicolumn{17}{c}{continued} \\
  \bottomrule\endfoot
  \bottomrule\endlastfoot
  \csvreader[%
    late after line=\\,%
    filter=\lengthtest{\QsqGeV pt<7.01 pt},%
  ]{tables/running_results2.csv}%
    {Q2GeV=\QsqGeV, t0Q2=\tnotQsq,
     33_mean=\resttm, 33_stat=\restts, 33_fit=\resttf, 33_scale=\resttsc, 33_ib=\resttib, 33_cq=\resttcq, 33_comb=\resttcomb,%
     88_mean=\reseem, 88_stat=\resees, 88_fit=\reseef, 88_scale=\reseesc, 88_ib=\reseeib, 88_cq=\reseecq, 88_comb=\reseecomb,%
     08_mean=\reszem, 08_stat=\reszes, 08_fit=\reszef, 08_scale=\reszesc, 08_ib=\reszeib, 08_cq=\reszecq, 08_comb=\reszecomb,%
     cc_mean=\resccm, cc_stat=\resccs, cc_fit=\resccf, cc_scale=\resccsc, cc_ib=\resccib, cc_cq=\rescccq, cc_comb=\rescccomb,%
     Dalpha_mean=\resggm, Dalpha_stat=\resggs, Dalpha_fit=\resggf, Dalpha_scale=\resggsc, Dalpha_cq=\resggcq, Dalpha_ib=\resggib, Dalpha_comb=\resggcomb,%
     Ds2thW_mean=\resZgm, Ds2thW_stat=\resZgs, Ds2thW_fit=\resZgf, Ds2thW_scale=\resZgsc, Ds2thW_cq=\resZgcq, Ds2thW_ib=\resZgib, Ds2thW_comb=\resZgcomb}%
    {& \QsqGeV & \tnotQsq
     & \resggm & \resggs & \resggf & \resggsc & \resggcq & \resggib & \resggcomb %
     & \resZgm & \resZgs & \resZgf & \resZgsc & \resZgcq & \resZgib & \resZgcomb}
\end{longtable}
\end{center}

\printbibliography

\end{document}